\pdfoutput=1

\documentclass[11pt,twoside,a4paper,cmspaper,final,collab]{cms-tdr}
\def\svnVersion{317871}\def\cmsCernNo{2013-037}\def\cmsCernDate{\today}\def\cmsMessage{Published in Physical Review D as \href{http://dx.doi.org/10.1103/PhysRevD.93.012001}{\doi{10.1103/PhysRevD.93.012001.}}}

\begin{document}\cmsNoteHeader{B2G-13-008}

\hyphenation{had-ron-i-za-tion}
\hyphenation{cal-or-i-me-ter}
\hyphenation{de-vices}
\RCS$Revision: 317851 $
\RCS$HeadURL: svn+ssh://svn.cern.ch/reps/tdr2/papers/B2G-13-008/trunk/B2G-13-008.tex $
\RCS$Id: B2G-13-008.tex 317851 2016-01-15 16:38:49Z rkogler $
\newlength\cmsFigWidth
\ifthenelse{\boolean{cms@external}}{\setlength\cmsFigWidth{0.85\columnwidth}}{\setlength\cmsFigWidth{0.4\textwidth}}
\ifthenelse{\boolean{cms@external}}{\providecommand{\cmsLeft}{top}\xspace}{\providecommand{\cmsLeft}{left}\xspace}
\ifthenelse{\boolean{cms@external}}{\providecommand{\cmsUpperLeft}{top}\xspace}{\providecommand{\cmsUpperLeft}{upper left}\xspace}
\ifthenelse{\boolean{cms@external}}{\providecommand{\cmsMiddle}{middle}\xspace}{\providecommand{\cmsMiddle}{upper right}\xspace}
\ifthenelse{\boolean{cms@external}}{\providecommand{\cmsRight}{bottom}\xspace}{\providecommand{\cmsRight}{right}\xspace}
\ifthenelse{\boolean{cms@external}}{\providecommand{\cmsBottom}{bottom}\xspace}{\providecommand{\cmsBottom}{bottom}\xspace}
\ifthenelse{\boolean{cms@external}}{\providecommand{\NA}{\ensuremath{\cdots}}\xspace}{\providecommand{\NA}{---\xspace}}
\newcommand{\Mttbar}{\ensuremath{M_{\ttbar}}\xspace}
\newcommand{\Mzp}{\ensuremath{M_{\PZpr}}\xspace}
\newcommand{\gKK}{\ensuremath{g_\mathrm{KK}}\xspace}
\newcommand{\GKK}{\ensuremath{G_\mathrm{{KK}}}\xspace}
\newcommand{\MgKK}{\ensuremath{M_{\gKK}}\xspace}
\newcommand{\rWzp}{\ensuremath{\Gamma_{\PZpr}/M_{\PZpr}}\xspace}

\newcommand{\mjet}{\ensuremath{M_{\text{jet}}}\xspace}
\newcommand{\mmin}{\ensuremath{M_{\text{min}}}\xspace}

\newcommand{\pb}{\unit{pb}\xspace}
\newcommand{\fb}{\unit{fb}\xspace}

\cmsNoteHeader{B2G-13-008}
\title{Search for resonant \texorpdfstring{\ttbar}{ttbar} production in proton-proton collisions
at \texorpdfstring{$\sqrt{s}=8\TeV$}{sqrt(s)=8 TeV}}

\date{\today}

\abstract{
A search is performed for the production of heavy resonances decaying into top-antitop quark pairs
in proton-proton collisions at $\sqrt{s}=8\TeV$.
Data used for the analyses were collected with the CMS detector and correspond to an integrated luminosity of
19.7\fbinv.
The search is performed using events with three different final states, defined by the number of leptons (electrons and muons)
from the $\ttbar \to \PW \cPqb \PW \cPqb$ decay.
The analyses are optimized for reconstruction of top quarks with high Lorentz boosts, where jet substructure techniques are used
to enhance the sensitivity.
Results are presented for all channels and a combination is performed.
No significant excess of events relative to the expected yield from standard model processes is observed.
Upper limits on the
production cross section of heavy resonances decaying to $\ttbar$ are calculated.
A narrow leptophobic topcolor \PZpr resonance with a mass below 2.4\TeV is excluded
at 95\% confidence level.
Limits are also derived for a broad \PZpr resonance with a 10\% width relative to the resonance mass, and a Kaluza--Klein excitation
of the gluon in the Randall--Sundrum model.
These are the most stringent limits to date on heavy resonances decaying into top-antitop quark pairs.
}

\hypersetup{%
pdfauthor={CMS Collaboration},%
pdftitle={Search for resonant ttbar production in proton-proton collisions
at sqrt(s)=8 TeV},%
pdfsubject={CMS},%
pdfkeywords={CMS, physics, B2G, top quark, BSM}}

\maketitle

\section{Introduction}
\label{sec:introduction}

The top quark is the heaviest known fundamental particle, with a mass
close to the electroweak scale.  It has a Yukawa
coupling to the Higgs potential close to unity,
and is, therefore, closely connected to
the hierarchy problem, where the largest corrections to the mass of
the Higgs boson arise from top-quark loops.  Studies of top-quark
production may provide further insight into the mechanism of
electroweak symmetry breaking, especially in the light of the recent
discovery of a Higgs boson~\cite{Aad:2012tfa,Chatrchyan:2012ufa,Chatrchyan:2013lba} and a precision measurement of its
mass~\cite{Aad:2014aba,Khachatryan:2014jba, Aad:2015zhl}.

Many theories beyond the
standard model (SM) predict the existence of heavy resonances, generically referred to as \PZpr, that preferentially
decay to \ttbar pairs and manifest themselves as a resonant component
on top of the SM \ttbar continuum production. Examples of such models include
colorons~\cite{theory_hill1,theory_hill_parke,theory_hill2}
including a leptophobic topcolor \PZpr~\cite{theory_jainharris},
extended gauge theories with massive color-singlet \PZpr
bosons~\cite{theory_rosner,theory_lynch,theory_carena},
axigluons~\cite{theory_frampton_glashow,theory_choudhury}, and models in which
a pseudoscalar Higgs boson may couple strongly to top quarks~\cite{theory_dicus}.
Furthermore, various extensions of the Randall--Sundrum model~\cite{theory_randall_sundrum}
with extra dimensions predict Kaluza--Klein (KK) excitations of
gluons $\gKK$~\cite{theory_agashe} or
gravitons $\GKK$~\cite{theory_davoudiasl}, both of
which can have enhanced couplings to \ttbar pairs.

Direct searches for heavy \ttbar resonances have been performed
at the Fermilab Tevatron and the CERN LHC colliders, with no evidence for such signals.
The experiments at the Tevatron 
have probed the mass range
up to about 900\GeV~\cite{cdf_ttbar_resonance1, cdf_ttbar_resonance2, cdf_ttbar_resonance3,
d0_ttbar_resonance1, cdf_ttbar_resonance4, d0_ttbar_resonance2}, using the leptophobic \PZpr model, 
and the LHC experiments have set subpicobarn limits
on the production cross section in the mass range of 1--3\TeV~\cite{cms_ttbar_resonance1,
atlas_ttbar_resonance1, atlas_ttbar_resonance2, cms_ttbar_resonance2, cms_ttbar_resonance3, cms_ttbar_resonance4}.

This paper presents a model-independent search for $\PZpr \to \ttbar \to \PWp \cPqb \PWm \cPaqb$ production,
where the leptonic and hadronic decay modes of the $\PW$ bosons are considered.
Unless otherwise indicated, the symbol \PZpr is used in the following
to refer to the resonance decaying to $\ttbar$, irrespective of the specific model.
This results in final states with two, one, or zero leptons, which are referred to as the dilepton, lepton+jets,
and all-hadronic channels, respectively.
The search is based on $\Pp\Pp$ collision data collected by the CMS
experiment at the LHC at a center-of-mass energy $\sqrt{s} = 8\TeV$, and corresponding
to an integrated luminosity of 19.7\fbinv.

The final state of the dilepton channel consists of two leptons of
opposite charge ($\Pe\Pe$, $\Pe \Pgm$, or $\Pgm\Pgm$) with
high transverse momentum (\pt), at least two jets from
the fragmentation of $\cPqb$ quarks,
and missing transverse momentum due to escaping neutrinos.
The final-state objects arising from decays of heavy $\ttbar$ resonances are collimated because of
the large Lorentz boosts of the top quark decay products.
Leptons from the $\PW$ boson decay are reconstructed in the proximity of jets from the
fragmentation of $\cPqb$ quarks.
Special selection criteria
are used to preserve high lepton selection efficiency for non-isolated leptons at high resonance masses.
The dominant
irreducible background is the \ttbar continuum production. Other
SM processes contributing to the background are single top quarks,
$\Z$+jets, and diboson production.

The final state considered in the lepton+jets channel consists of one
high-$\pt$ lepton ($\Pe$ or $\Pgm$),
at least two jets, of which at least one jet is identified to arise from
the fragmentation of a $\cPqb$ quark, and missing transverse momentum.
As in the case of the dilepton analysis, special selection criteria are used to identify non-isolated leptons at
high resonance masses.
A top-quark tagging algorithm, referred to as a $\cPqt$ tagging algorithm, is applied to identify fully hadronic decays of the
type $\cPqt \to \PW  \cPqb \to \cPq \cPaq^\prime \cPqb$ merged into one single jet. The use of
the $\cPqt$ tagging algorithm enhances the sensitivity of this channel at high resonance masses
by about 30\%--40\%, and leaves \ttbar continuum production as the dominant irreducible background.
A bottom-quark tagging algorithm is also used to create regions enhanced in signal for the analysis.

The all-hadronic channel considers events with a dijet topology, where two wide jets are 
selected and required to be consistent with the decay of a top quark. Two separate regions are explored:
a search region sensitive to \PZpr masses $M_{\PZpr}$ below 1\TeV,
where Cambridge-Aachen (CA) jets~\cite{CACluster1,CACluster2} with a distance parameter
of $R=1.5$ are considered, and a search region for high resonance masses, using
CA jets with $R=0.8$.
Two distinct $\cPqt$ tagging algorithms~\cite{Kaplan:2008ie,Plehn:2010st} are used for these two regions.
In both regions the
dominant background from
non-top quark multijet production can be reduced considerably by requiring one identified
subjet in each of the two top quark candidates to be consistent with
the fragmentation of a $\cPqb$ or $\cPqc$ quark, leaving irreducible SM \ttbar
continuum production as the dominant background.

Except for the non-top multijet backgrounds in the all-hadronic channels, the
shapes of all SM backgrounds are estimated from simulation.
The total yield of the simulated samples is obtained with a binned maximum
likelihood fit to the reconstructed \ttbar invariant mass ($\Mttbar$) distributions.
A limit on the production cross section of heavy resonances is extracted by performing
a template-based statistical evaluation of the $\Mttbar$ distributions
of all channels.

This paper is organized as follows:
Section~\ref{sec:detector} gives a description of the CMS detector.
The reconstruction and identification of electrons, muons, and jets is described in Section~\ref{sec:reco}.
Section~\ref{sec:reco} also gives an overview of the $\cPqt$ tagging algorithms used.
The data sets and simulated Monte Carlo (MC) samples used in the analysis are given
in Section~\ref{sec:datasets}.
Section~\ref{sec:selection} describes the event selection for the three different channels.
Systematic uncertainties are discussed in
Section~\ref{sec:systematics}, while
Section~\ref{sec:background} describes the evaluation of the SM background processes.
The statistical analysis and the results are given in Section~\ref{sec:results},
and a summary is given in Section~\ref{sec:conclusions}.

\section{CMS detector \label{sec:detector}}

The central feature of the CMS detector is a superconducting solenoid
of 6\unit{m} internal diameter, providing a magnetic field of
3.8\unit{T}. Within the solenoid volume are a silicon
pixel and strip tracker, a lead tungstate crystal electromagnetic
calorimeter, and a brass and scintillator hadron calorimeter,
each composed of a barrel and two endcap sections. In addition
to the barrel and endcap detectors, CMS has extensive forward
calorimetry.
Muons are detected by four layers of gas-ionization detectors embedded
in the steel flux-return yoke of the magnet.
The inner tracker measures charged particle
trajectories within the pseudorapidity range $\abs{\eta} < 2.5$, and
provides an impact parameter resolution of approximately 15\mum.
A two-stage trigger system selects
$\Pp\Pp$ collision events of interest for use in physics analyses.
A more detailed description of the CMS detector, together with a
definition of the coordinate system used and the relevant kinematic
variables, can be found in Ref.~\cite{Chatrchyan:2008zzk}.

\section{Event reconstruction}
\label{sec:reco}
The CMS experiment uses a particle-flow (PF) based event
reconstruction~\cite{CMS-PAS-PFT-09-001, CMS-PAS-PFT-10-001},
which aggregates input from all subdetectors.
This information includes charged-particle tracks from the tracking system and deposited
energy from the electromagnetic and hadronic calorimeters, taking advantage of excellent
granularity of the sub-systems.
Particles are classified as electrons, muons, photons, charged hadrons, and neutral hadrons.
Primary vertices are reconstructed
using a deterministic annealing filter algorithm~\cite{Chatrchyan:2014fea}.
The vertex with the largest squared sum of the associated track $\pt$ values
is taken to be the primary event vertex.

Electrons are reconstructed in the pseudorapidity range $\abs{\eta} < 2.5$,
by combining tracking information with energy deposits in the
electromagnetic calorimeter~\cite{electronreco, Chatrchyan:2013dga}.
Electron candidates are required to originate from the primary event vertex.
Electrons are identified using information on the shower shape, the
track quality, and the spatial match between the track and
electromagnetic cluster, and the fraction of total cluster energy in the
hadron calorimeter. Electron candidates that
are consistent with originating from photon conversions in the
detector material are rejected.

Muons are detected and measured in the pseudorapidity range $\abs{\eta} < 2.4$
using the information collected in the muon and tracker
detectors~\cite{Chatrchyan:2012xi}. Tracks from muon candidates
must be consistent with a muon originating
from the primary event vertex and satisfy track fit quality
requirements.

Since the top-quark decay products can be collimated at high values of $\Mzp$,
no isolation requirements on the leptons are imposed in either the trigger or offline selections.

The missing transverse momentum vector \ptvecmiss is defined as the projection on the
plane perpendicular to the beams of the negative vector sum of the momenta of all
reconstructed particles in an event. Its magnitude is referred to as \ETmiss.

Particle-flow candidates are
clustered into jets using the \FASTJET 3.0 software
package~\cite{FastJet}.
Charged hadrons associated with other event vertices than the primary event vertex
are removed prior to jet clustering.
All jets are required to satisfy $\abs{\eta} < 2.4$.
The dilepton and lepton+jets analyses use jets obtained by the anti-\kt
jet-clustering algorithm~\cite{Cacciari:2008gp} with a distance parameter
of 0.5 (AK5 jets). If a lepton candidate (electron or muon) is found within ${\Delta R < 0.5}$
of an AK5 jet, its four-momentum is subtracted from that of the jet.
In this paper the unmodified term `jet' will refer to these AK5 jets.
The all-hadronic analyses use the CA
jet-clustering algorithm~\cite{CACluster1,CACluster2} with distance
parameters of 0.8 (CA8 jets) and 1.5 (CA15 jets) for the analyses at
high and low values of the \ttbar invariant mass, respectively.
The CA algorithm has been chosen because of its use in the declustering 
of jets for the identification of jet substructure.
The CA8 jets are also employed in the lepton+jets analysis to identify
the hadronic decay of top quarks with high $\pt$ in the hemisphere
opposite to one defined by the momentum vector of the lepton.
All jets contain neutral particles from additional collisions in the beam crossing (pileup).
The extra contribution is subtracted based on the
average expectation of the pileup in the jet catchment area~\cite{Cacciari:2008gn}.
 This is done by calculating a correction for the average offset energy density in each event 
 as function of the number of primary vertices~\cite{Chatrchyan:2011ds}. 
Jets are identified as originating from the fragmentation of a $\cPqb$
or $\cPqc$ quark
by the combined secondary vertex algorithm (CSV).
The loose and medium operating points are used,
which were chosen to have a misidentification
probability of 10\% and 1\%, respectively, for tagging light-parton jets with an average 
\pt of about 80\GeV.
The efficiency for the medium operating point varies between 70\%--75\% 
for jet $\pt$ in the range 50--100\GeV, where it reaches a plateau. 
Above 200\GeV the efficiency decreases gradually to 
about 60\% for $\pt$ values of 500\GeV~\cite{Chatrchyan:2012jua}. 
All jets are required to satisfy quality selections to remove calorimeter noise and
other sources of fake jets~\cite{CMS-PAS-JME-10-003}.
Events are required to also satisfy selection criteria to remove calorimeter noise
from $\MET$ signals as described in Ref.~\cite{Chatrchyan:2011tn}.

The structure of CA jets is used to distinguish hadronically
decaying top quarks merged into a single jet from light quark or gluon jets.
For CA8 jets the CMS $\cPqt$ tagging algorithm is used~\cite{JME-09-001, JME-13-007},
which is based on an algorithm studied in Ref.~\cite{Kaplan:2008ie}.
Only jets with $\pt > 400\GeV$
are considered, as at lower momenta the decay products of the hadronically decaying
top quark are rarely merged into a single jet.
The algorithm attempts to split the merged jets into subjets. In the process, soft and
wide-angle particles relative to the parent in the clustering are ignored,
enhancing the separation into subjets.
CA8 jets that pass the CMS $\cPqt$ tagging algorithm (CA8 $\cPqt$-tagged jets)
are required to have at least three subjets.
The mass of the jet has to satisfy the condition $140 < \mjet < 250\GeV$, the minimum pairwise
mass $\mmin$ of the three highest $\pt$ subjets  is required to be greater than $50\GeV$, and
the $N$-subjettiness~\cite{Thaler:2010tr, Thaler:2011gf} ratio $\tau_{32} \equiv \tau_3/\tau_2$
must be smaller than the value of 0.7, which has been obtained from optimization studies.
The $N$-subjettiness observable $\tau_N$,
defined through the relation
\[
\tau_N = \frac{1}{d_0} \sum_i p_{\mathrm{T},i} \min\left[\Delta R_{1,i}, \Delta R_{2,i}, \cdots, \Delta R_{N,i}\right] ,
\]
is a measure of the consistency of a CA jet with
$N$ or fewer subjets, where $i$ is a sum over all jet constituents, and
the $\Delta R$ terms represent distances between a given constituent $i$ and one of the $N$ candidate
subjet axes. The quantity $d_0$ is a normalization constant.

The HEPTopTagger~\cite{Plehn:2010st} algorithm is applied to CA15 jets. The larger
distance parameter allows the identification of hadronic decays of top quarks with
intermediate transverse momenta, $\pt > 200\GeV$.
The CA15 jet is decomposed according to the last clustering step of the CA algorithm.
Subjets are identified by an iterative procedure: when undoing the last
clustering of the jet into two subjets, the mass of
the heavier subjet is required to be between 30\GeV and
80\% of the mass of the
original jet.
The algorithm fails if fewer than three subjets are found.
If three or more subjets are reconstructed, jet constituents are
reclustered using the CA algorithm with filtering~\cite{Butterworth:2008iy}
until there are exactly three subjets. Additional
criteria are applied to the invariant mass calculated from the three subjets and
the pairwise masses using combinations of the three subjets
to reject jets from light quarks or gluons~\cite{JME-13-007}.
Jets identified by the {\sc HEPTopTagger} are referred to as CA15 $\cPqt$-tagged jets~\cite{Khachatryan:2015axa}.

In the all-hadronic channel, additional discriminating power against
background processes is obtained from the application of the CSV algorithm
to the subjets of the CA jets.
A CA jet is considered to be $\cPqb$-tagged if the subjet with the highest discriminator value
satisfies the requirement for the medium operating point. This has an efficiency of about 65\% and a
misidentification probability of approximately 5\%.
In the following, this algorithm will be called subjet $\cPqb$ tagging.
Its performance has been studied in data, and shows a gain in efficiency for
boosted topologies with respect to the
standard $\cPqb$ tagging algorithm~\cite{CMS-PAS-BTV-13-001}.  The same study also compared
the $\cPqb$-quark efficiency in data and simulated events, and established
that the measured data-to-simulation scale factor for $\cPqb$-tagged subjets is the
same as for unmerged $\cPqb$ jets.

\section{Trigger and data sets}
\label{sec:datasets}

Dilepton events are collected with single-lepton triggers.
Events for the $\Pe\Pe$ channel are selected using a single electron trigger with a
\pt threshold of 80\GeV and an efficiency of 90\%. In all cases, no isolation requirement is applied to the leptons.
Similarly, $\Pe\Pgm$ and $\Pgm\Pgm$ events are recorded with a trigger requiring
a single muon with $\pt > 40$\GeV and $\abs{\eta}<2.1$.
The efficiency for this trigger is 95\% for muons measured within $\abs{\eta}<0.9$,
85\% if they are measured within $0.9<\abs{\eta}<1.2$ and 83\% for $1.2<\abs{\eta}<2.1$.

The data used in the lepton+jets channel also rely on single lepton
triggers. The trigger for electron events requires one electron with $\pt > 35$\GeV in
conjunction with two jets that have $\pt >100$ and $25$\GeV, respectively.
The trigger for muon events is the same one used in the
dilepton analysis.
In both cases, no isolation requirement is applied to the leptons.
A 10\% increase in the signal efficiency at $\Mzp=2\TeV$ is gained
in the electron channel by including
events that are triggered by a single jet with $\pt>320\GeV$.
The events recovered by the single-jet trigger contain an
electron merged in a jet, which can not
be resolved at the trigger level.
The resulting trigger efficiency is 90\% for events with a leading (highest $\pt$)
jet with $\pt<320\GeV$. Above this value the trigger shows a turn-on behavior 
and is fully efficient above a value of $350\GeV$.

The all-hadronic data sample is based on two different
triggers. The first requires the
scalar sum of the $\pt$ of jets ($\HT$) to be greater
than 750$\GeV$, with an efficiency of 95\% or higher after
the analysis selection.
The second requires four
jets with $\pt>50\GeV$
at trigger level, used to gain efficiency in the low mass regime with
$\Mzp<1\TeV$. The efficiency of this trigger is 50\% for events with
the fourth leading jet having $\pt>50\GeV$, and increases to 100\% for
jets with $\pt>100\GeV$.

The total integrated luminosity associated with the datasets is 19.7\fbinv, except for the
four-jet data set, which corresponds to an integrated luminosity of
18.3\fbinv. The lower integrated luminosity in the latter case
is due to the unavailability of the four-jet trigger at the start of data taking.
The efficiencies for all triggers are well modeled by the simulation.

The $\PZpr\to\ttbar$ process is simulated
using the \MADGRAPH 4.4~\cite{madgraph4} event generator, which produces a generic
high-mass resonance with the same left- and right-handed couplings to fermions as the
SM $\Z$ boson. Higher-order parton radiations are calculated for
up to three extra partons at tree level.
The simulation is performed for masses $\Mzp$
of 0.75, 1, 1.25, 1.5, 2, and 3\TeV,
and for relative decay widths $\rWzp$ of 1\% (narrow-width) and 10\% (wide-width).
Kaluza--Klein gluon excitations are simulated
using \PYTHIA 8~\cite{pythia8},
where interference effects with SM $\ttbar$ production are neglected.
The widths of the $\gKK$ signals
are about 15\%--20\% of the resonance mass.
For visualization purposes,
the signal samples are scaled to an arbitrary cross section of 1\pb.
This is about a factor of 30 larger than the cross section expected from the
narrow-width topcolor \PZpr model.

\begin{figure}[htb]
\centering
\includegraphics[width=0.48\textwidth]{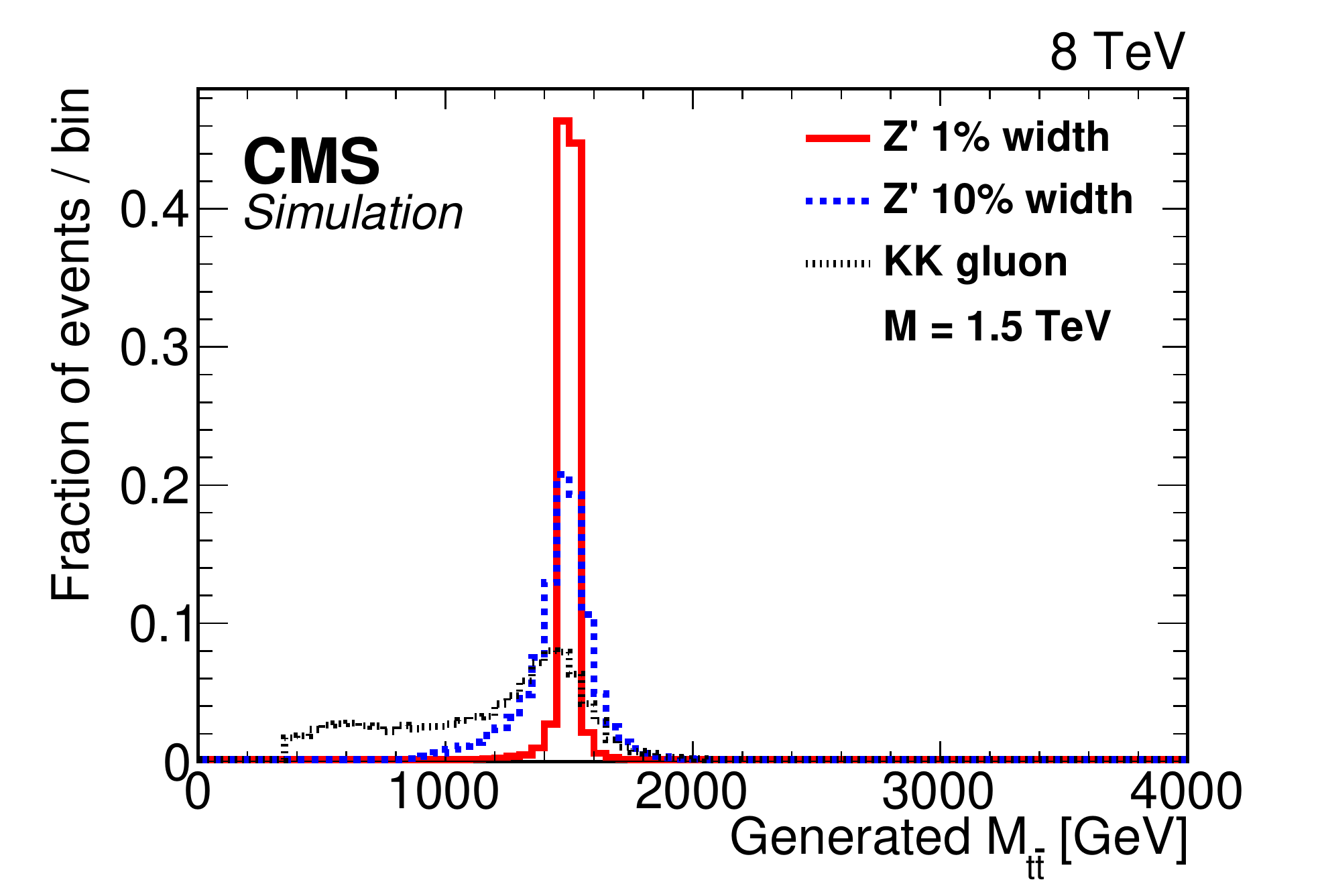}
\includegraphics[width=0.48\textwidth]{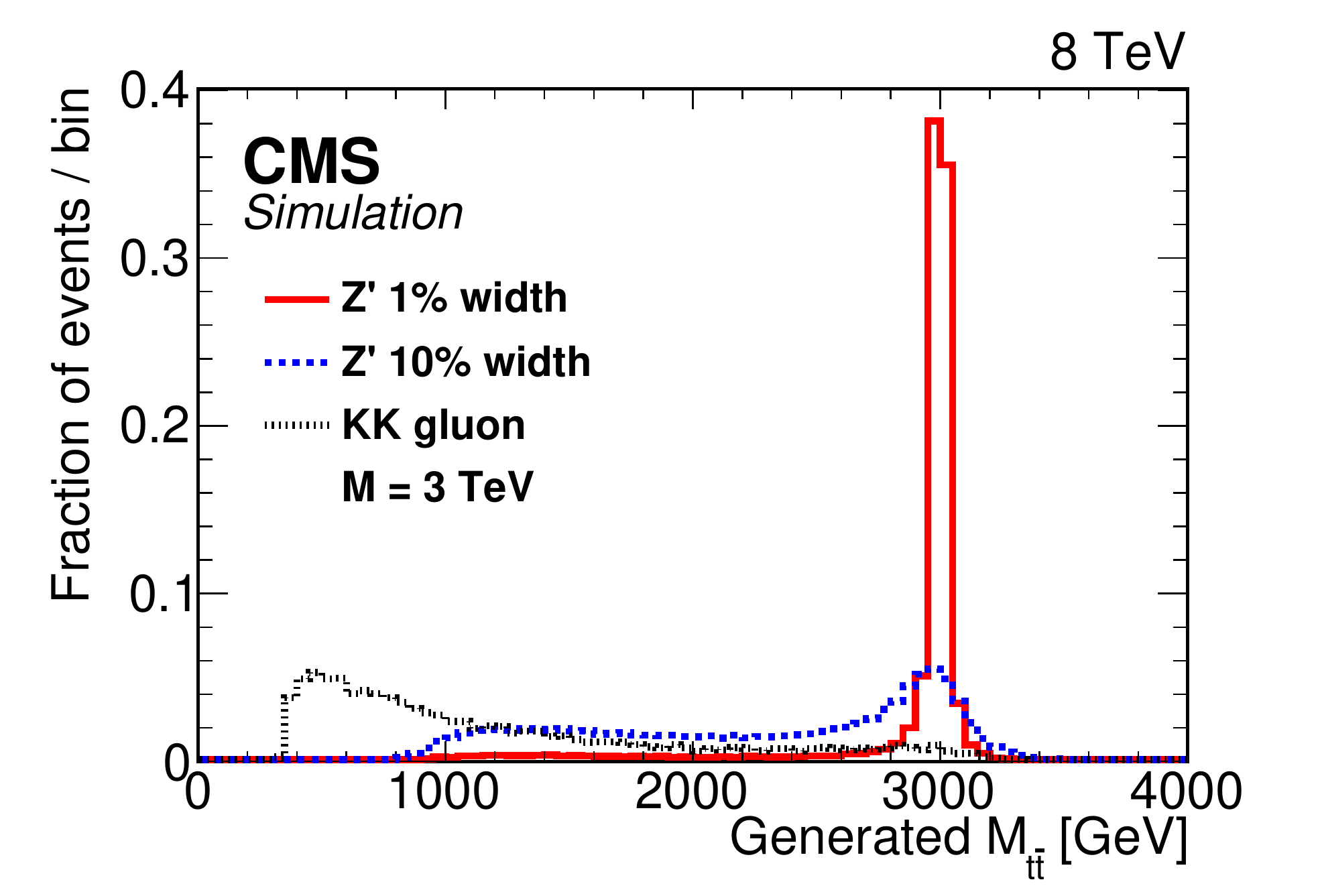}
\caption{The invariant mass distributions for different signal models, as described in the text,
for (\cmsLeft) 1.5\TeV and (\cmsRight) 3\TeV masses.}
\label{fig:genmass}
\end{figure}

Figure~\ref{fig:genmass} shows the invariant mass of the $\ttbar$ quark system at
the parton level for the \PZpr and $\gKK$ samples for two different
invariant masses, 1.5 and 3\TeV. The samples at 1.5\TeV show a
peaked structure, characteristic of a narrow-width resonance. The
samples at 3\TeV exhibit a significant tail at low invariant mass values due
to the interplay between the available partonic center-of-mass energy and the width
of the resonance; this is most pronounced for $\gKK$ because of its very
large width.
Above 3\TeV, the resonant hypothesis for the signal
samples is not valid anymore, thus signals with masses above 3\TeV
are not considered in this paper.

Top-quark events, produced via the strong and electroweak interactions, are simulated using
the next-to-leading-order (NLO) generator \POWHEG 1.0~\cite{Nason:2004rx,Frixione:2007vw,Alioli:2010xd}. The $\PW(\to\ell\nu)$+jets and
\Z/$\gamma^*(\to\ell\ell)$+jets processes are simulated using \MADGRAPH 5.1~\cite{madgraph5}, and the
diboson production processes (\PW\PW, \PW\Z, and $\Z\Z$) are simulated
using \PYTHIA 6.424~\cite{pythia6}.
Simulated QCD multijet samples produced with \MADGRAPH are used to validate the estimation of 
the multijet background from data.

All of the samples produced with \MADGRAPH are interfaced
to \PYTHIA for parton showering and fragmentation. The MLM algorithm~\cite{mlm}
is applied during the parton matching to avoid double counting of partons.
The \MADGRAPH samples use the CTEQ6L~\cite{cteq} parton distribution functions (PDF).
For the \POWHEG \ttbar sample, the CT10~\cite{CT10} PDF set is utilized,
whereas the single top quark processes are produced with the CTEQ6M PDF set.
The most recent \PYTHIA 6 Z2* tune is used to model the underlying event activity. 
It is derived from the Z1 tune~\cite{Field:2010bc},
which uses the CTEQ5L parton distribution set, whereas Z2* uses CTEQ6L~\cite{Pumplin:2002vw}.

The leading-order (LO) cross sections for the topcolor \PZpr signal
are taken from Ref.~\cite{theory_jainharris},
whereas for $\gKK$ production, calculations from Ref.~\cite{theory_agashe} are used.
However, both cross sections are multiplied by a factor of 1.3 to approximate NLO effects~\cite{ZprimeKfactor}.
The normalizations of the background samples are taken
from the NLO+next-to-next-to-leading logarithms (NNLL) calculation for
the single top quark production~\cite{singletop_xsec},
the next-to-next-to-leading order (NNLO) calculations for
$\PW(\to\ell\nu)$+jets and \Z/$\gamma^*(\to\ell\ell)$+jets~\cite{fewz1, fewz2, fewz3},
and the NLO calculation for diboson
production~\cite{diboson_xsec}. The normalization for the continuum \ttbar background uses
NNLO calculations~\cite{ttbar_xsec_new}.
However, by comparing the number of simulated and data events in control regions,
we determine additional cross section scale factors.
This is discussed in Section~\ref{sec:selection}.

A detailed simulation of particle propagation through the CMS
apparatus and detector response is performed
with \GEANTfour v9.2~\cite{Agostinelli2003250}.
For all simulated samples, the hard interaction collision is
overlaid with a number of simulated minimum bias collisions. The
resulting events are weighted to reproduce the pileup distribution
measured in data.
The same event reconstruction software is used for data and simulated events.
The resolutions and efficiencies for reconstructed objects are corrected
to match those measured in
data~\cite{electronreco, Chatrchyan:2012xi, Chatrchyan:2011tn, Chatrchyan:2012jua, JME-13-007}.

\section{Reconstruction of \texorpdfstring{\ttbar}{ttbar} events}
\label{sec:selection}

\subsection{Dilepton channel}

In the dilepton channel, the selection is based on the assumption
that both $\PW$ bosons decay leptonically.
The selection requires two leptons and at least two jets.
The lepton and the $\cPqb$ quark from the decay of a
highly Lorentz-boosted top quark are usually not well separated,
resulting in a non-isolated lepton that partially or fully
overlaps with the $\cPqb$-quark jet.

Offline, the following selection requirements are applied.
In the $\Pe\Pe$ channel, events are required to have two electrons with
$\pt>85\GeV$ and $\pt>20\GeV$, each within $\abs{\eta}<2.5$.
In the $\Pe\Pgm$ channel, there must be a muon with
$\pt>45\GeV$ with $\abs{\eta}<2.1$ and an electron with $\pt>20\GeV$
with $\abs{\eta}<2.5$.
Events in the $\Pgm\Pgm$ channel should contain two muons with $\pt>45\GeV$
and $\pt>20\GeV$ with $\abs{\eta}<2.1$ and
$\abs{\eta}<2.4$, respectively.
In all three channels, the two lepton candidates must have opposite charge.
The invariant mass of the lepton candidates in the $\Pe\Pe$ and $\Pgm\Pgm$ channels
must be $M_{\ell\ell}>12\GeV$ and outside the mass window of
$76<M_{\ell\ell}<106\GeV$. These selections reduce the contribution from
the production of low-mass resonances and from on-shell $\Z$ boson production.
Events are required to contain at least two jets with
$\pt>100\GeV$ and $\pt>50\GeV$ within $\abs{\eta}<2.5$.

\begin{figure}[htb]
\centering
\includegraphics[width=0.48\textwidth]{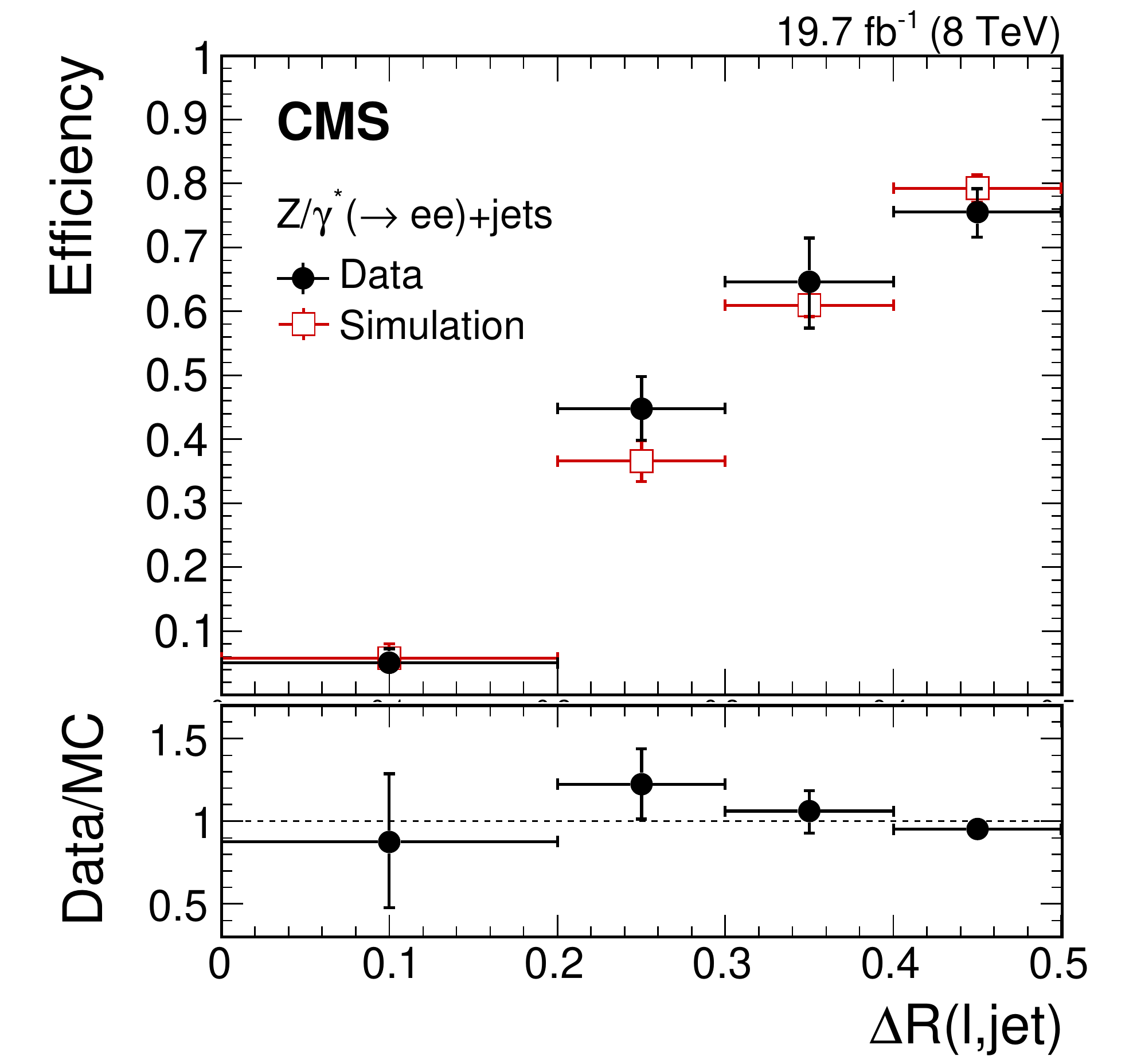}
\includegraphics[width=0.48\textwidth]{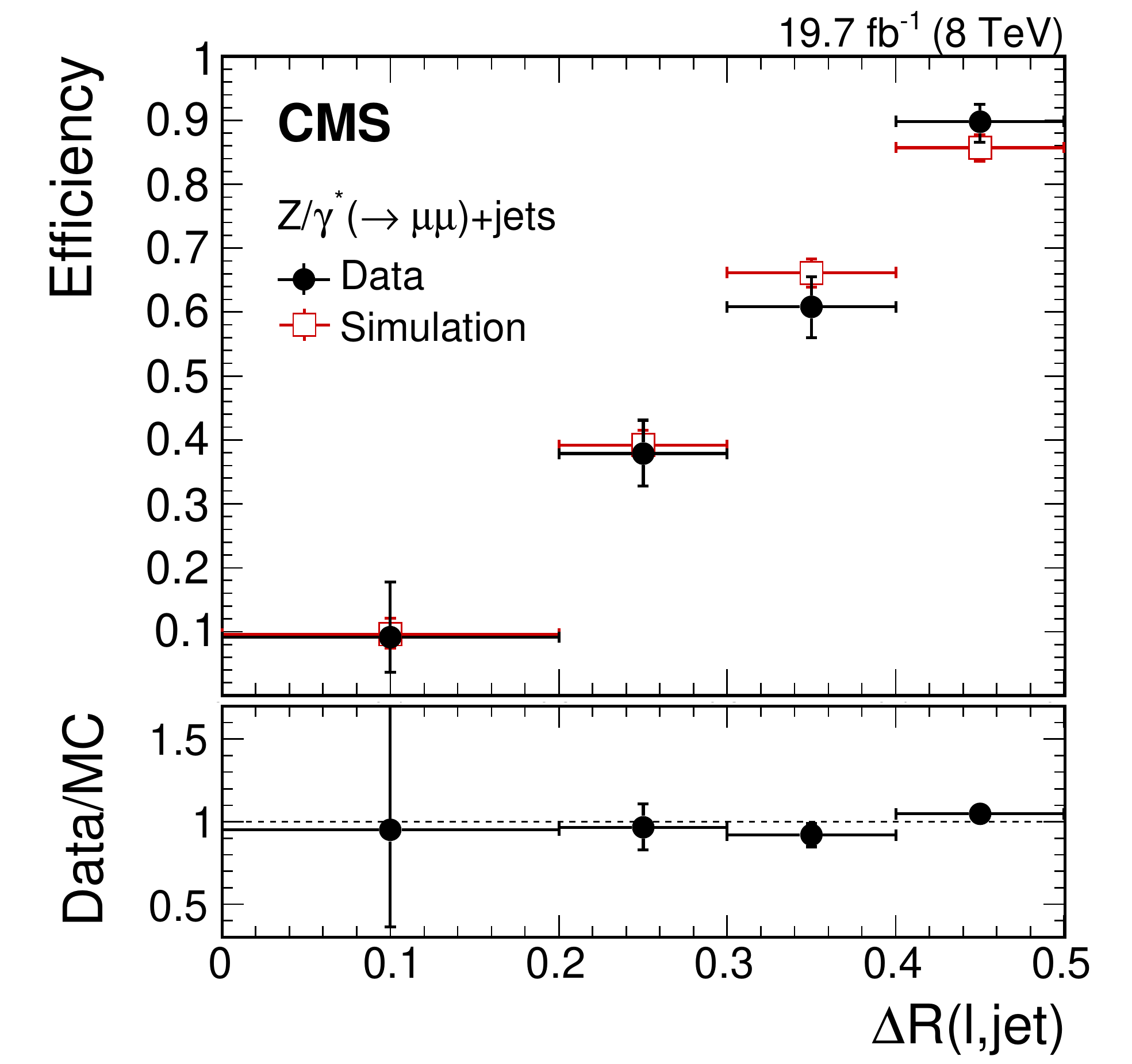}
\caption{Efficiency of the two-dimensional isolation requirement for data and simulated
events for the electron (\cmsLeft) and muon (\cmsRight) selection, as measured in a sample of $\Z/\gamma^*(\to\ell\ell)$+jets.
The ratio of the efficiencies in data to simulation is shown at the bottom of each panel.
\label{fig:2dcut_eff}
}
\end{figure}
Signal events are selected with a two-dimensional isolation
variable that is efficient at high top-quark boosts yet reduces
multijet backgrounds. This two-dimensional isolation requires
$\DR(\ell, \text{jet})>0.5$ or $p_{\text{T},\text{rel}}(\ell, \text{jet})>15\GeV$, where
$\DR(\ell, \text{jet})$ is the distance in $(\eta,\phi)$ between the lepton
and the nearest jet, and $p_{\text{T},\text{rel}}(\ell, \text{jet})$
is the transverse momentum of the lepton with respect to the axis of
the closest jet.
In calculating these quantities only jets with $\pt >30\GeV$ are considered.
The efficiency of the two-dimensional isolation requirement
has been studied in data and simulation in a
$\Z/\gamma^*(\to\ell\ell)$+jets sample. In this sample, one lepton, passing isolation
criteria, is used as a tag, and the other lepton is used as a probe to study the efficiency.
The dilepton invariant mass is used to determine the number of events passing
and failing the two-dimensional isolation criteria.
Figure~\ref{fig:2dcut_eff} shows
the efficiency as function of $\DR(\ell, \text{jet})$ for data and
simulated events. At small $\DR$ separation between the lepton and the closest jet, the
efficiency for electrons is 5\%, increasing to 75\% for larger $\DR(\ell, \text{jet})$. The
corresponding values are 10\% to 90\% for muons. The efficiency is well described by the simulation and
no correction is needed.
The relative efficiency of the two-dimensional isolation requirement for 
the 1.5\TeV signal is approximately
$70\%$ independent of width.

A requirement of $\MET>30\GeV$ in the $\Pe\Pe$ and $\Pgm\Pgm$ channels additionally reduces
the contribution from multijet and $\Z/\gamma^*(\to\ell\ell)$+jets production.
Given the presence of two $\cPqb$ quarks in the events, a logical OR
of two $\cPqb$ tagging algorithms is used:
at least one of the
two leading jets is required to be tagged as a $\cPqb$-quark jet by the
CSV algorithm at the medium working point or both leading jets
must be tagged using the loose working point of the CSV algorithm. After these requirements, the sample contains about 90\% $\ttbar$ background.

\begin{figure*}[tbp]
\centering
\includegraphics[width=0.41\textwidth]{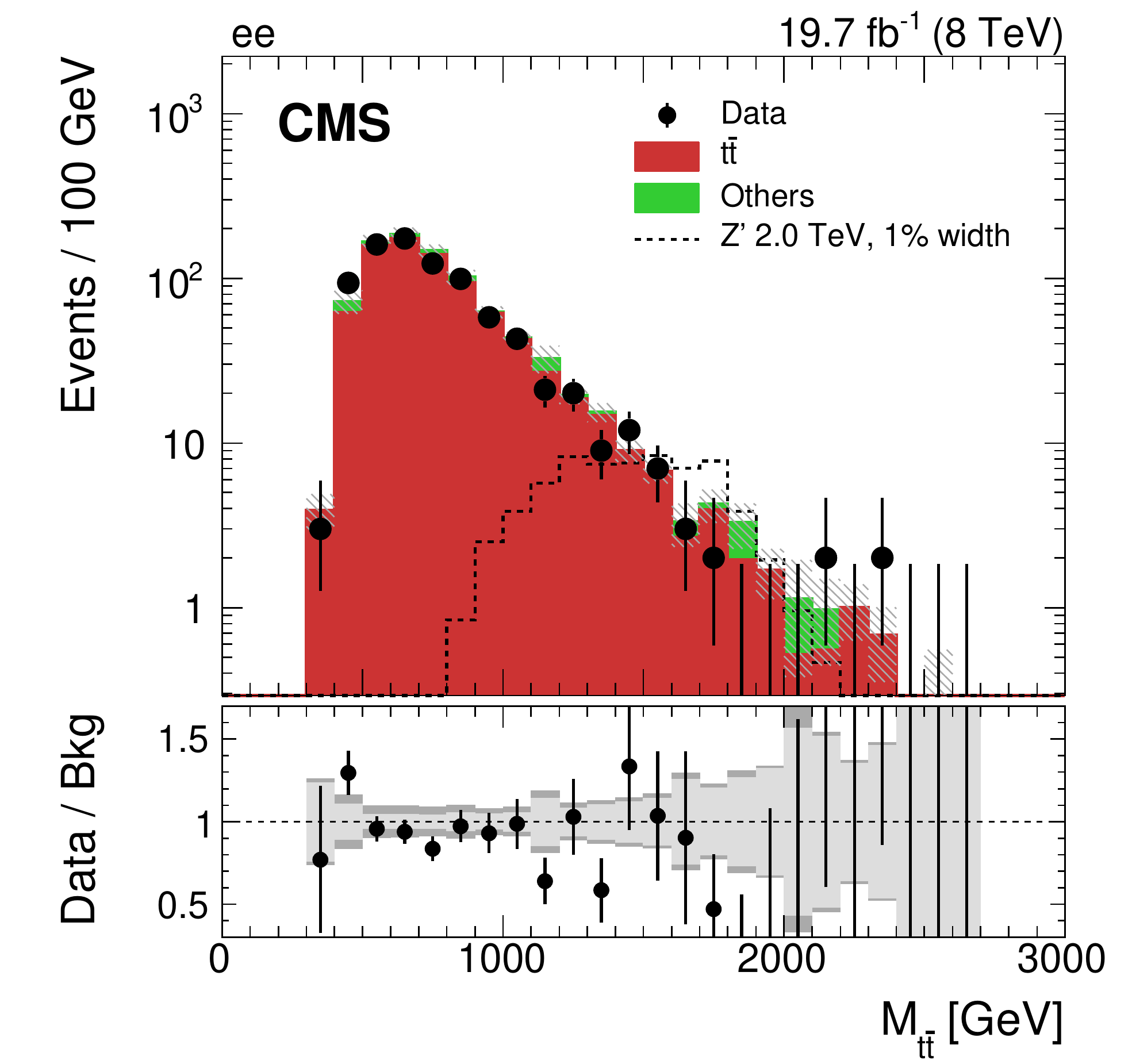}
\includegraphics[width=0.41\textwidth]{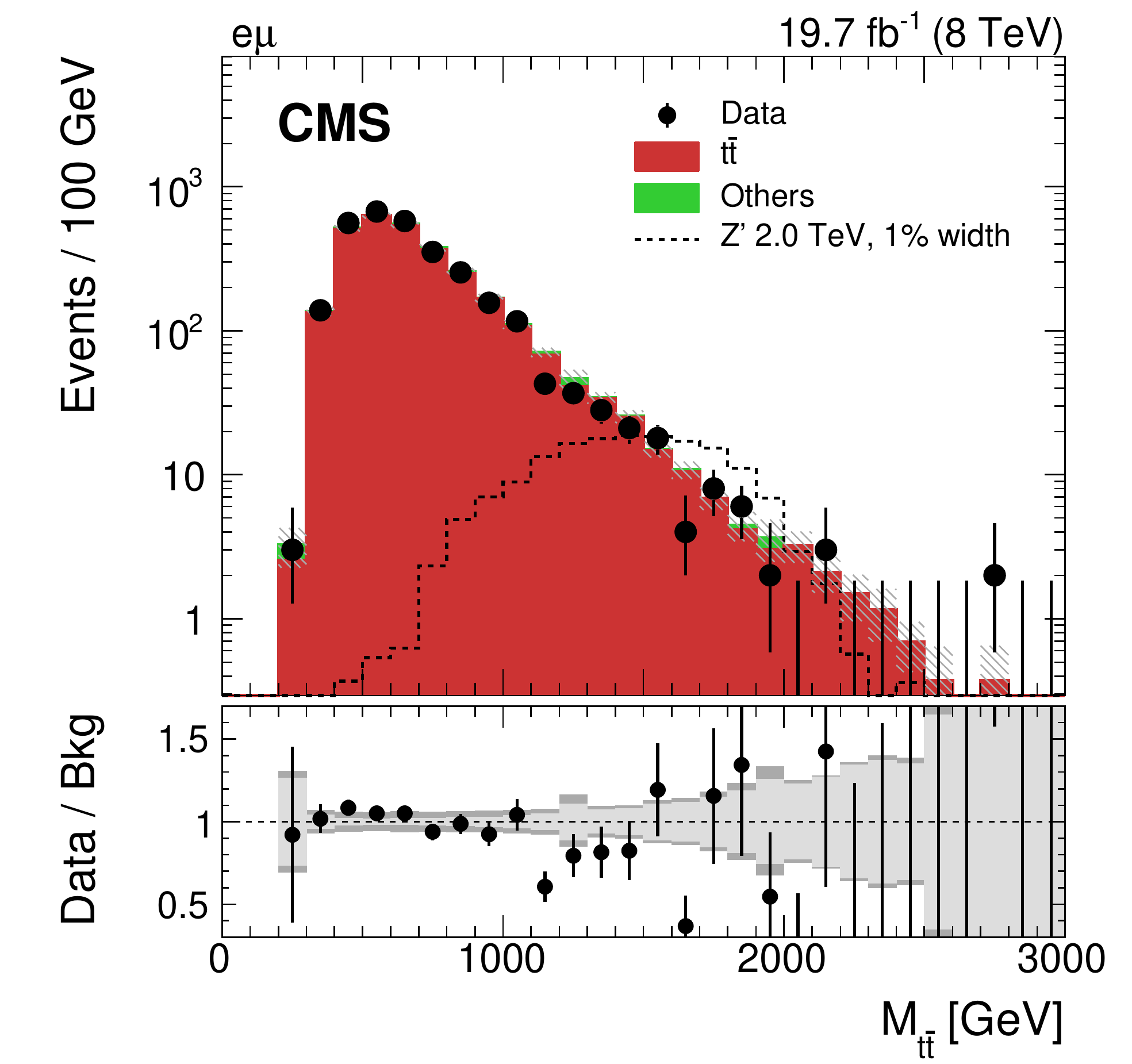}
\includegraphics[width=0.41\textwidth]{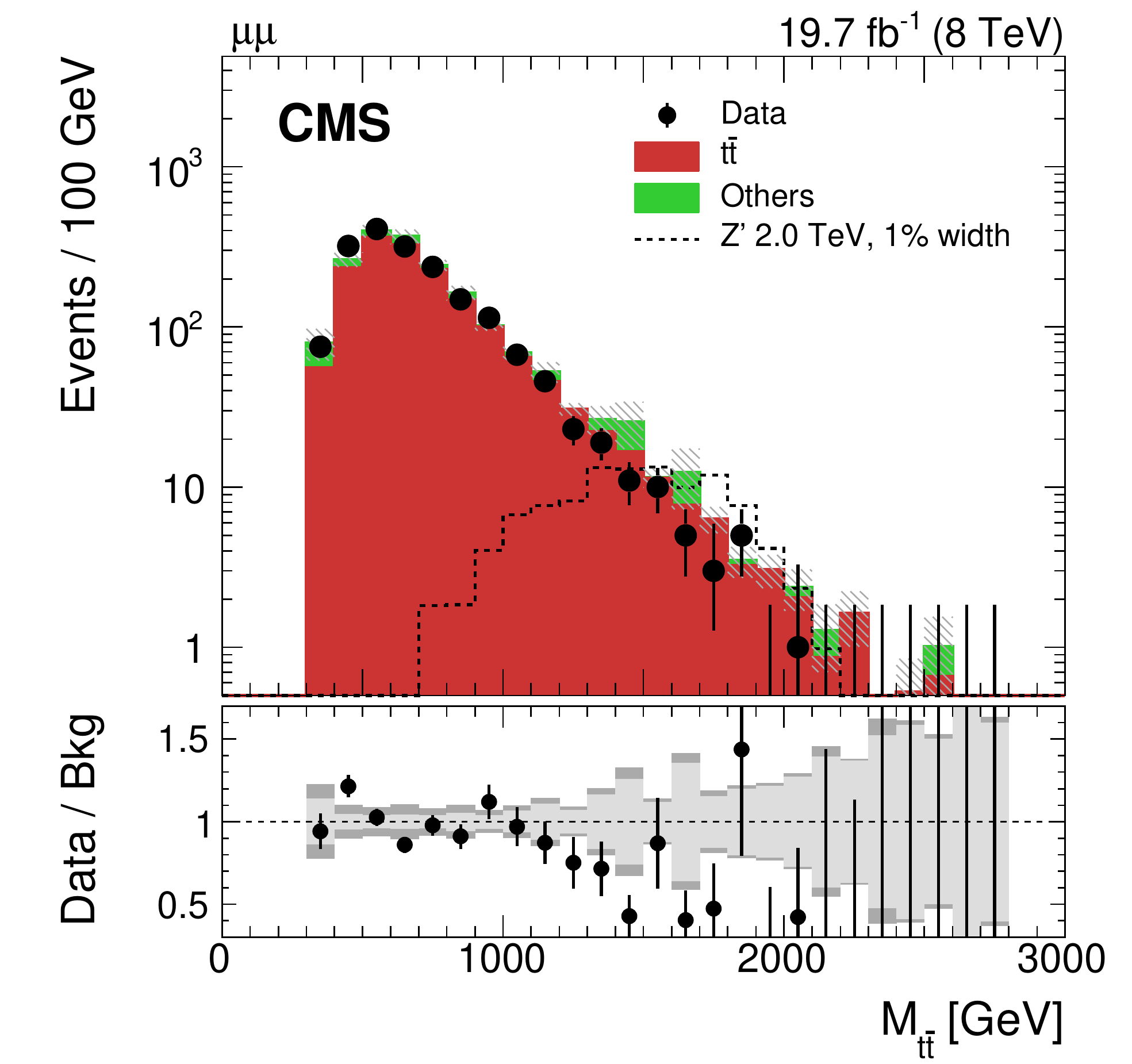}
\caption{Reconstructed invariant mass of the $\ttbar$ pair
         in the $\Pe\Pe$ (upper left), $\Pe\Pgm$ (upper right), and $\Pgm\Pgm$ (bottom) channels for data and simulated events.
         Each background process is scaled by a factor
         derived from a maximum likelihood fit to data.
         The expected distribution from a \PZpr signal with $\Mzp = 2\TeV$ and $\rWzp = 0.01$, normalized to a cross section of 1\pb, is also shown.
         The uncertainty associated with the background expectation includes all the statistical and
         systematic uncertainties.
         For bins with no events in data but with nonzero background expectation, 
         vertical lines are shown indicating the 68\% confidence level coverage 
         corresponding to an observation of zero events.
         The data-to-background ratio is shown in the bottom panel of each figure.
         For the ratio plot, the statistical uncertainty is shown in light gray, while
         the total uncertainty, which is the quadratic sum of the statistical and systematic uncertainties, is shown in dark gray.
         There is a systematic disagreement observed in the high-mass region that is accommodated by the renormalization and factorization scale uncertainty.
         \label{fig:mttbar_dilepton}}
\end{figure*}
The boosted nature of the signal events provides an additional handle for
further reduction of the \ttbar background:
the separation in $\DR$ between each lepton and its nearest jet.
Requiring $\DR(\ell_1, \text{jet})<1.2$ and
$\DR(\ell_2, \; \text{jet})<1.5$,
where $\ell_1$ and $\ell_2$ denote the leading and sub-leading leptons,
reduces the \ttbar background contribution by more than a factor of 2,
while the loss for a \PZpr signal with mass of 1.5\TeV is about 10\%.
Additionally, the region with $\DR(\ell_2, \text{jet})>1.5$
is dominated by events from continuum \ttbar production
and provides an independent sample to check the \ttbar background normalization.
The contamination from resonant \ttbar production
is expected to be less than 0.2\% in this sample.
The normalization of the \ttbar background is found to be compatible
with the SM expectation using the NNLO cross section calculations,
and good agreement between the $\Pe\Pe$, $\Pe\Pgm$, and $\Pgm\Pgm$
channels is observed.

The resonant nature of the signal is exploited by constructing a mass variable from the
four-momenta of the two leading leptons, the two leading jets, and the neutrinos,
which approximates the invariant mass of the \ttbar system.
For the momentum components $p_x$ and $p_y$ of the pair of neutrinos, the $x$- and
$y$-components of \ptvecmiss are used, and the $p_z$ component of each neutrino is set to zero.

Figure~\ref{fig:mttbar_dilepton} shows the \Mttbar distributions for the dilepton channel.
The expected distribution
from a \PZpr signal with $\Mzp = 2\TeV$ is also shown.
Good agreement between the data and the SM background expectation is found.

\subsection{Lepton+jets channel}

The selection in the lepton+jets channel is based on events with
one $\PW$ boson decaying leptonically, $\PW \to \ell \nu$,
and the other one decaying hadronically, $\PW \to \cPq \cPaq^\prime$.
It requires one lepton (electron or muon) and at least two jets with high $\pt$,
including events with non-isolated leptons and
merged jets arising from decays of two high-\pt top quarks.

Events are required to have exactly one electron with $\pt > 35 \GeV$ and $\abs{\eta} < 2.5$, or one muon with $\pt > 45\GeV$
and $\abs{\eta}<2.1$.
The reconstructed lepton has to be consistent with originating
from the primary event vertex.
In order to avoid overlap with the dilepton sample,
events with a second reconstructed lepton are removed.
All events must have at least two jets with $\pt>150\GeV$ and $\pt>50\GeV$,
both with $\abs{\eta}<2.4$.
In order to ensure that there is no overlap with the all-hadronic channel,
events with two or more CA8 $\cPqt$-tagged jets are rejected.
To reduce the background from multijet production, events are required to have
$\MET>50\GeV$ and the scalar sum of the
lepton $\pt$ and $\MET$ has to be larger than 150\GeV.

A further reduction of the multijet background contribution is achieved by applying
a similar two-dimensional isolation criterion as described for the dilepton channel.
It is applied for both the electron and muon channels, requiring
$\DR(\ell, \text{jet})>0.5$ or $p_{\text{T},\text{rel}}(\ell, \text{jet})>25\GeV$,
where only jets with $\pt >25\GeV$ are considered when calculating these quantities.

In addition, in the electron channel topological requirements are imposed that ensure
that \ptvecmiss does not point along the transverse direction of the electron or the
leading jet~\cite{cms_ttbar_resonance2},
\[| \Delta\phi ( \{ \Pe \,\text{or}\, \text{jet} \} , \,\ptvecmiss ) - 1.5| < \ETmiss / 50\GeV\, \]
with \ETmiss measured in \GeVns.
The efficiency of this requirement is above 95\% for all signal samples,
while the background from multijet production is reduced significantly.

The \ttbar system is reconstructed by assigning the four-vectors of the reconstructed
final-state objects to either the leptonic or hadronic leg of the \ttbar decay.
This is done by constructing a two-term $\chi^2$ function, based on the
masses of the reconstructed \ttbar candidates~\cite{cms_ttbar_resonance2, cms_ttbar_resonance4}.
For each event, the hypothesis with the smallest $\chi^2$ value is chosen.
In case a CA8 $\cPqt$-tagged jet is found in the event,
this jet is used for the hadronic leg of the \ttbar decay and all jets with
$\DR < 1.3$ from the $\cPqt$-tagged jet are removed from the list of
possible hypotheses.
Events are required to have a minimum value of $\chi^2$ smaller than 50,
which reduces the contribution from background processes and
enhances the sensitivity of the search.
In the electron channel, the transverse momentum of the reconstructed
leptonic leg of the top-quark decay is required to be larger than 140\GeV, to suppress
the background from multijet production to a negligible level.

Events are categorized based on the lepton flavor and on the number
of CA8 $\cPqt$-tagged jets. In case no CA8 $\cPqt$-tagged jet is found, the events are further
split into two categories, depending if any jets are identified as
originating from the fragmentation of a $\cPqb$ quark using the medium working point
of the CSV algorithm. This has a tagging efficiency of 65\% per jet~\cite{Chatrchyan:2012jua}.

\begin{figure}[tbh]
\centering
\includegraphics[width=0.45\textwidth]{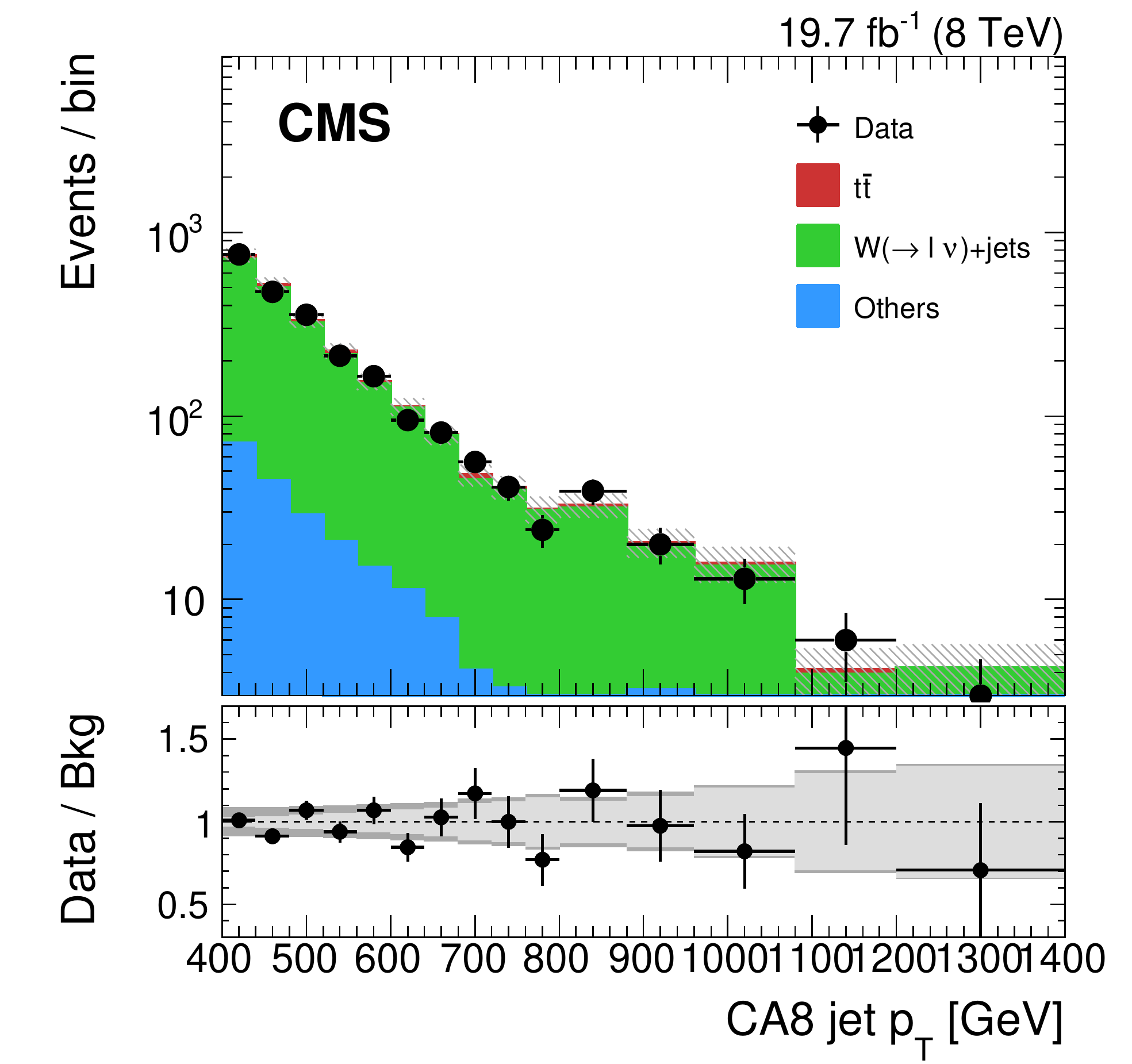}
\includegraphics[width=0.45\textwidth]{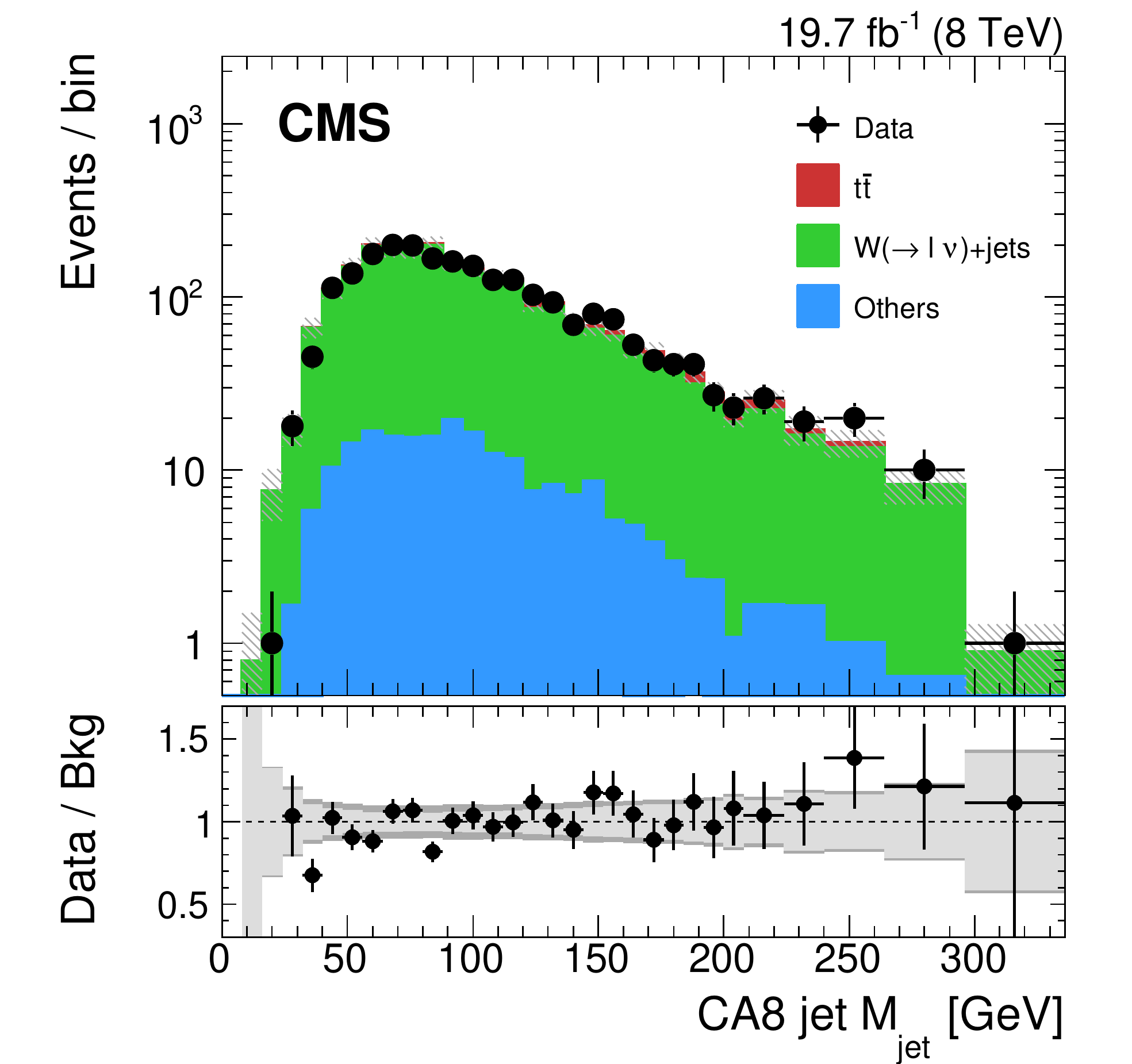}
\caption{Distribution of $\pt$ (\cmsLeft) and the jet mass (\cmsRight) of the leading
CA8 jet in a $\PW$+jets control region, used to obtain the mistag rate of CA8 $\cPqt$-tagged jets.
The horizontal error bars indicate the bin width.
The data-to-background ratio is shown in the bottom panel of each figure.
For the ratio plot, the statistical uncertainty is shown in light gray, while
         the total uncertainty, which is the quadratic sum of the statistical and
         systematic uncertainties, is shown in dark gray.
\label{fig:mistag_wjet_control}}
\end{figure}
An independent control sample is used to validate the mistag rate of 
CA8 $\cPqt$-tagged jets in the $\PW$+jets sample.
This sample is obtained by inverting the
$\chi^2$ criterion, using the leptonic leg of the \ttbar decay hypothesis only,
and requiring that no jet has been tagged as a $\cPqb$-quark jet by the loose
operating point of the CSV algorithm~\cite{Chatrchyan:2012jua}.
This removes most of the $\ttbar$ contamination while retaining events from
$\PW$+jets production.
Figure~\ref{fig:mistag_wjet_control} shows the $\pt$ and mass of the
leading CA8 jet in this sample. These jets are used to determine the mistag rate
of the CA8 $\cPqt$-tagged jets in data and simulated events that
also contain a lepton. Such events have a higher fraction of jets from
quark fragmentation than
the non-top-quark multijet background for the all-hadronic channel, which
has a higher fraction of jets from gluon fragmentation. Good agreement
is observed, with a mistag rate of 1.2\% in data,
and a data-to-simulation ratio of $0.83 \pm 0.21$. This factor is used to scale
simulated events containing a misidentified top-quark jet.

\begin{figure*}[tbp]
 \centering
\includegraphics[width=0.45\textwidth]{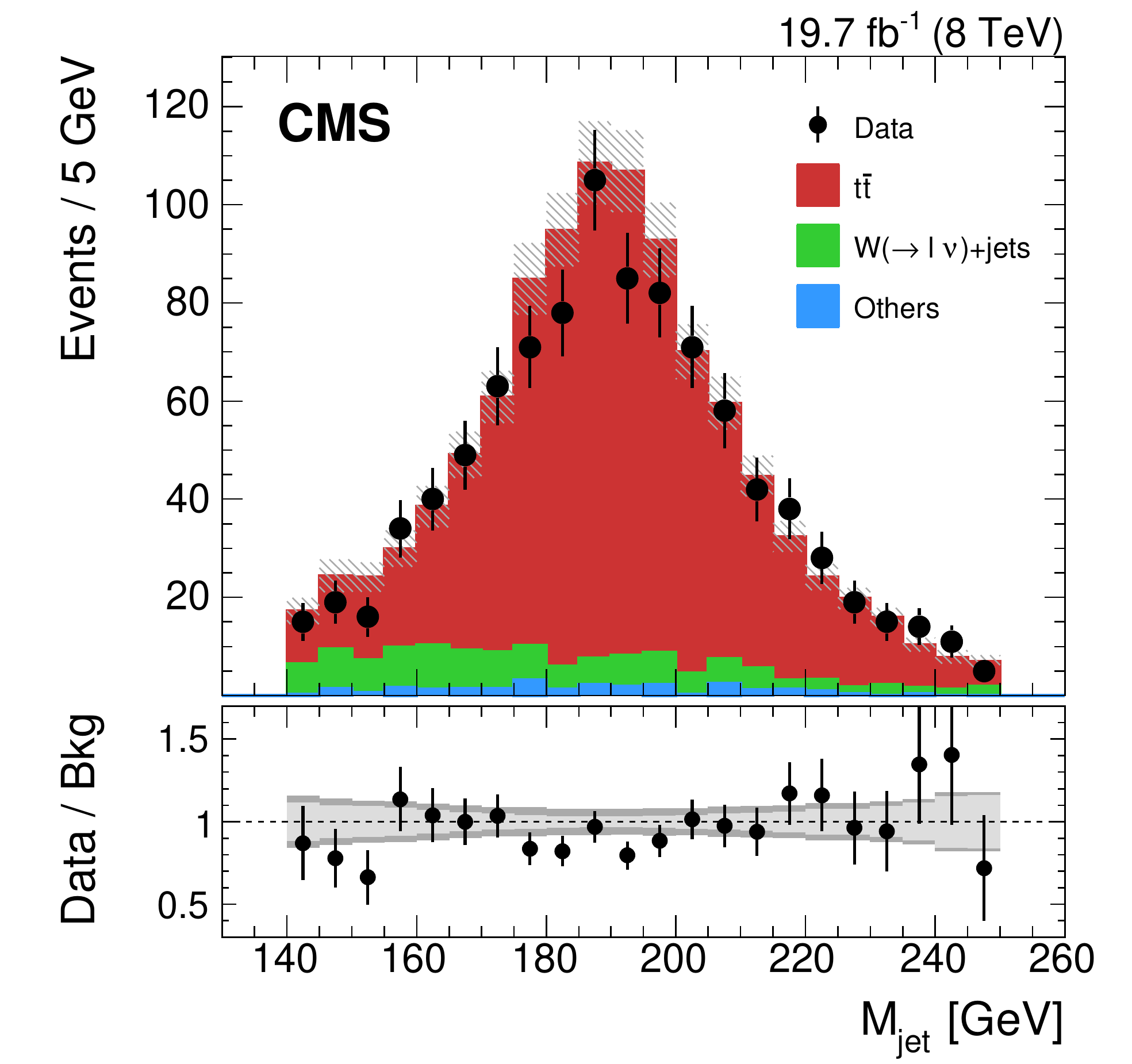}
\includegraphics[width=0.45\textwidth]{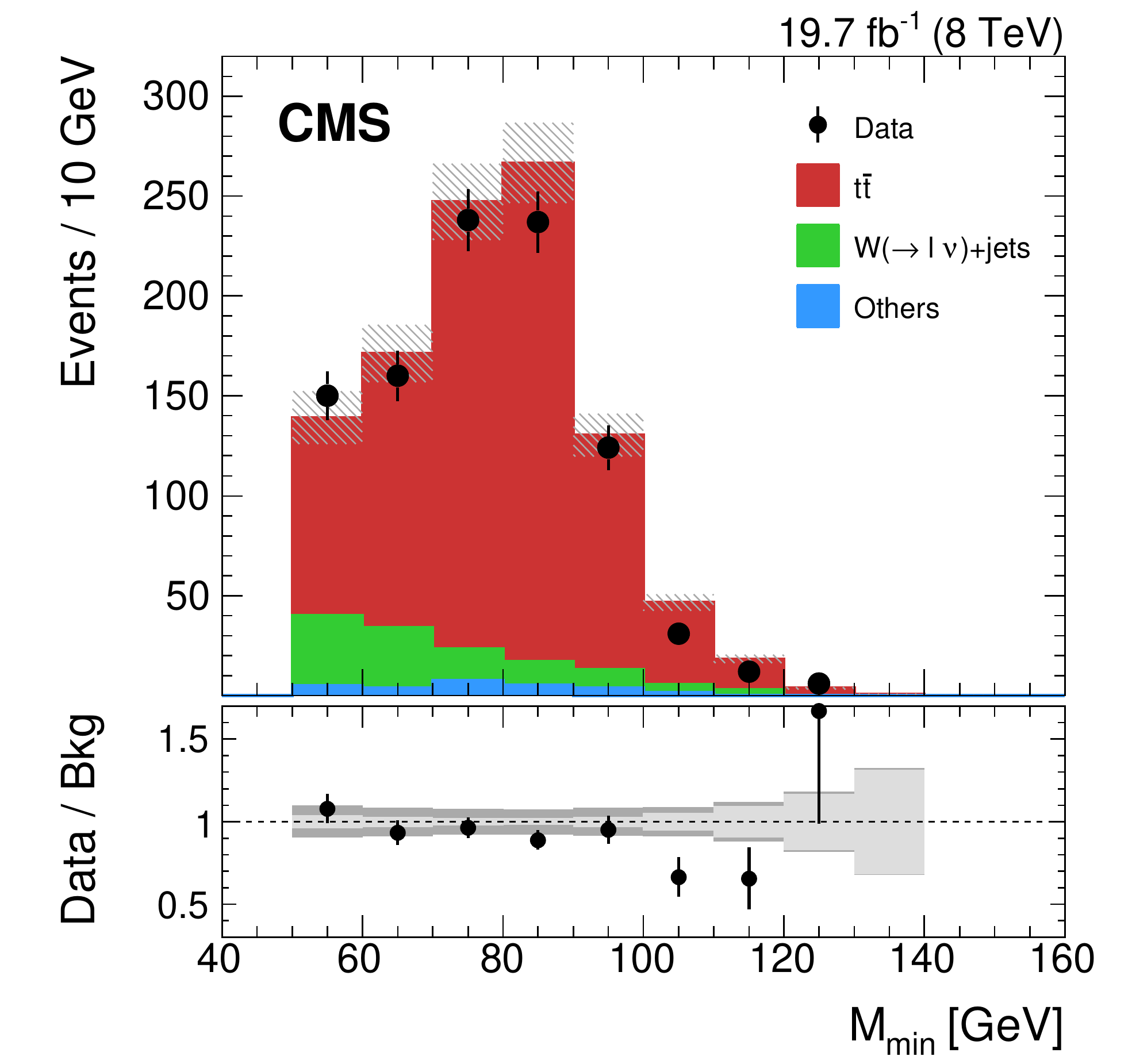}
\includegraphics[width=0.45\textwidth]{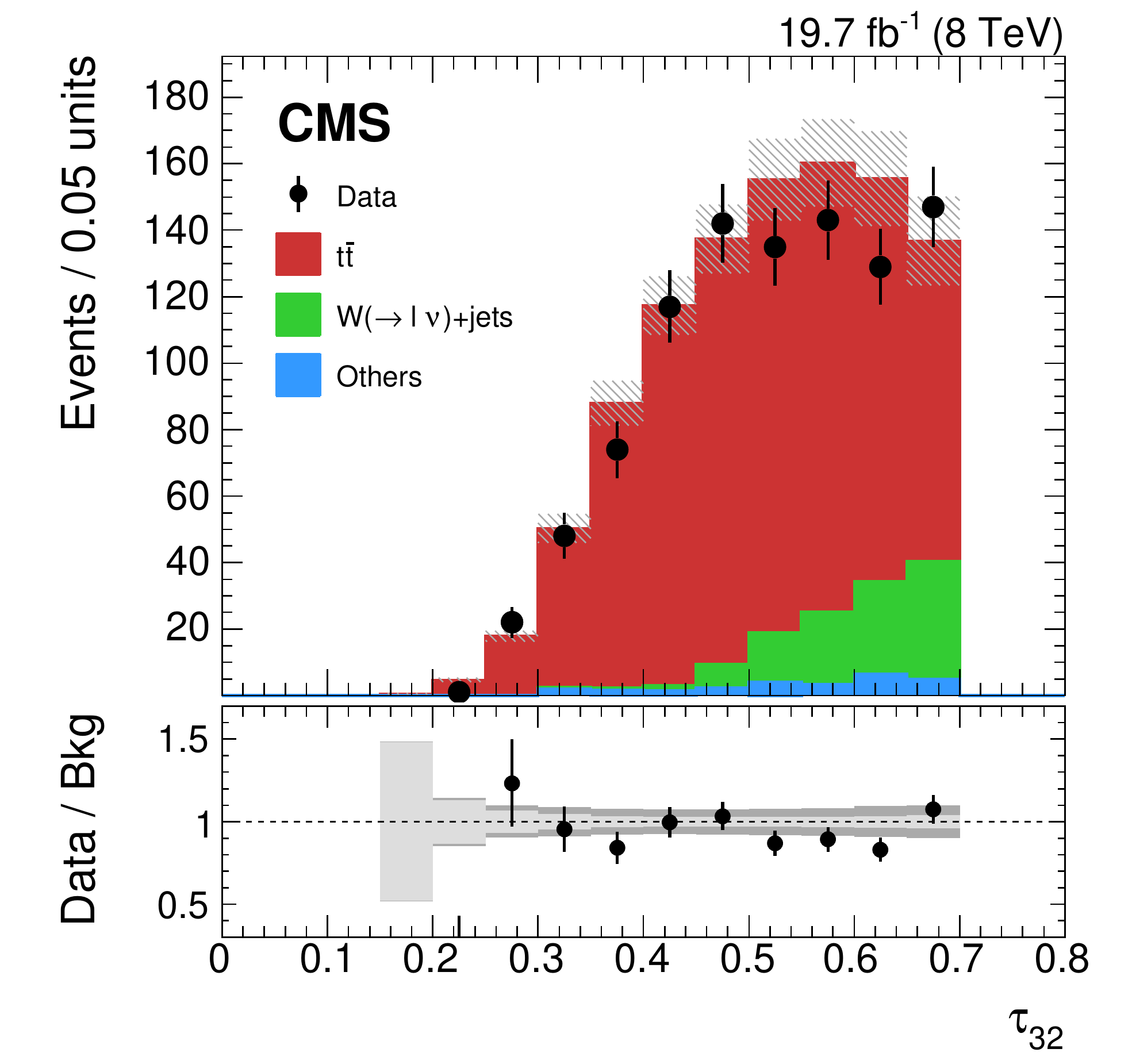}
 \caption{Distributions of the jet mass (upper left), minimum pairwise mass \mmin (upper right),
 and the ratio $\tau_{32}$ (bottom) for CA8 $\cPqt$-tagged jets
 in the lepton+jets channel.
 The SM backgrounds are scaled by a factor
 derived from the maximum likelihood fit to data.
 The uncertainty associated with the background expectation includes all the statistical and
 systematic uncertainties.
 The data-to-background ratio is shown in the bottom panel of each figure.
 For the ratio plot, the statistical uncertainty is shown in light gray, while
 the total uncertainty, which is the quadratic sum of the statistical and systematic uncertainties, is shown in dark gray.
 \label{fig:toptags}}
\end{figure*}
To validate the CA8 $\cPqt$ tagging efficiency,
distributions of the jet mass, minimum pairwise mass, and the ratio $\tau_{32}$
for CA8 $\cPqt$-tagged jets are shown in the lepton+jets channel for data and
simulated events in Fig.~\ref{fig:toptags}, where all SM components are
normalized to the output of the maximum likelihood fit.
Differences between the distributions in data and simulation lead to different efficiencies
of the $\cPqt$ tagging algorithm. In order to account for these differences,
a correction factor for simulated events is derived from a combined
maximum likelihood fit. The fit is performed by comparing the yields in categories of
events which pass and fail the CA8 $\cPqt$ tagging selection criteria,
as explained in Section~\ref{sec:background}.
The scale factor for $\cPqt$-tagged jets is estimated to be $0.94\pm0.03$, reflecting the 
somewhat better resolution of the reconstructed mass of CA8 jets in simulation.

\begin{figure*}[tbp]
\centering
\includegraphics[width=0.41\textwidth]{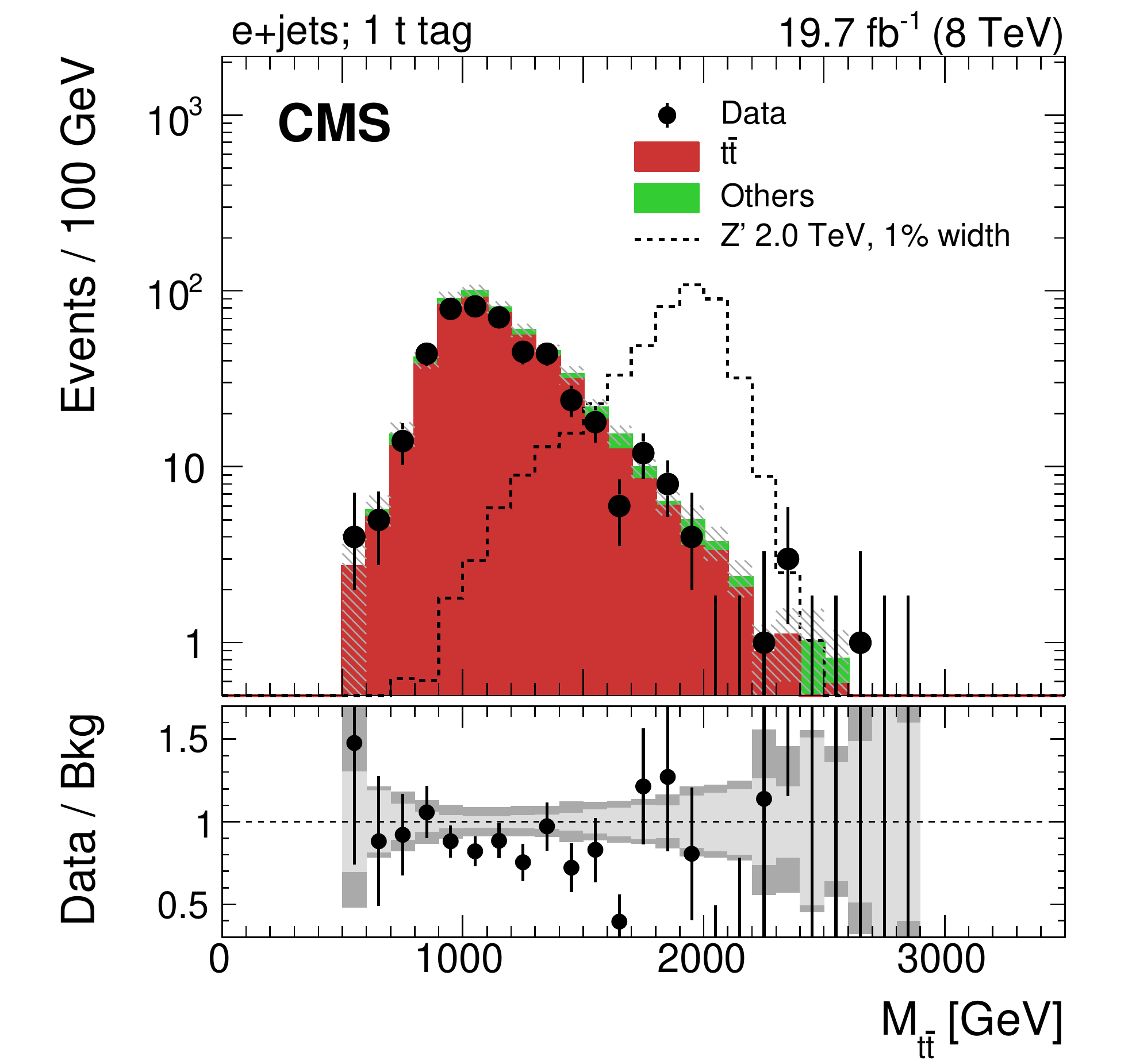}
\includegraphics[width=0.41\textwidth]{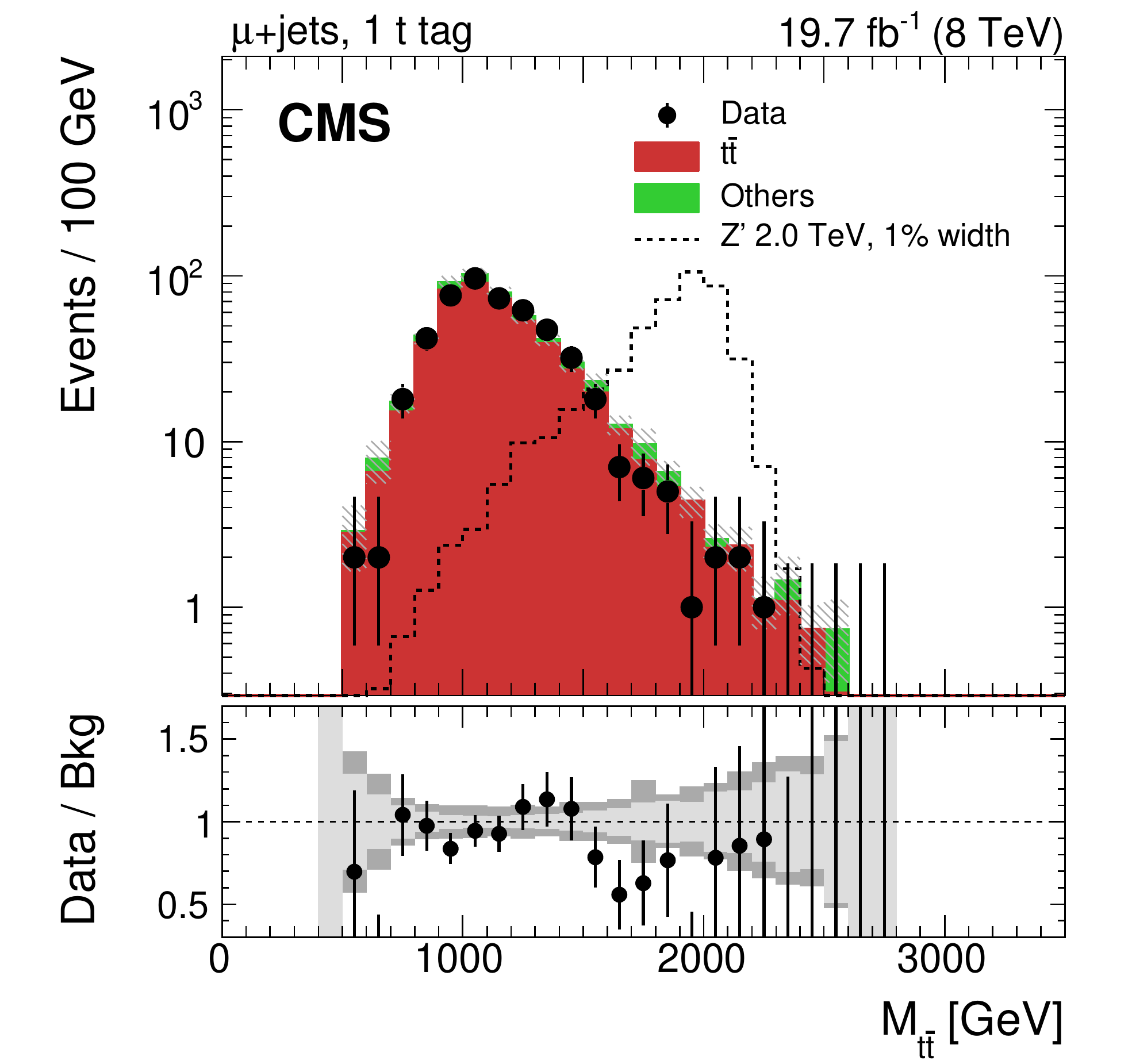}
\includegraphics[width=0.41\textwidth]{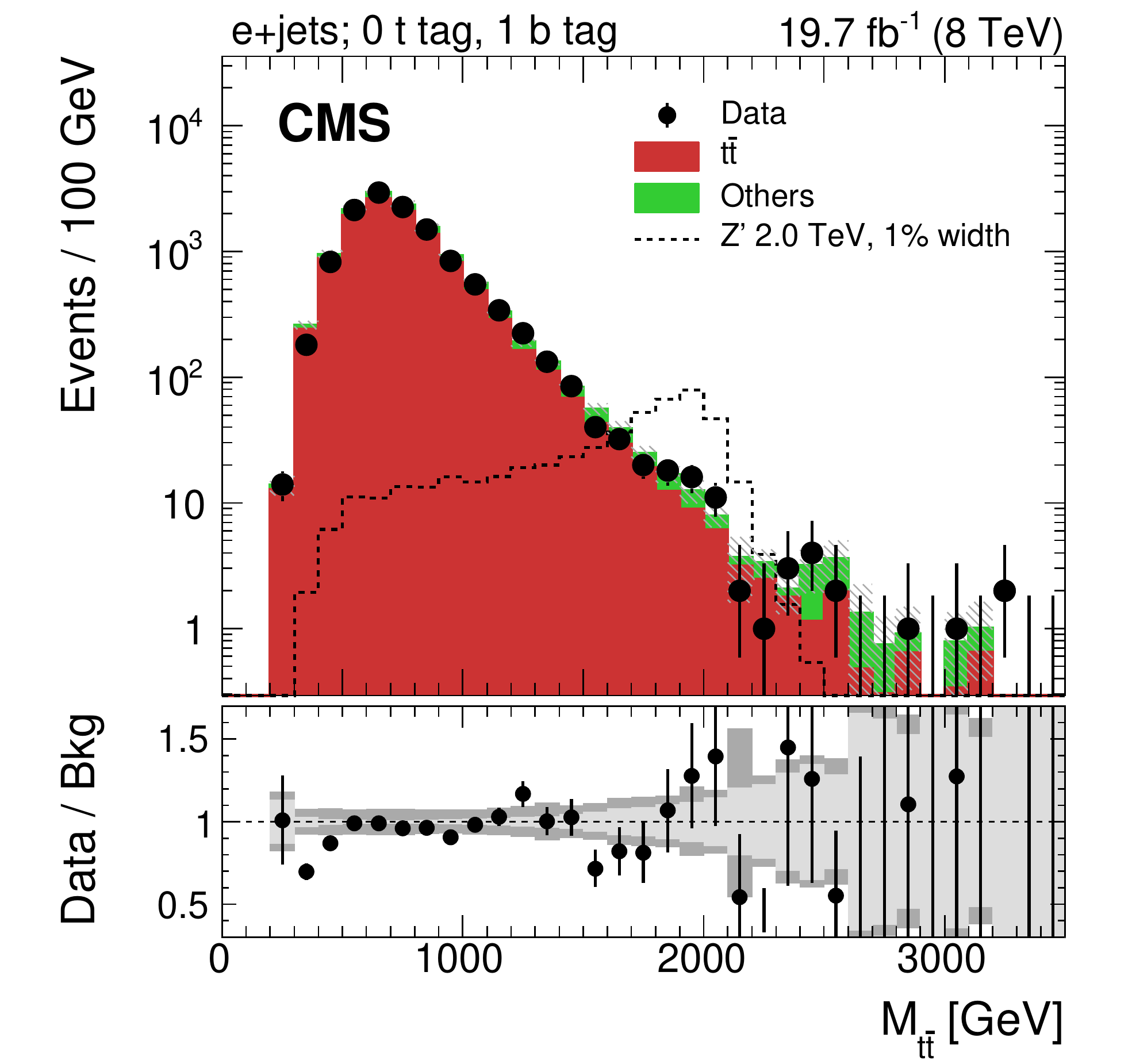}
\includegraphics[width=0.41\textwidth]{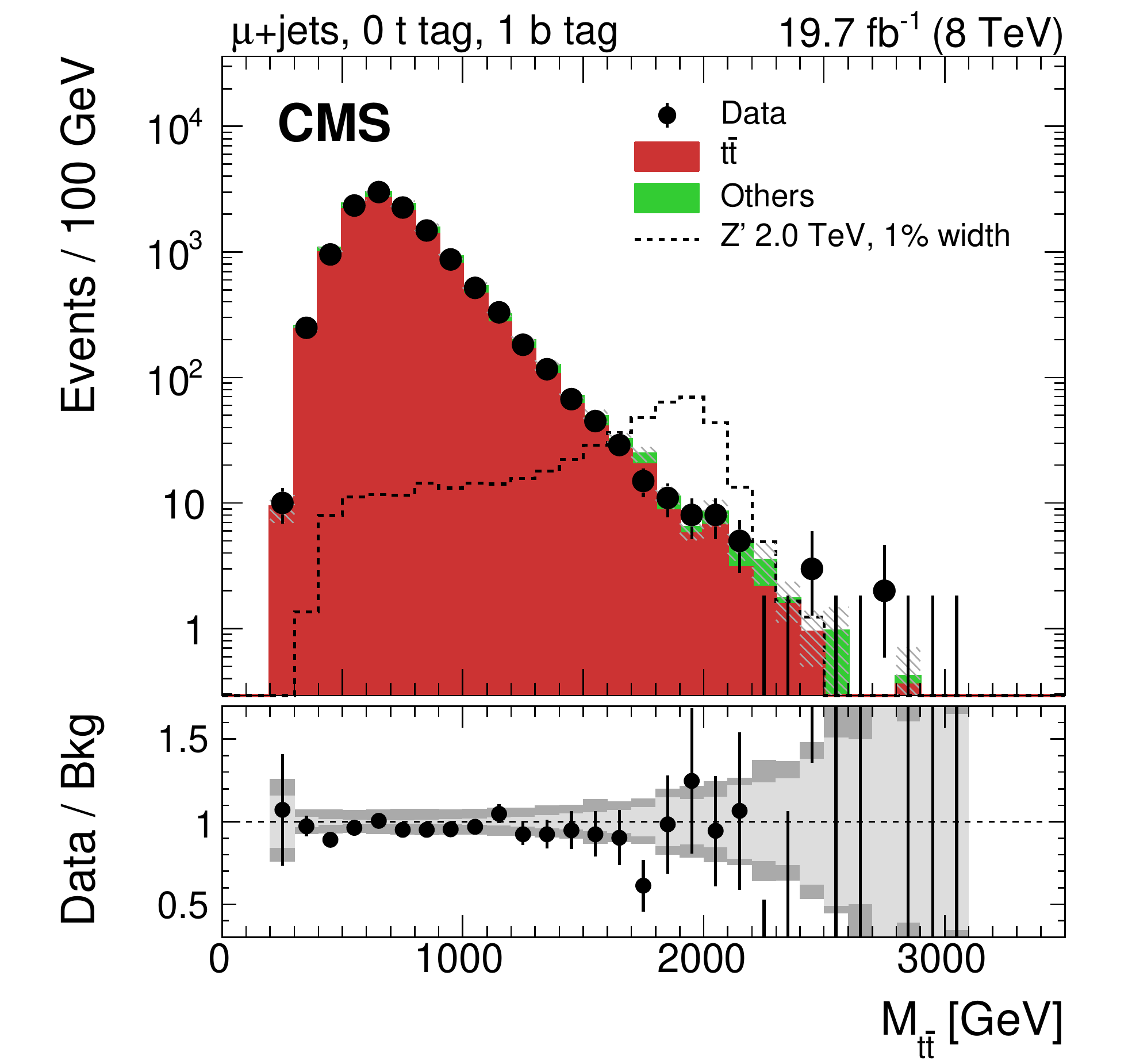}
\includegraphics[width=0.41\textwidth]{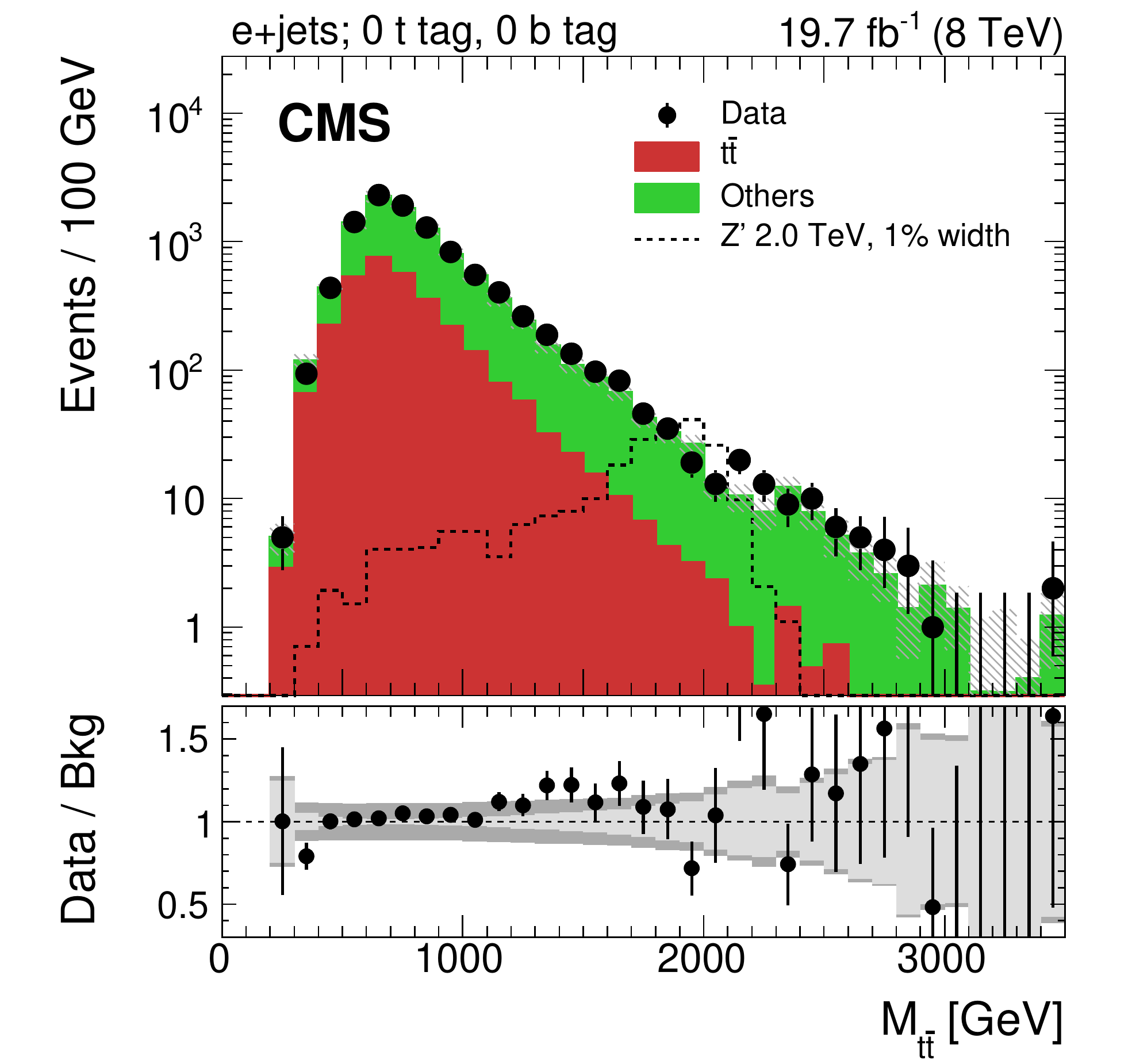}
\includegraphics[width=0.41\textwidth]{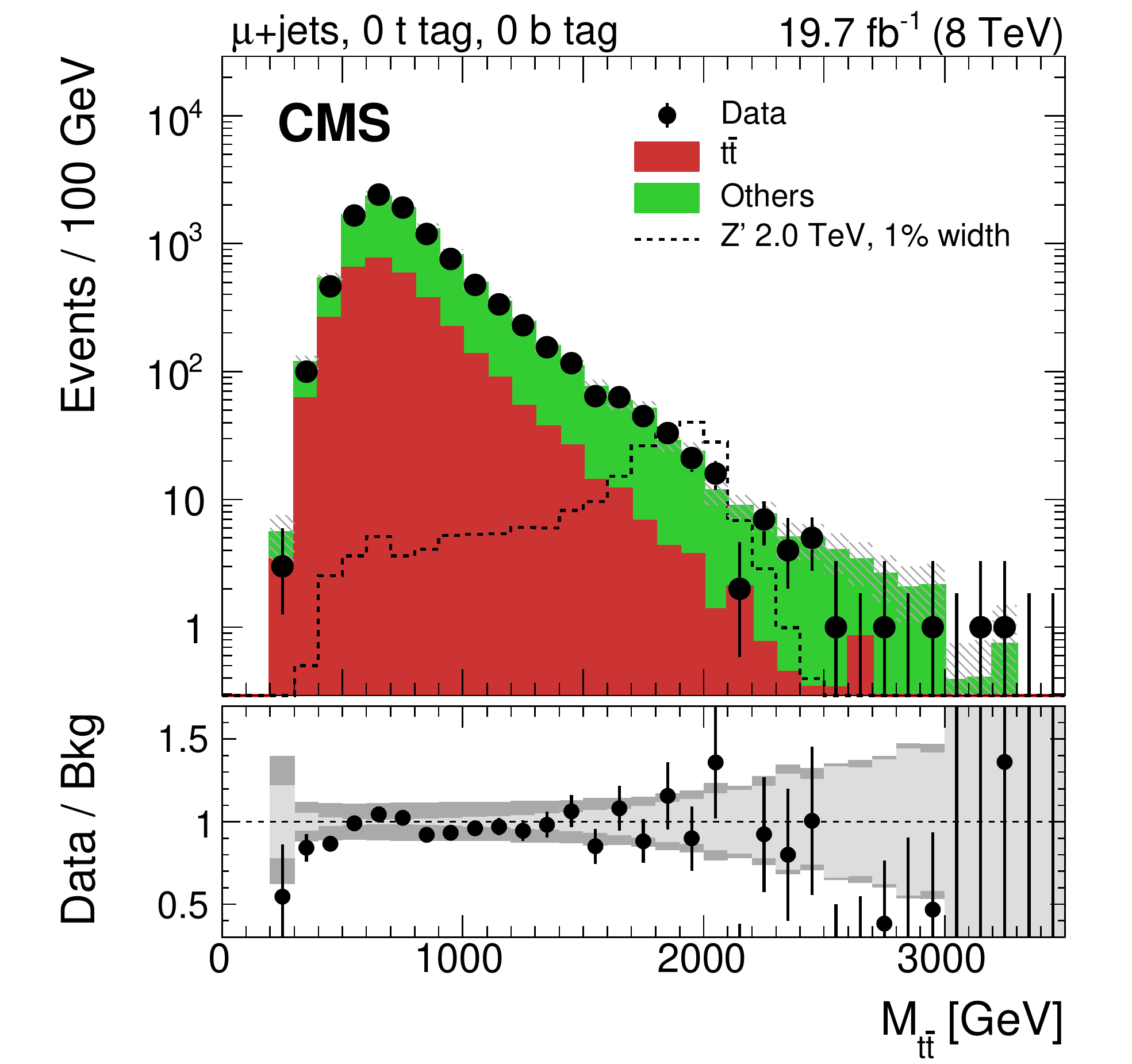}
\caption{Invariant mass of the reconstructed $\ttbar$ pair in data and simulation
         in the electron+jets (left column) and muon+jets (right column) channels.
         Events are separated into three categories: one CA8 $\cPqt$-tagged jet (top row),
	 no CA8 $\cPqt$-tagged jet and at least one $\cPqb$ tag (middle row), and no CA8 $\cPqt$-tagged jet
	 and no $\cPqb$ tag (bottom row).
         Each background process is scaled by a factor
         derived from the maximum likelihood fit to data.
         The expected distribution from a \PZpr signal with $\Mzp = 2\TeV$ and $\rWzp = 0.01$, normalized to a cross section of 1\pb, is also shown.
         The uncertainty associated with the background expectation includes all statistical and
         systematic uncertainties.
         For bins with no events in data but with non-zero background expectation, 
         vertical lines are shown indicating the 68\% confidence level coverage 
         corresponding to an observation of zero events.
         The data-to-background ratio is shown in the bottom panel of each figure.
         For the ratio plot, the statistical uncertainty is shown in light gray, while
         the total uncertainty, which is the quadratic sum of the statistical and systematic uncertainties, is shown in dark gray.
         \label{fig:mttbar_ljets}}
\end{figure*}

The reconstructed top-quark candidates are used to calculate the $\ttbar$ invariant mass. 
The events are divided into six categories,
three for each lepton+jets channel, so in total six \Mttbar distributions are obtained.
The three categories for each lepton flavor are: events with one CA8 $\cPqt$-tagged jet, 
events without a CA8 $\cPqt$-tagged jet but at least one $\cPqb$ tag, and events with neither 
a CA8 $\cPqt$-tagged jet nor a $\cPqb$ tag.
All six distributions are shown in Fig.~\ref{fig:mttbar_ljets}.

\subsection{All-hadronic channel}

When the top quark has large $\pt$ and decays hadronically,
all decay products frequently merge into a single jet.
Events with high \ttbar invariant mass, where both quarks decay
hadronically, thus effectively result in a dijet topology.
This forms the basis of the selection
in the all-hadronic channel. Two exclusive selections are
made, one optimized for higher resonance masses, and
one optimized for lower resonance masses where the decay products
are still somewhat collimated.

To satisfy the high-mass selection, events are required
to have two CA8 $\cPqt$-tagged jets with $\pt>400\GeV$ and rapidity
$\abs{y}<2.4$. The two jets have to be separated in azimuthal angle by
$\abs{\Delta\phi}>2.1$ radians.
The rapidity difference between the two leading jets is also used
to divide the events into two categories ($\abs{\Delta y}<1.0$
and $\abs{\Delta y}>1.0$), since the QCD multijet background with light-quark and gluon
final states dominantly populates the $\abs{\Delta y}>1.0$ category,
whereas the \PZpr signal with a mass of 2\TeV is equally split between
the two.
The two categories are further subdivided depending on the
number of CA8 jets containing a \cPqb-tagged subjet: zero, one, or two. This results
in six exclusive search regions, with the highest sensitivity in the
categories with two \cPqb-tagged CA8 jets.

The low-mass selection is applied to events failing the
high-mass selection and is designed to gain sensitivity in regions where
the decay products are less collimated.
Events are selected if two CA15 $\cPqt$-tagged jets with $\pt>200\GeV$ and
$\abs{y}<2.4$ are found.
The sample is split into events with $\HT<800\GeV$ and $\HT>800\GeV$, where $\HT$ is
defined as the scalar sum of jet $\pt$, including all jets with $\pt > 50$\GeV.
The sample is further categorized according to the number of \cPqb-tagged CA15 jets.

In order to estimate the background for the all-hadronic analysis, an
approach based on control samples in data is applied. A sideband is selected by inverting the
CA8 $\cPqt$ tagging minimum mass requirement on one of the jets in the dijet
sample. For the low-mass analysis, the CA15 $\cPqt$~tagging selection criteria based on the subjet
invariant mass and pairwise masses are inverted. The other leading CA jet in the event provides a 
kinematically unbiased ensemble of
non-top-quark jets to measure the mistag rate. This mistag rate is
then applied to the events where exactly one jet passes the
$\cPqt$~tagging selection. These events have a higher gluon fraction than
the events used to derive the mistag rate for the lepton+jets
analysis, as mentioned above.

The misidentification probability, $r$, for a non-$\cPqt$-quark CA jet to be identified as a $\cPqt$-tagged CA jet,
is parametrized by three variables, the jet
$\pt$, the $N$-subjettiness ratio $\tau_{32}$, and the jet $\cPqb$ tagging discriminant $\beta$ (the output of the CSV algorithm described above),
$r = r(\pt,\tau_{32}, \beta)$. The variable $\tau_{32}$ is not used for
the low-mass analysis because it does not enhance the sensitivity of
the search.
In this case, the mistag rate is parametrized as a function of two variables only,
using the same procedure. The mistag rate is binned, defined as
$r_{i,j,k}$. To estimate the non-top-quark multijet background arising from
mistagging light jets, a
four-dimensional array of counts $N_{\alpha,i,j,k}$ is measured in the single t-tagged data sample,
where $\alpha$ is the bin of the variable of interest
(in this case, $\Mttbar$), given by
$N_\alpha = \sum_{a=1}^{N_\text{jets}} N_{\alpha,i,j,k}\, r_{i,j,k}$,
where the indices $i,j,k$ are the bins in $\pt$, $\tau_{32}$, and
$\beta$ in which jet ``$a$'' lies.
The four-dimensional parametrization properly accounts for correlated
and uncorrelated statistical uncertainties.
The uncertainty in each
bin of the predicted mistagged distribution has two parts:
one arises from the misidentification probability, and the other from the
number of jets in the ensemble; they are uncorrelated and are added
in quadrature.  The details of this procedure are given
in Appendix~\ref{sec:appendix_allhad_errors}.

\begin{figure}[htb]
\centering
\includegraphics[width=0.45\textwidth]{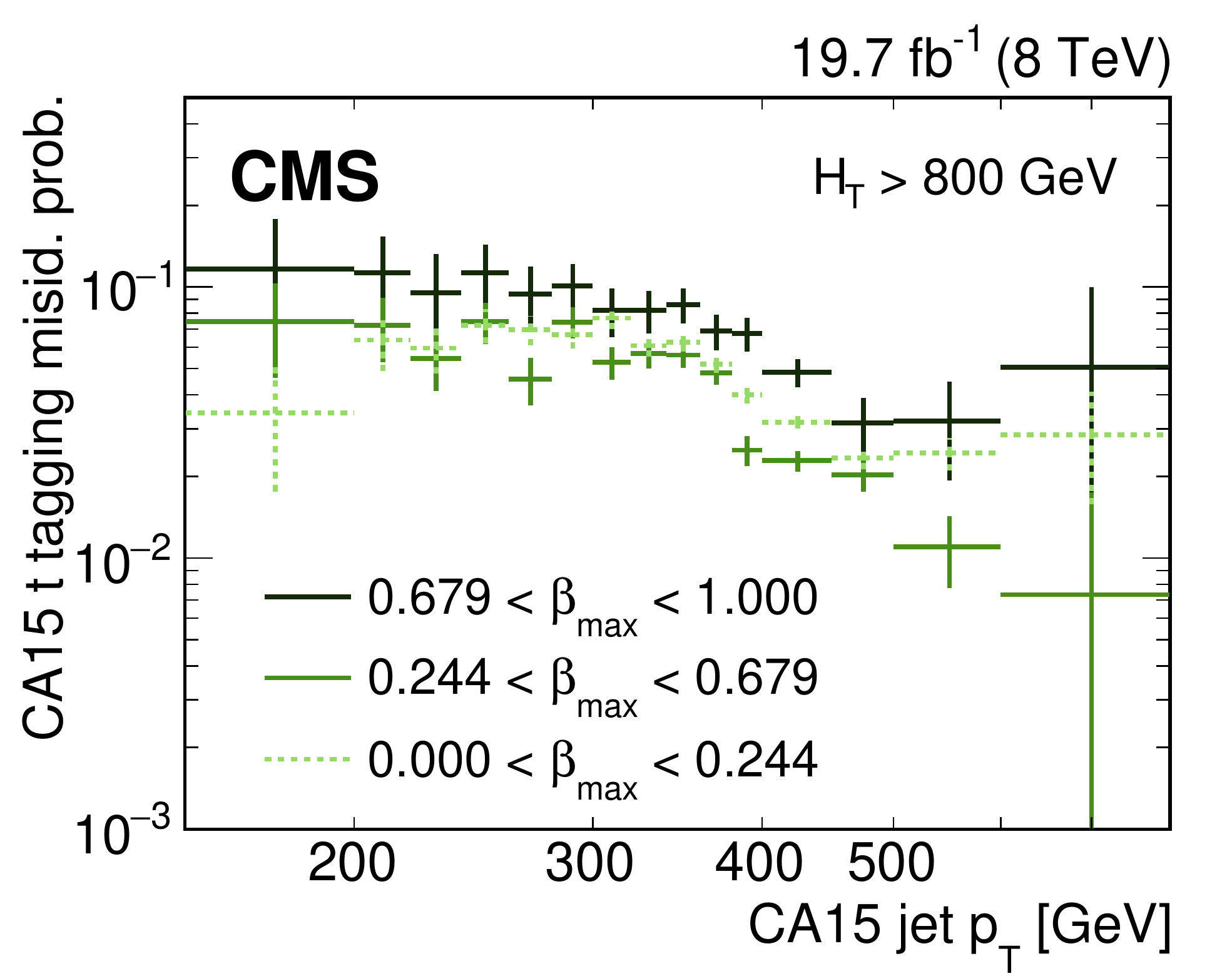}
\includegraphics[width=0.45\textwidth]{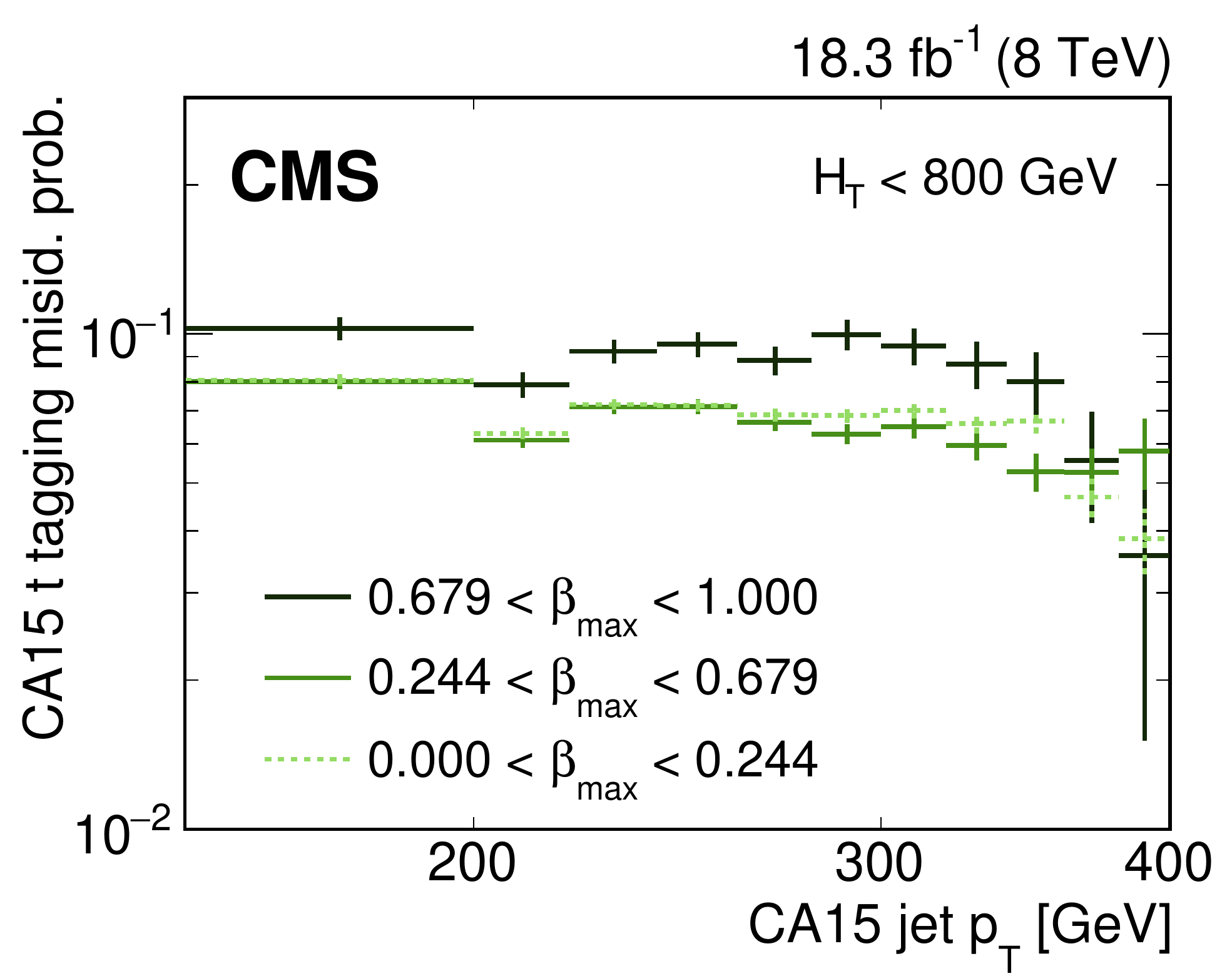}
\caption{Misidentification probability for CA15 jets to be tagged as
  top-quark jets for different $\beta$ values for $\HT > 800 \GeV$ (\cmsLeft) and $\HT < 800 \GeV$ (\cmsRight) in the low-mass all-hadronic analysis. The horizontal error bars indicate the bin width.
\label{fig:mistagHEP}
}
\end{figure}

\begin{figure}[htbp]
\centering
\includegraphics[width=0.45\textwidth]{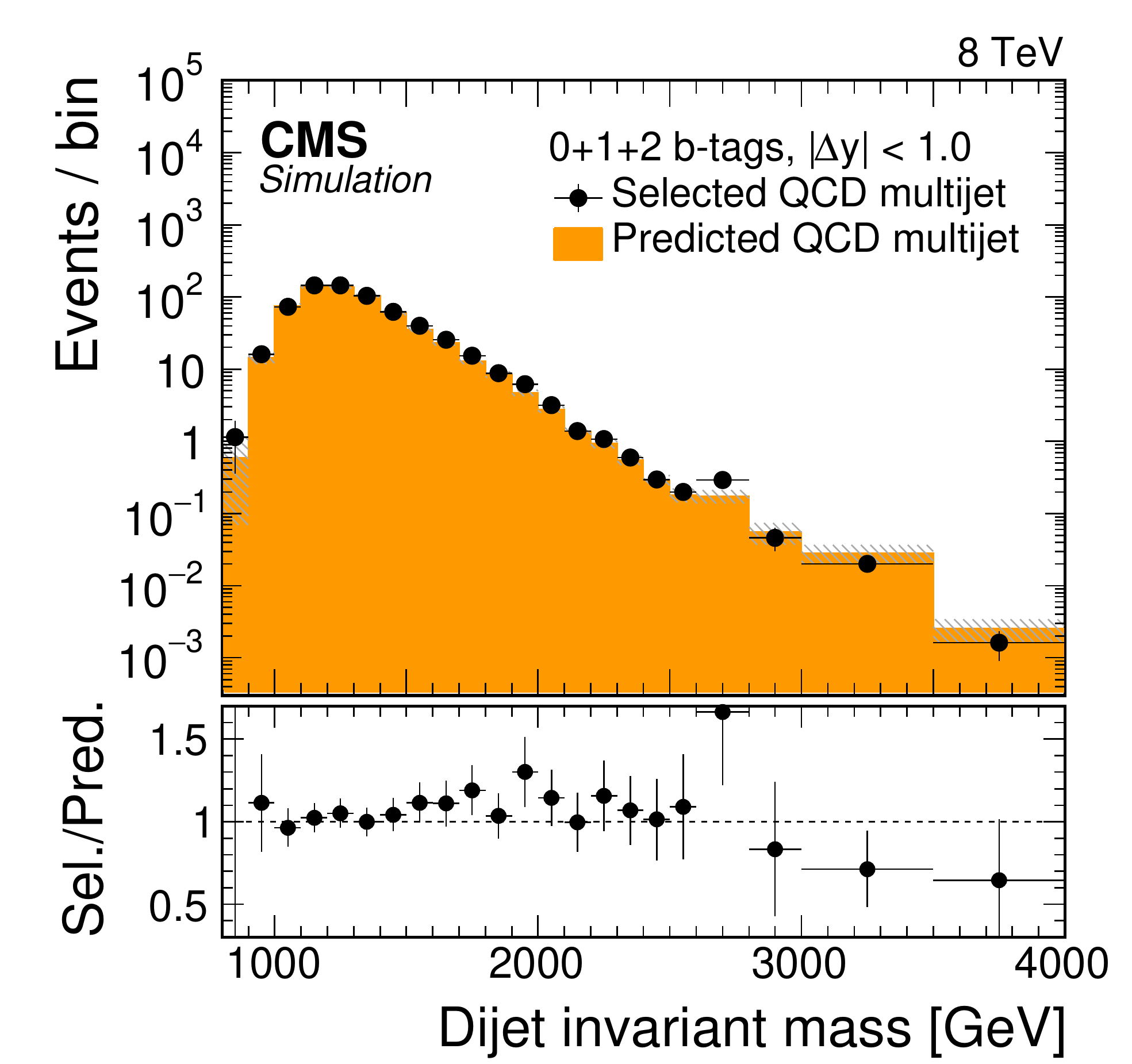}
\includegraphics[width=0.45\textwidth]{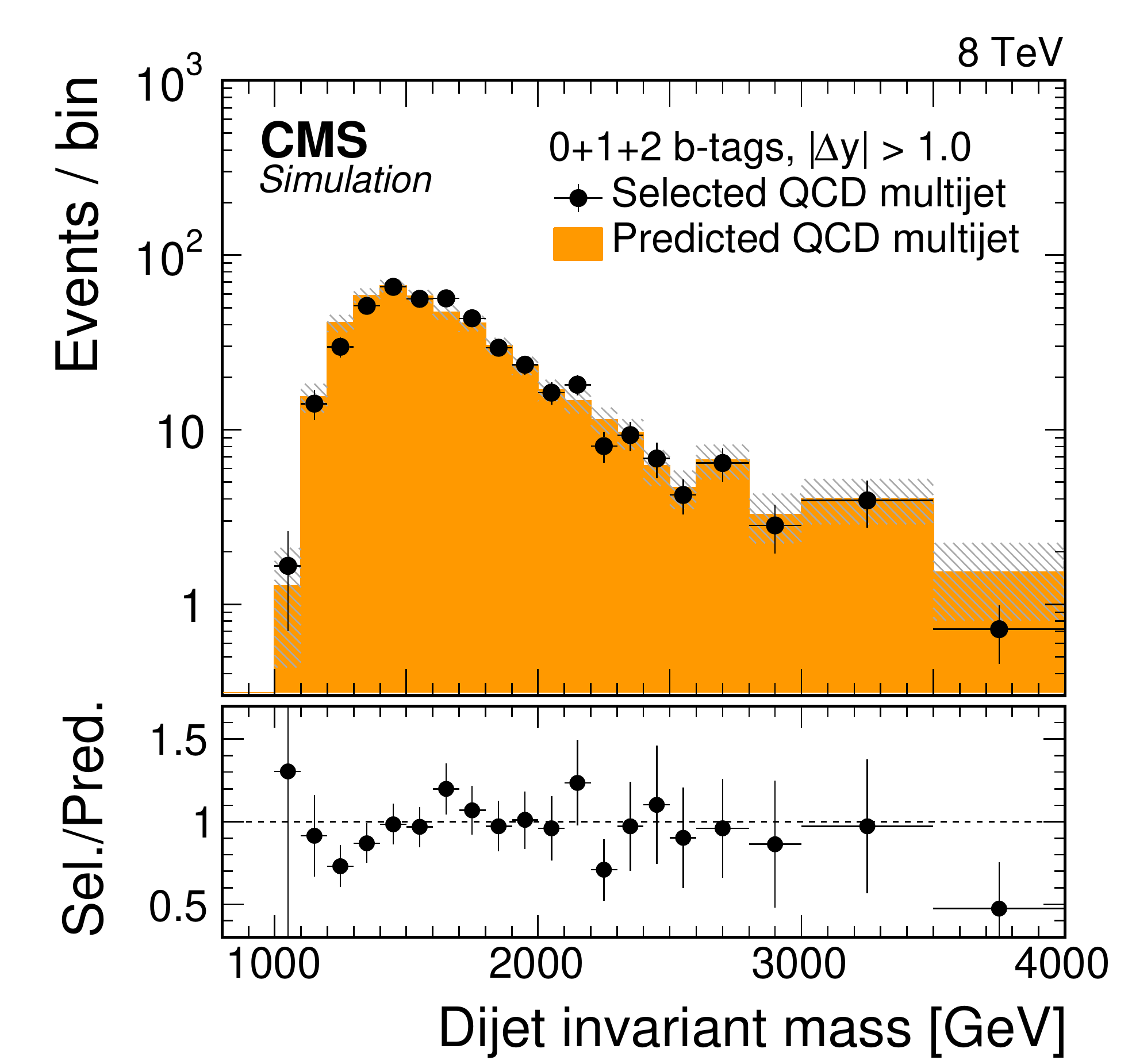}
\caption{Results of the validation test for the high-mass all-hadronic analysis,
using simulated QCD multijet events, to validate the data-driven background method used to estimate the QCD multijet contribution.
Events are shown without any selection or division applied based on the number of identified \cPqb-tagged jets for
$\abs{\Delta y} < 1.0$ (\cmsLeft) and $\abs{\Delta y} > 1.0$ (\cmsRight).
The points show the selected QCD multijet events in the signal region,
with the horizontal error bars indicating the bin width. The solid histogram shows the predicted
number of QCD multijet events using the misidentification probability for CA8 $\cPqt$-tagged jets
measured in a statistically independent sideband region.
The statistical uncertainty is shown as a shaded region.
The ratio of selected to predicted events is shown in the bottom panel of each figure.
}
\label{fig:allhad_closuretest}
\end{figure}

\begin{figure}[htbp]
\centering
\includegraphics[width=0.45\textwidth]{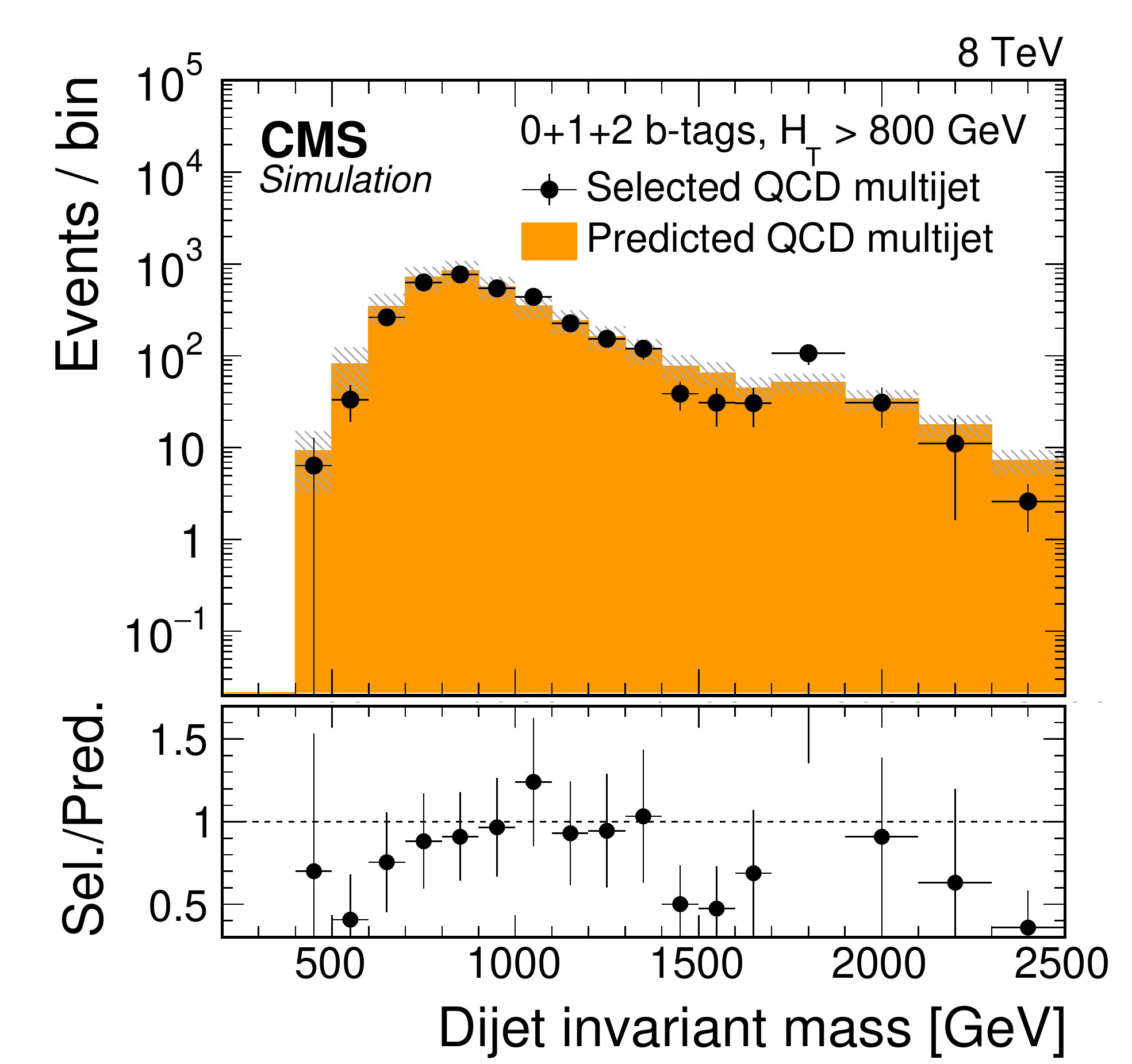}
\includegraphics[width=0.45\textwidth]{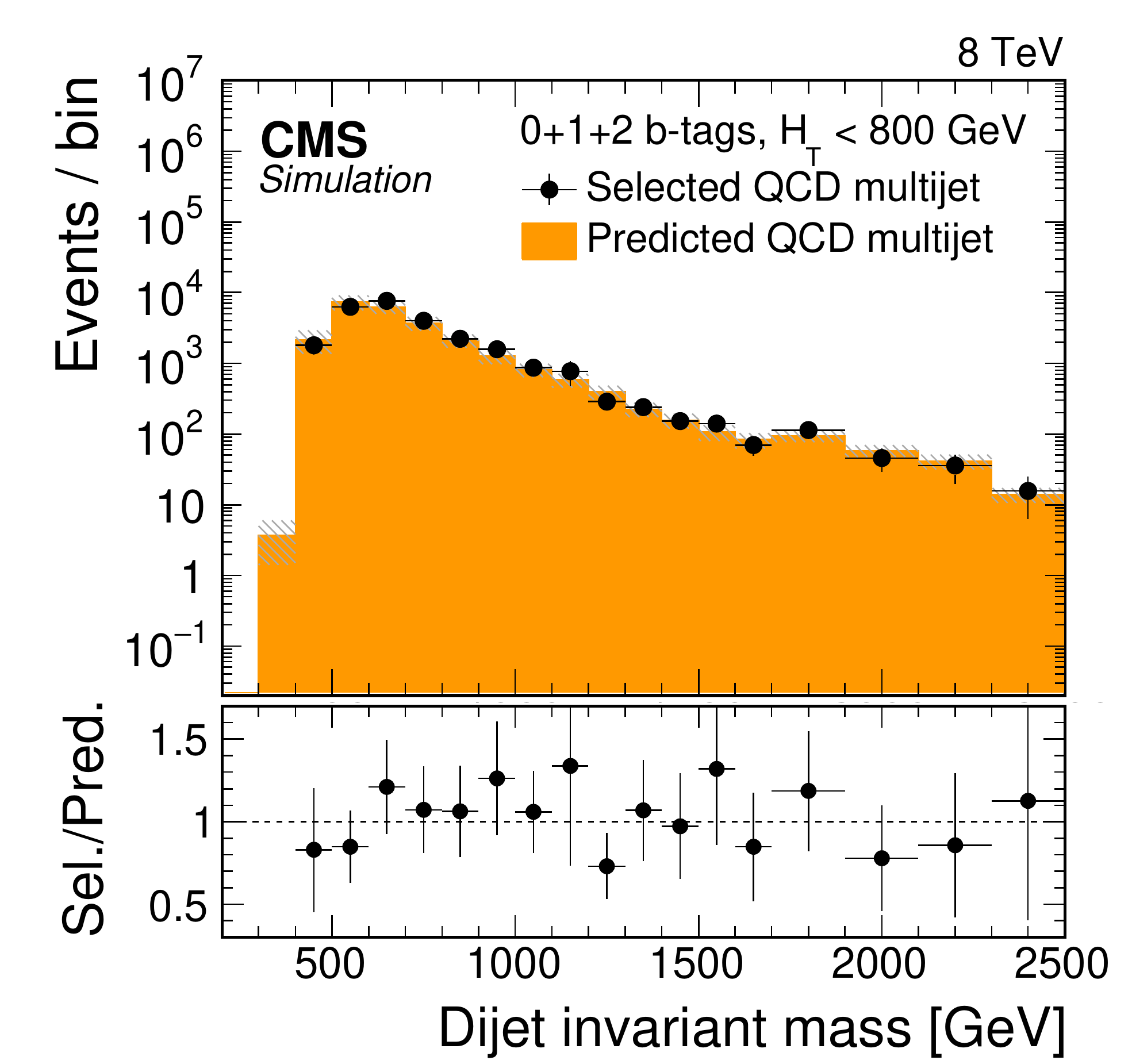}
\caption{Results of the validation test for the low-mass all-hadronic analysis,
using simulated QCD multijet events, to validate the data-driven background method used to estimate the QCD multijet contribution for
events with $\HT>800\GeV$ (\cmsLeft), and events with $\HT<800\GeV$ (\cmsRight).
The points show the selected QCD multijet events in the signal region,
with the horizontal error bars indicating the bin width. The solid histogram shows the predicted
number of QCD multijet events using the misidentification probability for CA15 $\cPqt$-tagged jets
measured in a statistically independent sideband region.
The statistical uncertainty is shown as a shaded region.
The ratio of selected to predicted events is shown in the bottom panel of each figure.
}
\label{fig:allhad_closuretestHEP}
\end{figure}

Figure~\ref{fig:mistagTau32} shows the CA8 $\cPqt$ tagging misidentification probability as a
function of the CA8 jet $\pt$ for different bins of $\tau_{32}$ and $\beta$. Figure~\ref{fig:mistagHEP} shows the CA15
$\cPqt$ tagging misidentification probability as a
function of the CA15 jet $\pt$ for different values of $\beta$ for the low-mass analysis.
Figures~\ref{fig:allhad_closuretest} and ~\ref{fig:allhad_closuretestHEP} show validation of this
procedure on QCD simulation in the various tagging categories for the high- and low-mass analysis, respectively.
Good agreement between the predicted and selected contribution from QCD multijet production
is observed in both analyses.

\begin{figure*}[htb]
\centering
\includegraphics[width=0.45\textwidth]{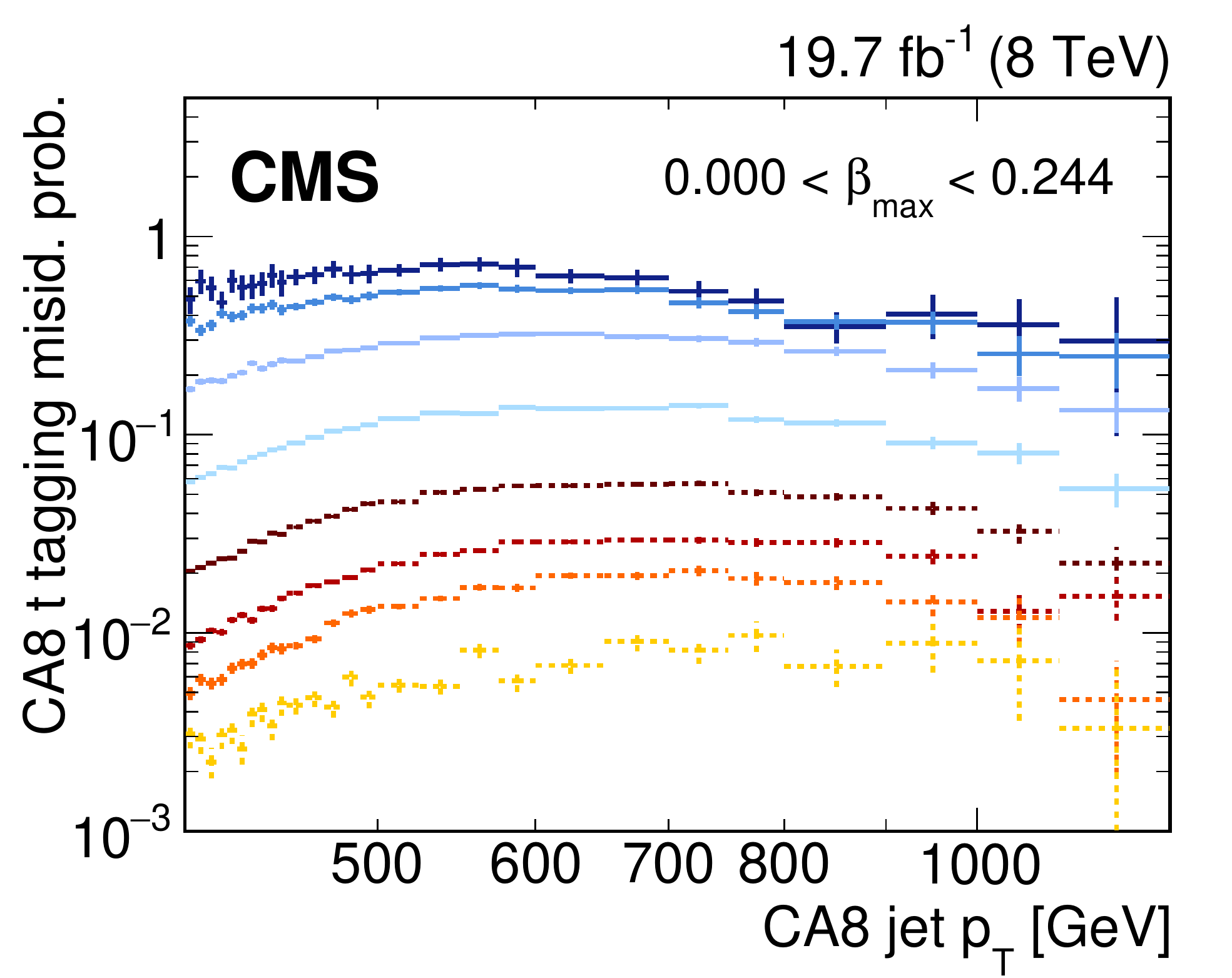}
\includegraphics[width=0.45\textwidth]{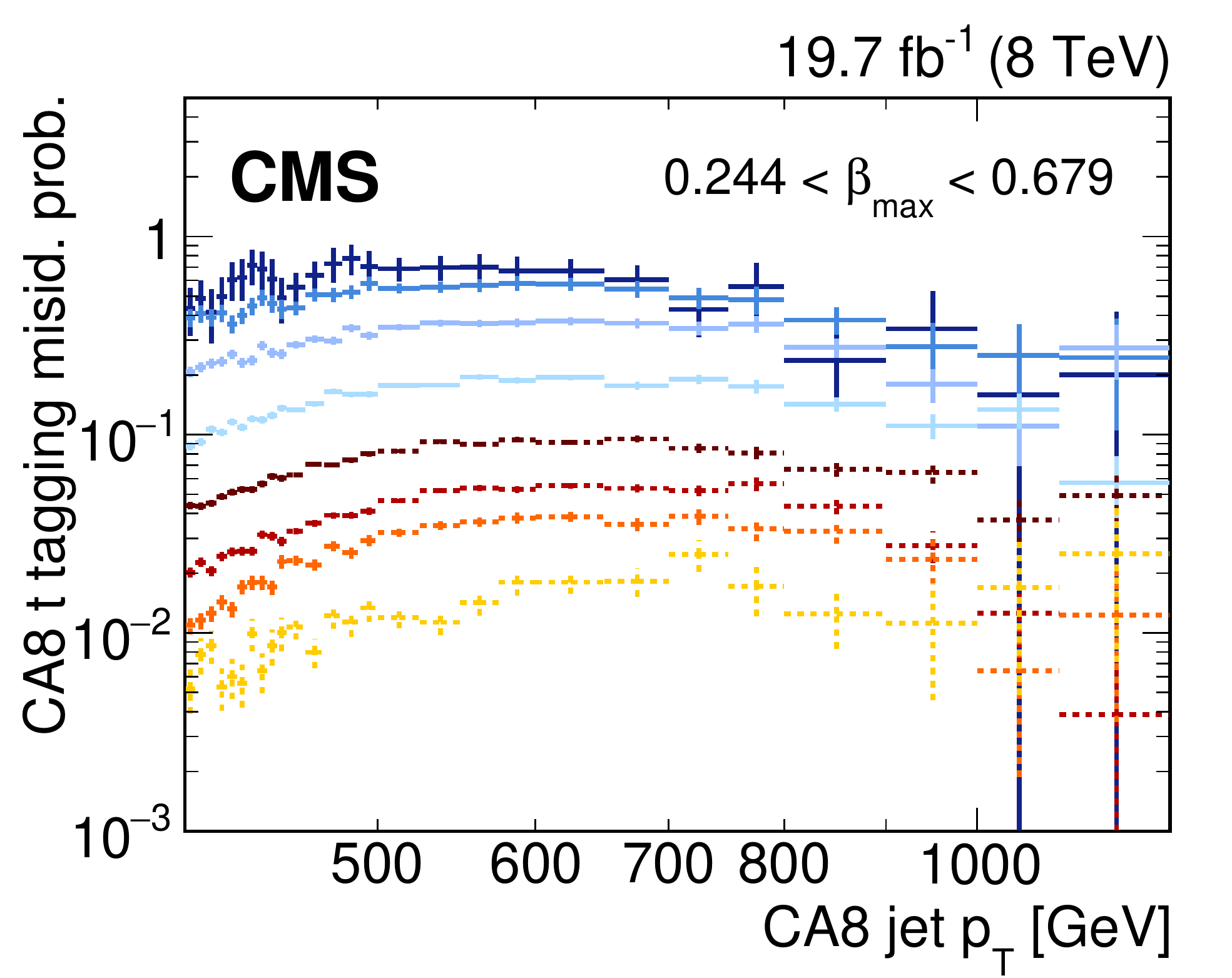} \\
\includegraphics[width=0.45\textwidth]{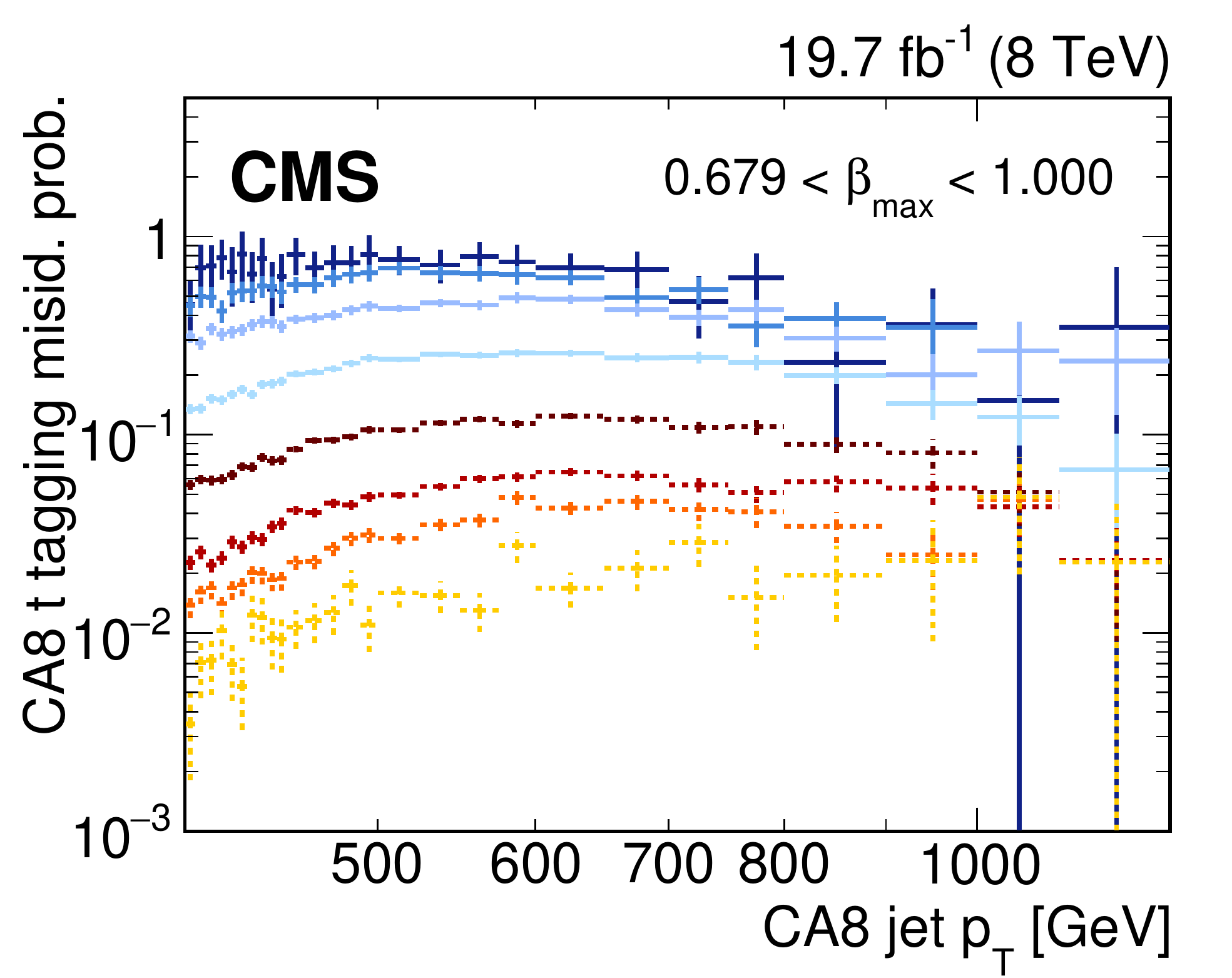}
\includegraphics[width=0.45\textwidth]{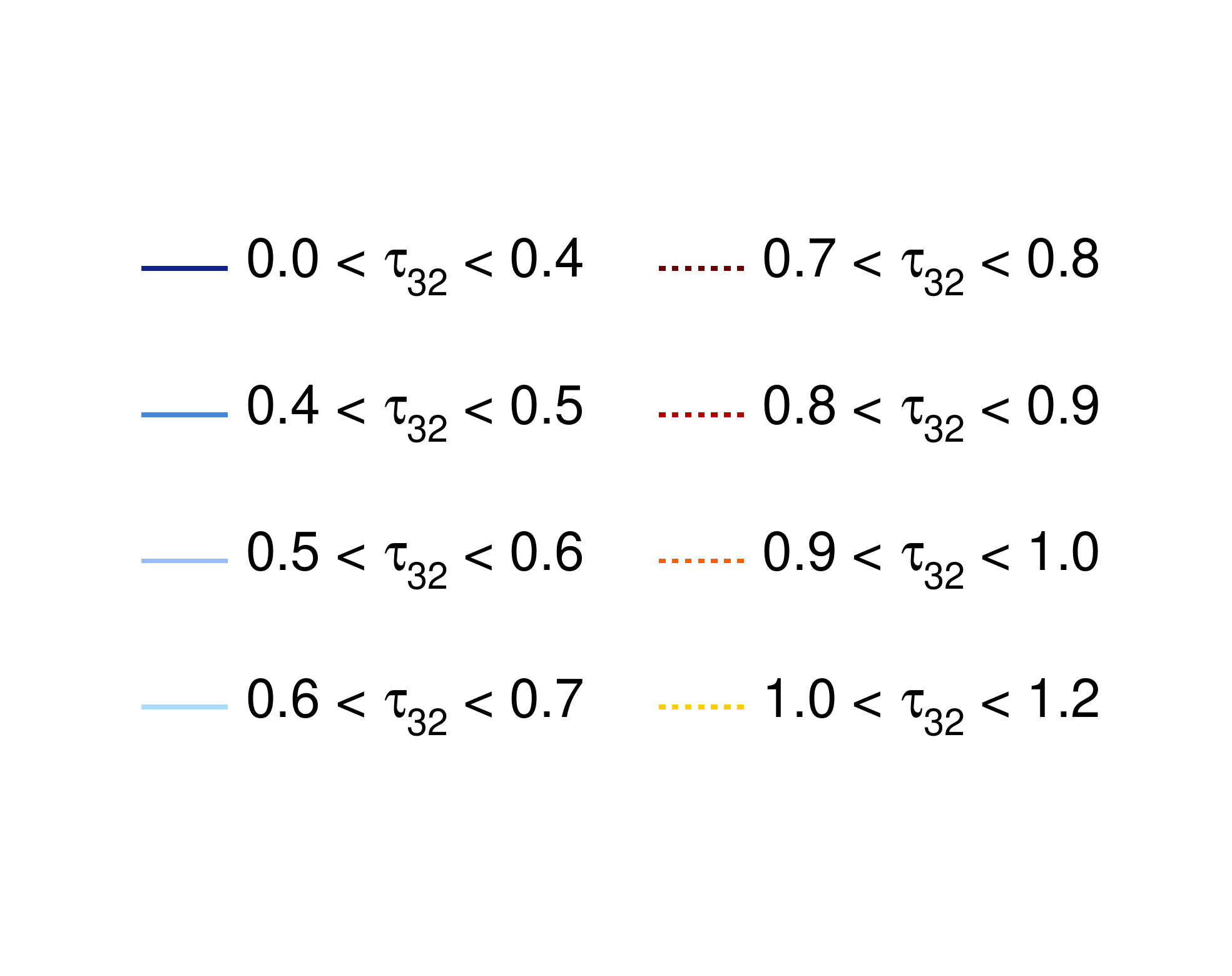}
\caption{Misidentification probability for CA8 jets to be tagged as top-quark jets for different
$\beta$ ranges and for different $\tau_{32}$ values in the high-mass all-hadronic analysis. The horizontal error bars indicate the bin width.
\label{fig:mistagTau32}
}
\end{figure*}

\begin{figure*}[tbp]
\centering
\includegraphics[width=0.41\textwidth]{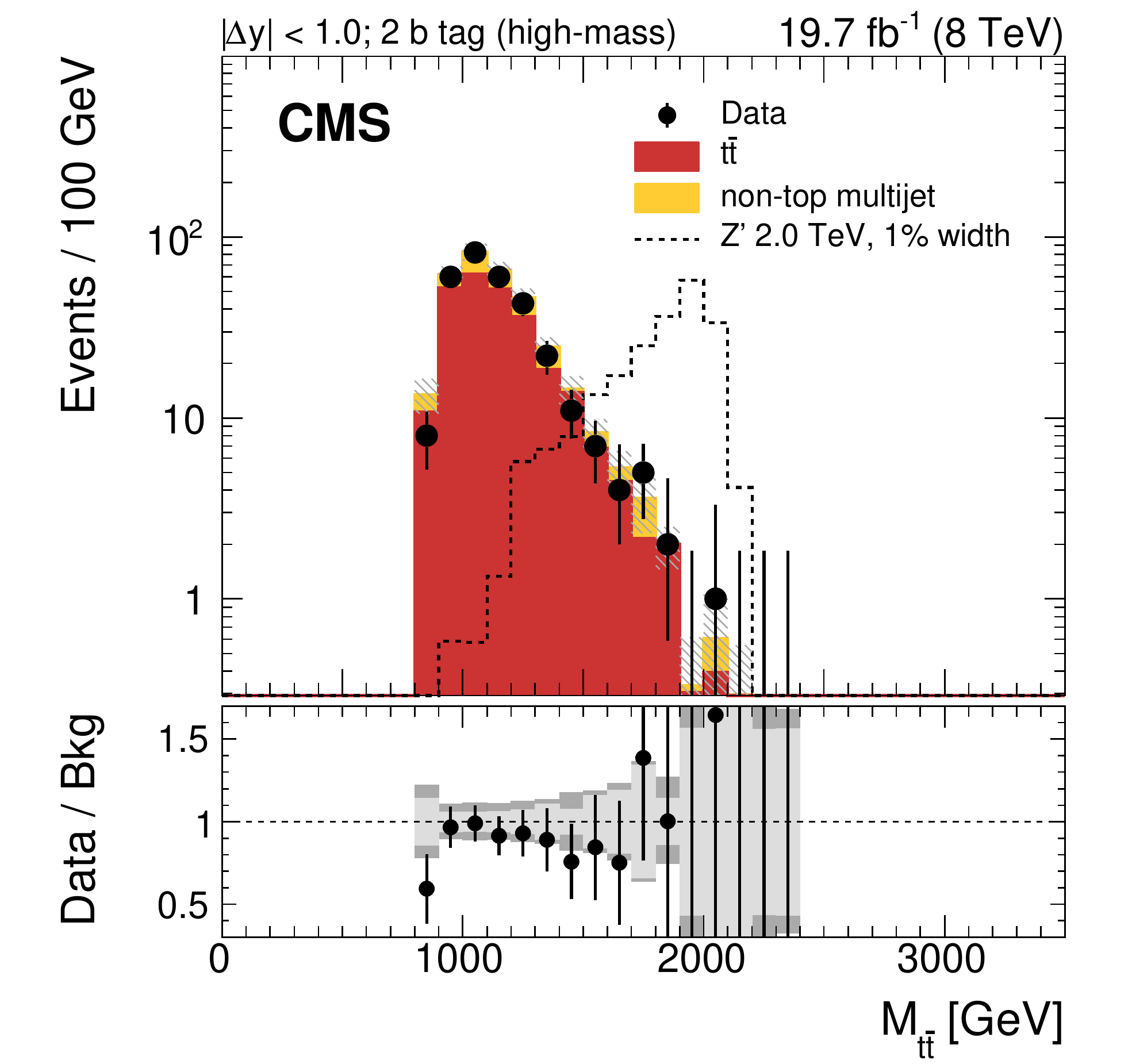}
\includegraphics[width=0.41\textwidth]{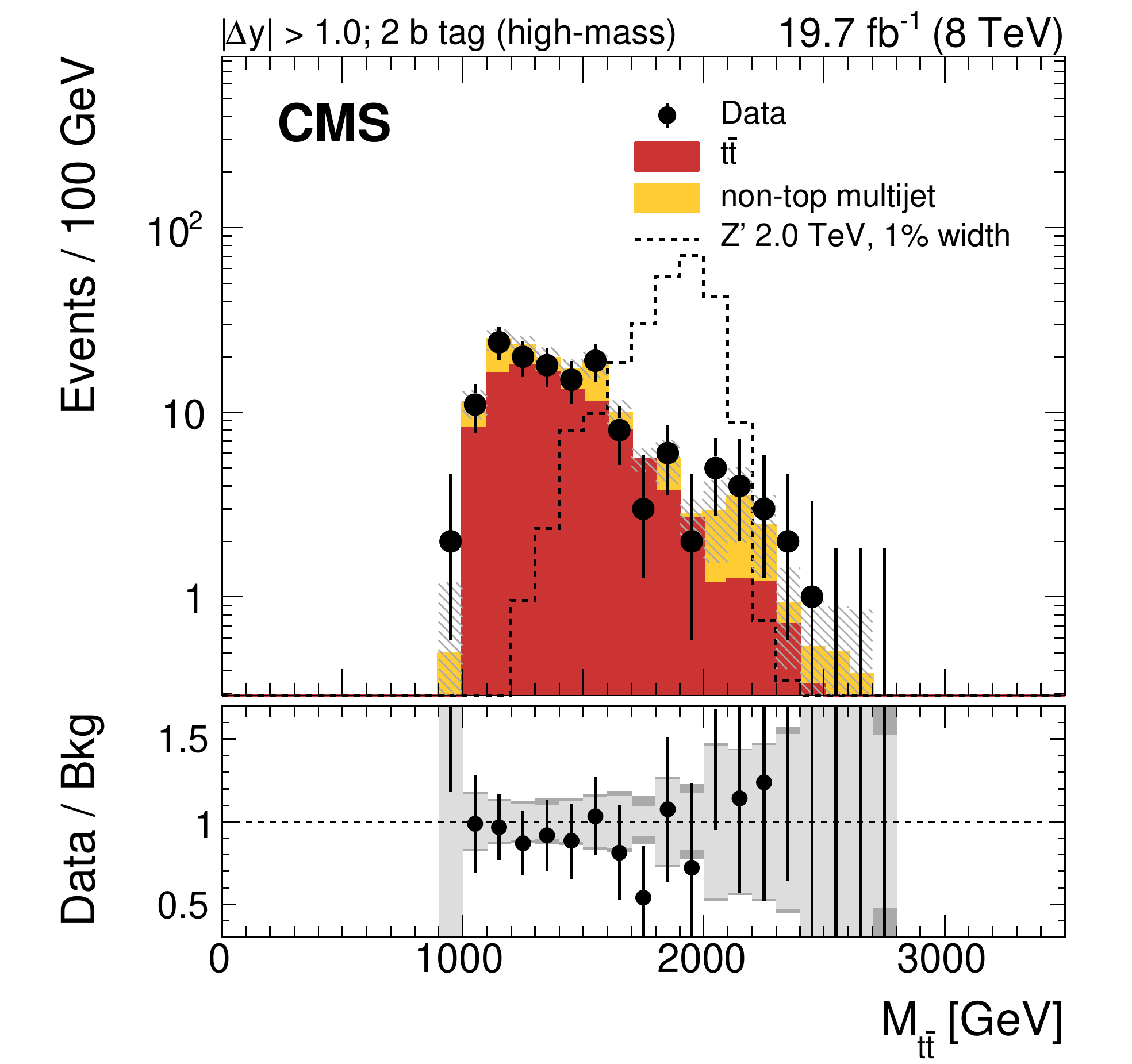}
\includegraphics[width=0.41\textwidth]{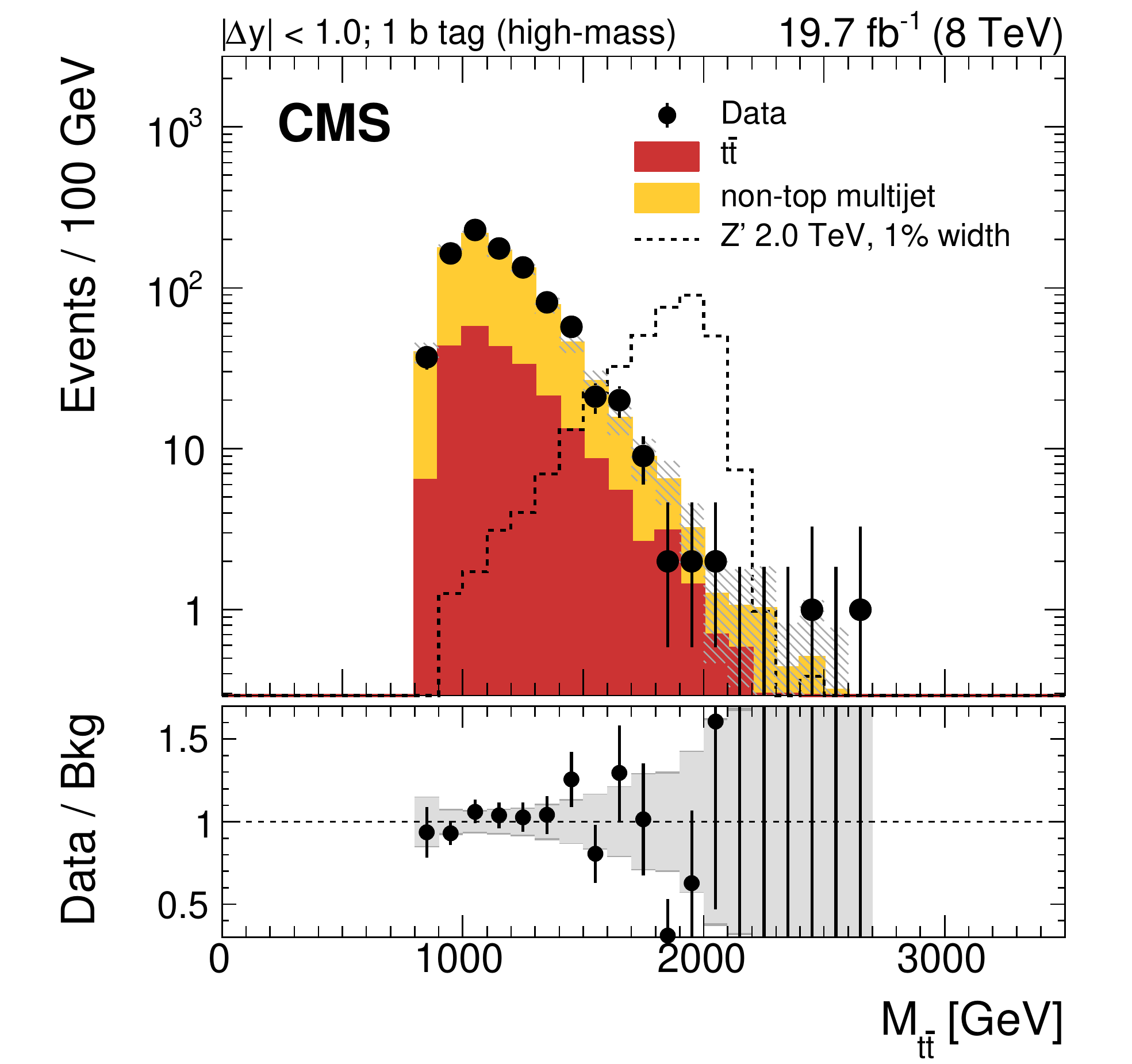}
\includegraphics[width=0.41\textwidth]{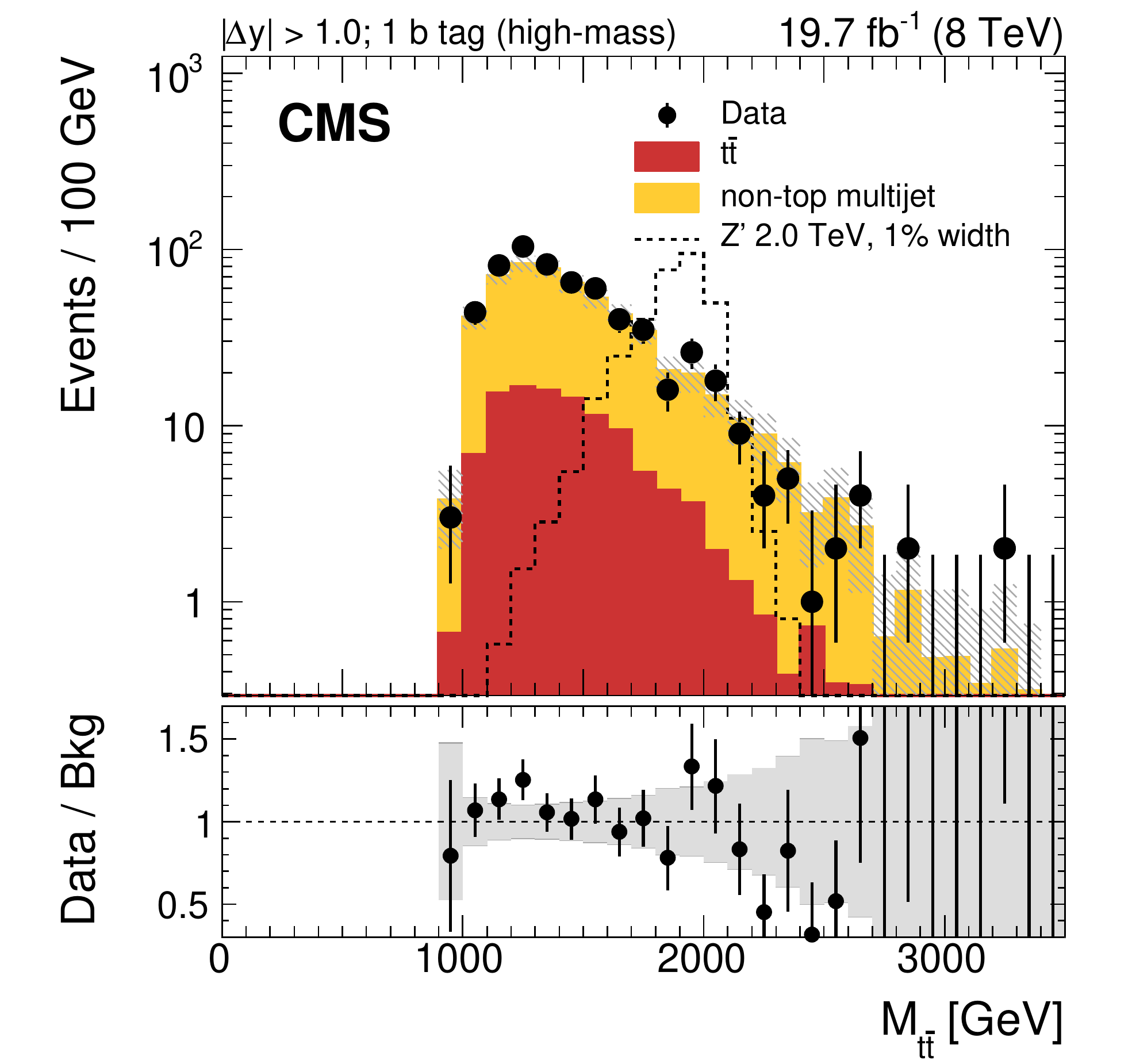}
\includegraphics[width=0.41\textwidth]{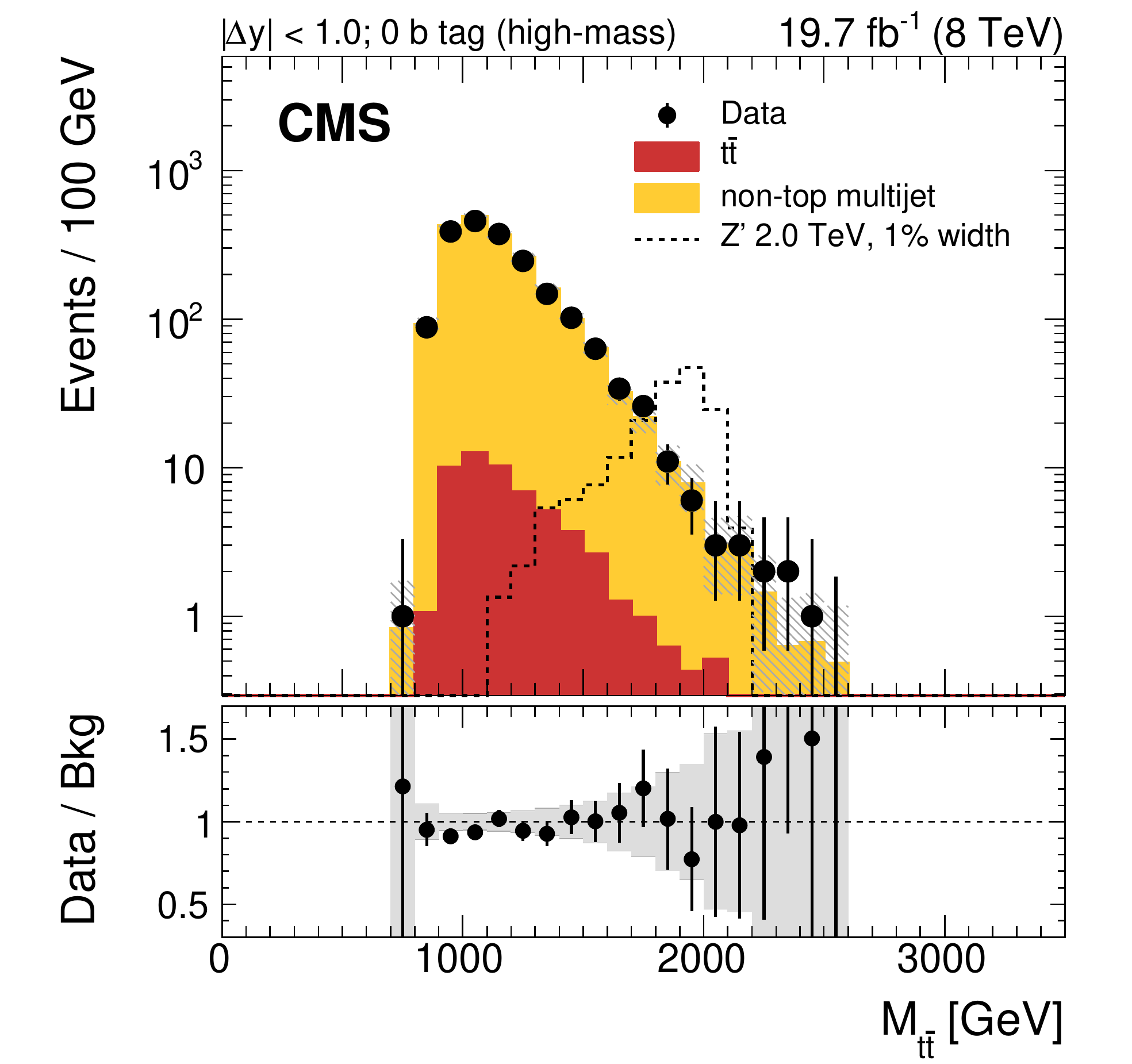}
\includegraphics[width=0.41\textwidth]{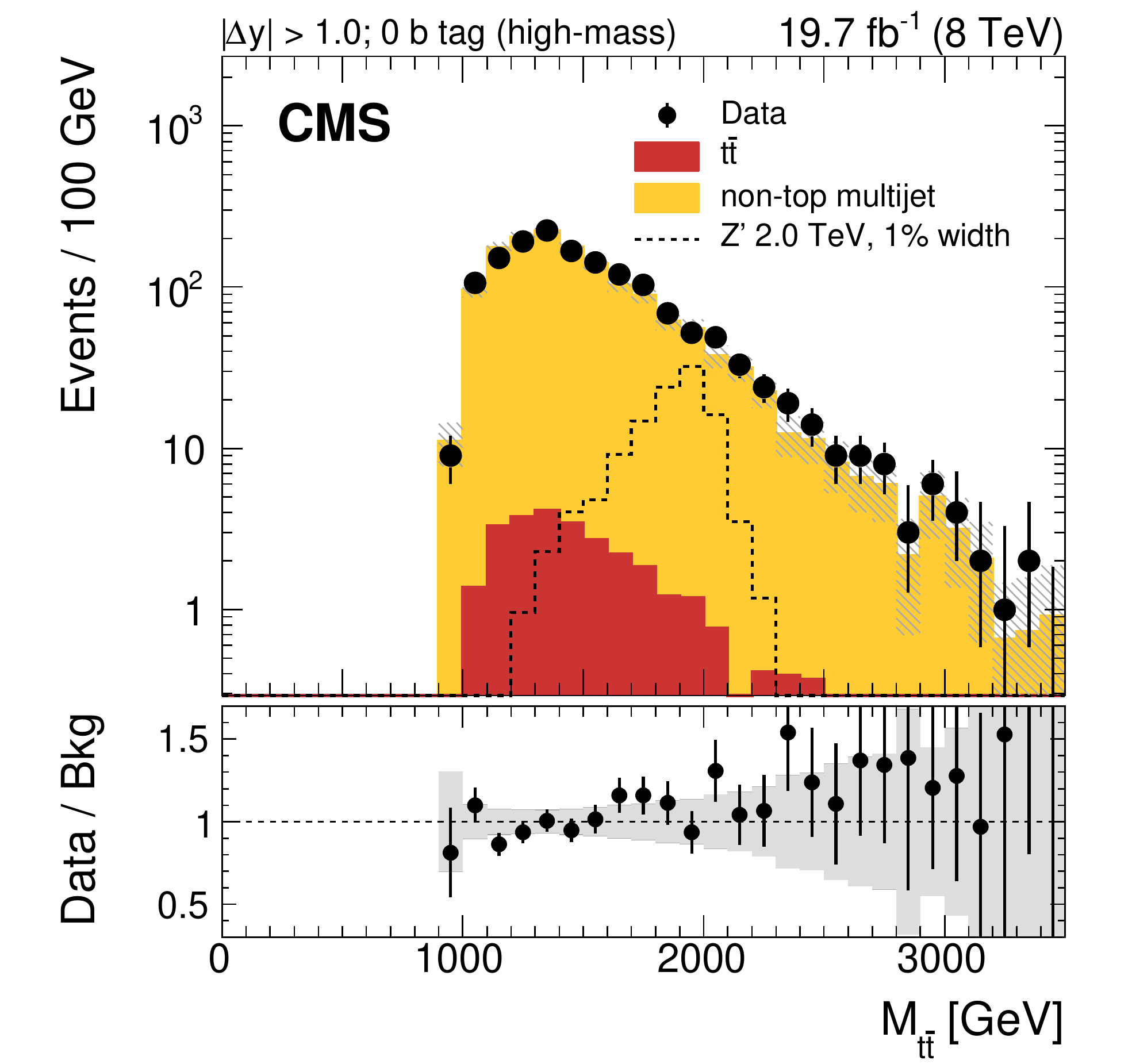}
\caption{Reconstructed invariant mass of the $\ttbar$ pair
         in the all-hadronic channel for data and simulated events passing the high-mass selection.
         Events are divided into six categories: events with two subjet \cPqb-tags and $\abs{\Delta y} < 1.0$ (upper left),
	 one subjet \cPqb tag and $\abs{\Delta y} < 1.0$ (middle left), no subjet
         \cPqb tag and $\abs{\Delta y} < 1.0$ (lower left), two subjet \cPqb tags and $\abs{\Delta y} > 1.0$ (upper right),
         one subjet \cPqb tag and $\abs{\Delta y} > 1.0$ (middle right), no subjet \cPqb tag and $\abs{\Delta y} > 1.0$ (lower right).
         The uncertainty associated with the background expectation includes all the statistical and
         systematic uncertainties.
         For bins with no events in data but with non-zero background expectation, 
         vertical lines are shown indicating the 68\% confidence level coverage 
         corresponding to an observation of zero events.
         The data-to-background ratio is shown in the bottom panel of each figure.
         For the ratio plot, the statistical uncertainty is shown in light gray, while
         the total uncertainty, which is the quadratic sum of the statistical and systematic uncertainties, is shown in dark gray. The expected distribution
         from a \PZpr signal with $\Mzp = 2\TeV$ and $\rWzp = 0.01$ is also shown, normalized to a cross section of 1\pb.
         \label{fig:mttbar_cmstt}}
\end{figure*}

\begin{figure}[tbp]
\centering
\includegraphics[width=0.41\textwidth]{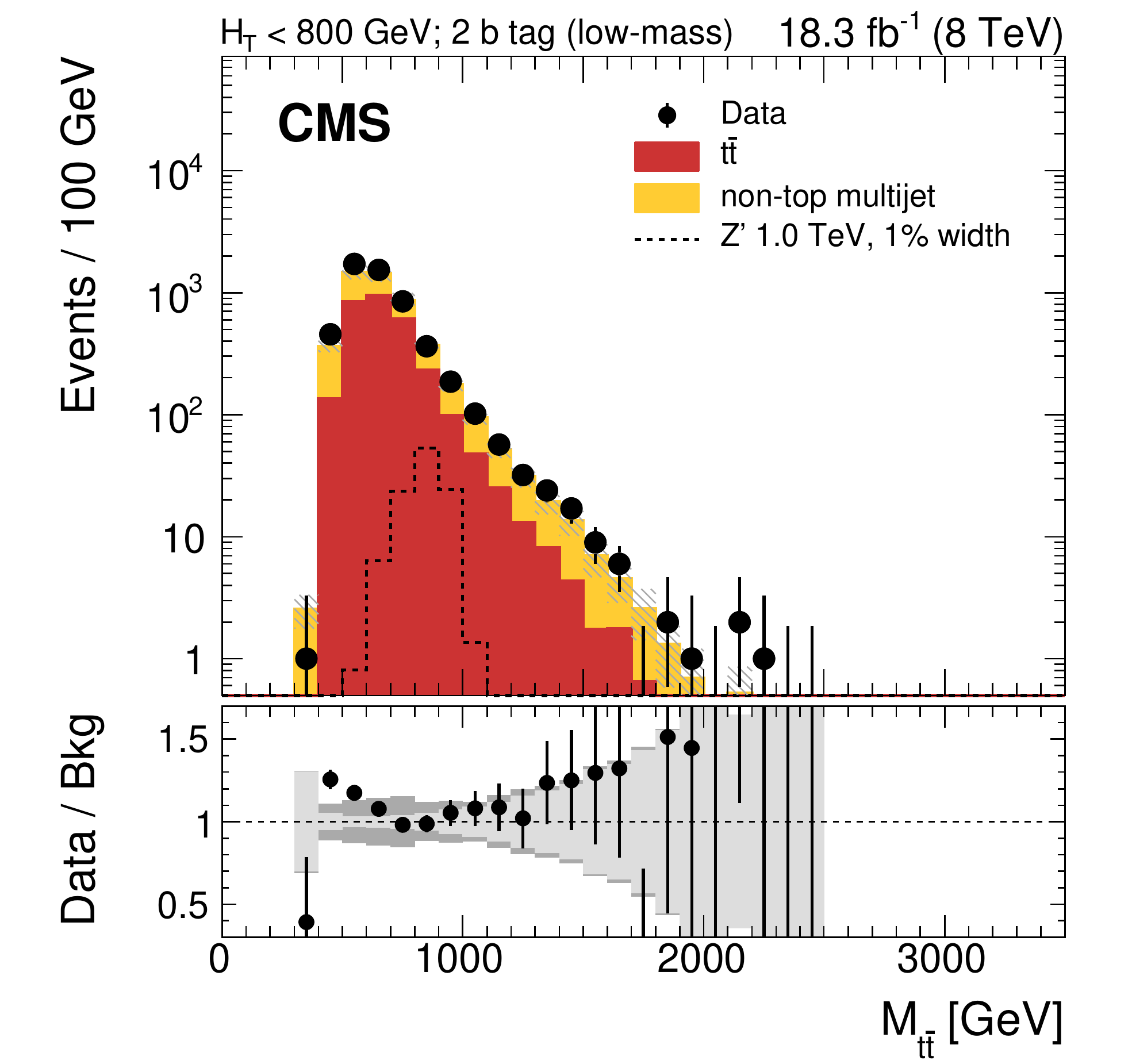}
\includegraphics[width=0.41\textwidth]{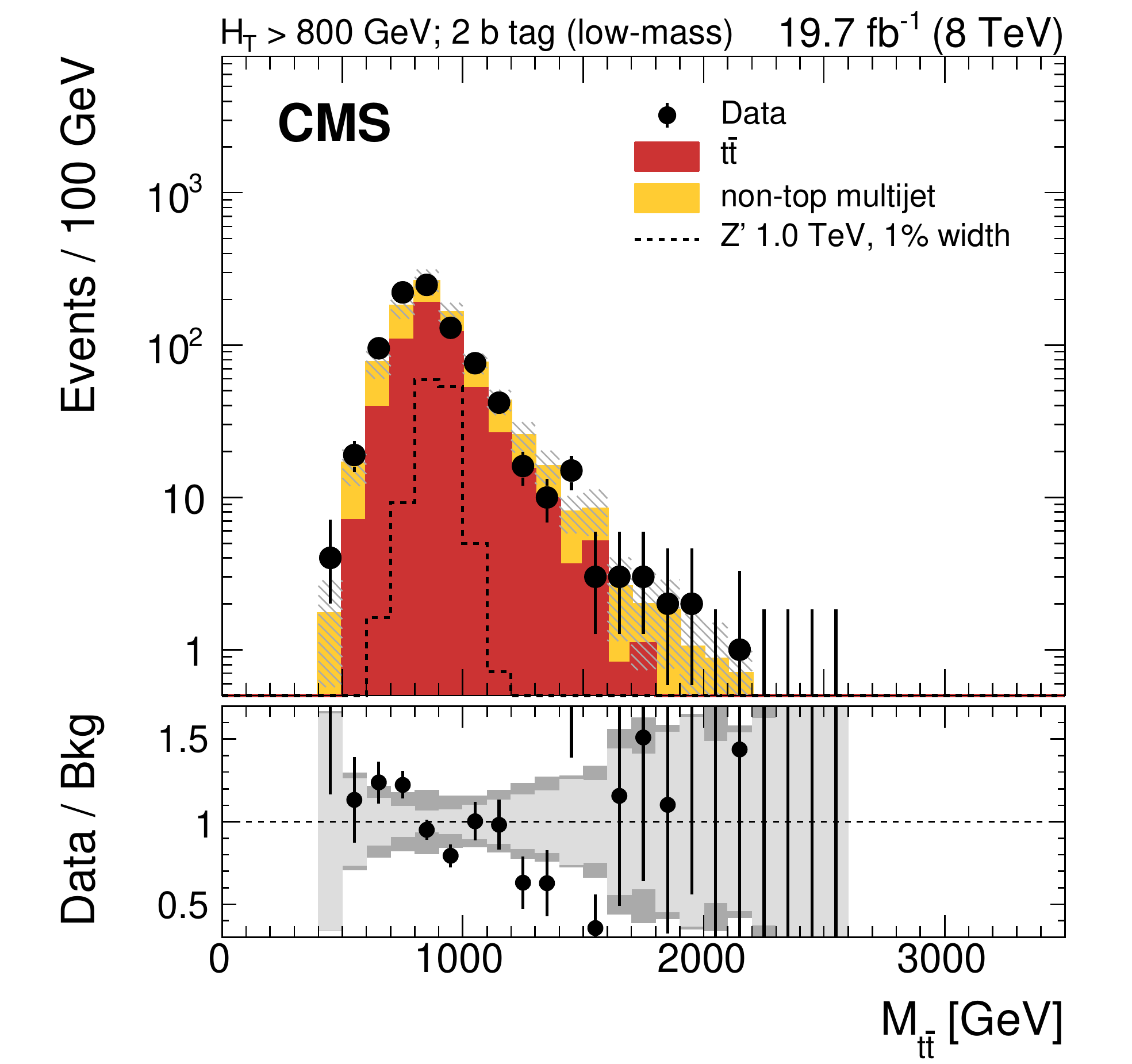}
\caption{Reconstructed invariant mass of the $\ttbar$ pair
         in the all-hadronic channel for data and simulated events passing the low-mass selection.
         Events with two subjet $\cPqb$ tags are shown, for $\HT<800\GeV$ (\cmsLeft) and
         $\HT>800\GeV$ (\cmsRight).
         The signal is normalized to a cross section of 1\unit{pb}.
         The uncertainty associated with the background expectation includes all the statistical and
         systematic uncertainties.
         For bins with no events in data but with non-zero background expectation, 
         vertical lines are shown indicating the 68\% confidence level coverage 
         corresponding to an observation of zero events.
         The data-to-background ratio is shown in the bottom panel of each figure.
         For the ratio plot, the statistical uncertainty is shown in light gray, while
         the total uncertainty, which is the quadratic sum of the statistical and systematic uncertainties, is shown in dark gray. The expected distribution
         from a \PZpr signal with $\Mzp = 1\TeV$ and $\rWzp = 0.01$ is also shown, normalized to a cross section of 1\pb.
         \label{fig:mttbar_htt}}
\end{figure}

The results of the high-mass selection in the all-hadronic channel are shown in
Fig.~\ref{fig:mttbar_cmstt} for events with $\abs{\Delta y} < 1.0$ and $\abs{\Delta y} > 1.0$
in the three $\cPqb$-tagged categories.
The distributions of \Mttbar obtained with the low-mass selection
are shown in Fig.~\ref{fig:mttbar_htt} for events with two subjet $\cPqb$ tags,
for $\HT>800\GeV$ and $\HT<800\GeV$.
The \ttbar background process is scaled by a factor
derived from the maximum likelihood fit to data as explained in
Section~\ref{sec:background}, and the
non-top-quark multijet background is obtained from data in a sideband region.

\section{Systematic uncertainties}
\label{sec:systematics}
The sources of systematic uncertainties considered in these analyses
are summarized in Table~\ref{tab:corr_systematics}.
Uncertainties originating from the same
source are assumed to be 100\% correlated between all channels.
The uncertainties can affect the normalization, the shape,
or both normalization and shape of the \Mttbar distribution.

\subsection{Uncertainties affecting the normalization}

The following systematic uncertainties in the normalization of the background processes are considered.
The uncertainty in the cross section for SM $\ttbar$ production is 15\%~\cite{CMS-PAS-TOP-12-028}.
Uncertainties in the production cross sections of $\PW$+jets are 9\%
for light-flavor jets~\cite{Khachatryan:2014uva} and 23\% for
heavy-flavor jets~\cite{Chatrchyan:2013uza}.
An uncertainty of 50\% is assigned to the cross section of $\Z$+jets production,
obtained by varying the renormalization and factorization scales simultaneously by
factors of 0.5 and 2.
The largest background contribution from single top quark production
originates from the $\cPqt\PW$ channel,
which has been measured with an accuracy of 23\%~\cite{Chatrchyan:2014tua};
this uncertainty is used for all electroweak single top production processes.
The uncertainty in diboson production is 20\%~\cite{Chatrchyan:2013oev, Chatrchyan:2014aqa}.

In addition, the following systematic uncertainties affect the normalization of all simulated processes,
including signal processes.
The uncertainty in the measurement of the integrated luminosity is $2.6\%$~\cite{lumi}.
The combined trigger used in the electron category in the lepton+jets
channel has an efficiency uncertainty of 1\%.
The uncertainty due to the single-muon trigger efficiency is 1\%, which affects the muon category in the
lepton+jets channel and the $\Pe\Pgm$ and $\Pgm\Pgm$ categories in the dilepton channel.

\subsection{Uncertainties affecting the shape}

Systematic uncertainties due to the electron identification are applied as a function of electron $\pt$ and $\eta$
to events with an identified electron in the dilepton and lepton+jets channels.
The uncertainty in the efficiency of the single electron trigger is applied
as a function of electron $\pt$ and $\eta$ and affects the $\Pe\Pe$ dilepton channel.
Systematic uncertainties due to the muon identification and trigger efficiencies are applied as a function of muon
$\pt$ and $\eta$ and affect events with a muon in the dilepton and lepton+jets channels.

The efficiencies of the $\HT$ and four-jet triggers are measured as a function of $\HT$ and
the $\pt$ of the fourth jet in the event, respectively. A correction is applied for the different behavior
between data and simulation in the region where the triggers are not fully efficient.
A systematic uncertainty of half the size of this correction is assigned, with a minimum of 2\%
in regions where the triggers are fully efficient as determined from MC studies of the efficiency.
These uncertainties affect the all-hadronic analyses.
Uncertainties in the jet energy scale and resolution are of the order
of a few percent, and are
functions of jet $\pt$ and $\eta$. These are taken into account in all channels. These
uncertainties are also propagated to the estimation of $\ptvecmiss$.
The systematic uncertainty associated with the pileup reweighting procedure is assumed to be fully
correlated among all channels and is evaluated by varying the minimum bias cross section.

Efficiencies and mistag rates of the $\cPqb$ tagging algorithm have been measured in
data and simulated events for jets~\cite{Chatrchyan:2012jua} and
subjets with a spatial separation between them
of $\Delta R > 0.3$~\cite{CMS-PAS-BTV-13-001}.
The corresponding uncertainty is correlated between the dilepton, lepton+jets, and
the low-mass category of the all-hadronic channel.
The high-mass selection in the all-hadronic channel uses subjet $\cPqb$ tagging in a
collimated region, where the subjets are separated by $\Delta R < 0.3$.
The applicability of the standard $\cPqb$ tagging correction factors and uncertainties is not
guaranteed in this kinematic regime, because of double-counting of
tracks from subjets, which are used in the $\cPqb$ tagging algorithm.
To account for this, the corresponding efficiency is measured
simultaneously with the derivation of the cross section limits as described in
Section~\ref{sec:background}, where
the efficiency is left unconstrained in the maximum likelihood fit
when deriving upper limits. This approach
allows for a consistent extraction directly from
the signal regions.
The same procedure is used for the efficiency of the CA8 $\cPqt$ tagging
algorithm, combined with the requirement on $\tau_{32}$.

The mistag rate of the CA8 $\cPqt$ tagging algorithm has been studied for
data and simulated events in a sideband region of the lepton+jets channel,
dominated by $\PW$+jets events (as described in Section~\ref{sec:selection}).
An uncertainty of 25\% is used for simulated events, which mostly affects the contribution
of events from $\PW$+jets processes in events with one misidentified top-quark jet
in the lepton+jets channel.
Misidentified CA8 and CA15 $\cPqt$-tagged jets are the
source of the QCD multijet background in the all-hadronic channel.
The background estimation is obtained from data in sideband regions and
the corresponding uncertainties are assumed to be fully
uncorrelated between individual bins of the \Mttbar distribution.
The efficiency of CA15 $\cPqt$-tagged jets has been studied in
data and simulated events~\cite{JME-13-007}.
The associated uncertainty affects only the low-mass selection of the all-hadronic channel.

In addition to the experimental uncertainties, the following uncertainties
affecting the predictions of the SM background processes are considered.
The effect due to missing higher-orders in the simulation of SM processes is
estimated by variations of the renormalization and factorization scales.
For the $\PW$+jets and $\ttbar$ simulated samples, the renormalization and
factorization scales are varied simultaneously by factors of 0.5 or 2.
The resulting uncertainty in continuum SM $\ttbar$ production affects all channels,
while the uncertainty in $\PW$+jets production affects only the lepton+jets channel.
The effect due to the uncertainty in extra hard-parton radiation is studied
by varying the jet matching threshold for simulated $\PW$+jets processes
by factors of 0.5 and 2. This uncertainty applies only to the lepton+jets channel.
All simulated signal and background events are reweighted according to the uncertainties
parametrized by the eigenvectors of the CTEQ6L and CT10 PDF sets.
The shifts produced by the individual eigenvectors are added in
quadrature in each bin of the \Mttbar distribution.
The resulting uncertainty is taken to be fully correlated among all channels.

\begin{table*}[thb]
\centering
\caption{Sources of systematic uncertainties and the channels they affect.
  The CA8 subjet b tagging uncertainty includes the uncertainties in both the efficiency and mistag rate.
  Uncorrelated uncertainties that apply to a given channel are
  marked by $\odot$.  Uncertainties correlated between channels
  are marked by $\oplus$.
  The uncertainties listed in the upper part of the table are used in the evaluation of
  the background normalization. 
  \label{tab:corr_systematics}}
\resizebox{\linewidth}{!}{
\begin{scotch}{l c | c c c c c}
\multirow{2}{*}{Source of uncertainty}   &  Prior                        & \multirow{2}{*}{$2\ell$}   &  \multirow{2}{*}{$\ell$+jets}  & Had. channel & Had. channel    \\
                                         &  uncertainty                  &                            &                                & high-mass    & low-mass        \\[0.1cm] \hline \hline \rule{-2pt}{12pt}
Integrated luminosity                       & 2.6\%                      & $\oplus$ & $\oplus$ & $\oplus$ & $\oplus$   \\
\ttbar cross section                        & 15\%                       & $\oplus$ & $\oplus$ & $\oplus$ & $\oplus$   \\
Single top quark cross section              & 23\%                       & $\oplus$ & $\oplus$ &          &            \\
Diboson cross section                       & 20\%                       & $\oplus$ & $\oplus$ &          &            \\
$\Z$+jets cross section                     & 50\%                       & $\oplus$ & $\oplus$ &          &            \\
$\PW$+jets (light flavor) cross section     &  9\%                       &          & $\odot$  &          &            \\
$\PW$+jets (heavy flavor) cross section     & 23\%                       &          & $\odot$  &          &            \\
Electron+jet trigger                      &  1\%                       &          & $\odot$  &          &            \\
\HT trigger                                 &  2\%                       &          &          & $\oplus$ & $\oplus$   \\
Four-jet trigger                            &  $\pm 1\sigma (\pt)$       &          &          &          & $\odot$    \\
Single-electron trigger                     &  $\pm 1\sigma (\pt,\eta)$  & $\odot$  &          &          &            \\
Single-muon trigger and id                  & $\pm 1\sigma (\pt,\eta)$   & $\oplus$ & $\oplus$ &          &            \\
Electron ID                                 & $\pm 1\sigma (\pt,\eta)$   & $\oplus$ & $\oplus$ &          &            \\
Jet energy scale                            & $\pm 1\sigma (\pt,\eta)$   & $\oplus$ & $\oplus$ & $\oplus$ & $\oplus$   \\
Jet energy resolution                       & $\pm 1\sigma (\eta)$       & $\oplus$ & $\oplus$ & $\oplus$ & $\oplus$   \\
Pileup uncertainty                          & $\pm 1\sigma$              & $\oplus$ & $\oplus$ & $\oplus$ & $\oplus$   \\
$\cPqb$ tagging efficiency$^{(\dagger)}$   & $\pm 1\sigma (\pt, \eta)$  & $\oplus$ & $\oplus$ &          & $\oplus$   \\
$\cPqb$ tagging mistag rate $^{(\dagger)}$ & $\pm 1\sigma (\pt, \eta)$  & $\oplus$ & $\oplus$ &          & $\oplus$   \\
CA8 subjet b tagging                       & unconstrained              &          &          & $\odot$  &            \\
CA8 $\cPqt$ tagged jet efficiency                            & unconstrained              &          & $\oplus$ & $\oplus$ &            \\
CA8 $\cPqt$-tagged jet mistag                      & $\pm 25\%$                 &          & $\odot$  &          &            \\
CA15 $\cPqt$-tagged jet efficiency                           & $\pm 1\sigma (\pt,\eta)$   &          &          &          & $\odot$    \\
QCD multijet background                     &  sideband                  &          &          & $\odot$  & $\odot$    \\[0.1cm]
MC statistical uncertainty                  &                            & $\odot$  & $\odot$  & $\odot$  & $\odot$\\
\hline \rule{-2pt}{12pt}
PDF uncertainty                             & $\pm 1\sigma$              & $\oplus$ & $\oplus$ & $\oplus$ & $\oplus$   \\
\ttbar ren. and fact. scales                          & 4$Q^2$ and 0.25$Q^2$       & $\oplus$ & $\oplus$ & $\oplus$ & $\oplus$   \\
$\PW$+jets ren. and fact. scales                      & 4$Q^2$ and 0.25$Q^2$       &          & $\odot$  &          &            \\
$\PW$+jets matching scale $\mu$                  & $2\mu$ and $0.5\mu$      &          & $\odot$  &          &            \\
\end{scotch}
}
\vspace{-0.2cm}
\footnotesize{$^{(\dagger)}$ AK5 and CA15 subjets} \\[0.2cm]
\end{table*}

The impact of the systematic uncertainties on the normalization of the total background
depends strongly on the channel considered. The following uncertainties are the dominant ones
in the two channels with the highest sensitivity, the category with
one CA8 $\cPqt$-tagged jet in the lepton+jets analysis and the category with two \cPqb-tagged jets in the high-mass
all-hadronic analysis.
The dominant uncertainty comes from missing higher orders in the simulation of the \ttbar background
and is on average 17\%. The uncertainty due to the uncertainties in the PDFs is 15\%, which is the
same size as the uncertainty in the total \ttbar cross section in the phase space considered.
The size of other experimental uncertainties, like the CA8 $\cPqt$ tagging efficiency,
the subjet $\cPqb$ tagging efficiency,
and uncertainties due to the jet energy scale and resolution, vary between 4\%--12\%.

\section{Estimation of the background normalization}
\label{sec:background}

The main source of irreducible background in all channels arises from
SM $\ttbar$ production.
In the lepton+jets channels, \PW+jets production contributes to events
without a CA8 $\cPqt$-tagged jet.
Single top quark, $\Z$+jets, and diboson production constitute small backgrounds overall,
and contribute to the dilepton and lepton+jets channels.
In the following, these processes are combined into a single ``others'' category.

Except for the non-top-quark multijet backgrounds in the all-hadronic channels, the
shapes of all SM backgrounds are estimated from simulation.
The total yield of the simulated samples is obtained with a maximum
likelihood fit to the $\Mttbar$ distributions.
Nuisance parameters are included in the fit to take into account the effect of 
systematic uncertainties. The parameters are constrained using log-normal 
probability density functions and are fitted simultaneously with the 
parameters corresponding to the background normalization.

Since there is no control sample of highly-boosted SM $\ttbar$ events
that is disjoint from the signal regions of this analysis, the maximum
likelihood fit is also used to extract the efficiency of the CA8 $\cPqt$
tagging and subjet $\cPqb$ tagging
algorithms simultaneously.
This is accomplished by separating the
sample into subsamples based on the tagging criteria, and allowing the
fit to find the best values for the nuisance parameters corresponding to these 
efficiencies. 
Higher-order calculations, as listed in Section~\ref{sec:datasets}, are used as
prior assumptions on the cross sections of each background process,
with the uncertainties discussed in Section~\ref{sec:systematics}.
No assumption on the scale factor for the CA8 $\cPqt$-tagged jets is made
and the corresponding nuisance parameter is left to float freely in the fit.
The same is true for the subjet $\cPqb$ tagging scale factor for the high-mass selection in the
all-hadronic channel.
This procedure effectively constrains the tagging efficiencies.
No signal hypothesis is used in this procedure.
Only the experimental uncertainties (see Section~\ref{sec:systematics}) are
included in the likelihood fit. Uncertainties due to scale and matching systematics, as
well as the uncertainties due to the PDF choice are not included.
The list of uncertainties considered is given in the upper part of Table~\ref{tab:corr_systematics}.

The fit converges with no parameter outside of two standard deviations
of the prior assumption.
A reduction of the uncertainty due to the \ttbar normalization is obtained
and a simultaneous measurement of the CA8 $\cPqt$ tagging and subjet
$\cPqb$ tagging efficiencies is performed.
These results do not change when including different signal hypotheses in the
maximum likelihood fit.

The best fit values are used to scale the predictions for the various background
processes. The measured CA8 $\cPqt$ tagging efficiency is used to scale simulated
events containing one CA8 $\cPqt$-tagged jet in the lepton+jets channel and
all categories of the high-mass all-hadronic channels.
The nuisance parameter for the subjet $\cPqb$ tagging scale factor is used to scale all
simulated events in the high-mass all-hadronic channels.
The numerical values for the background normalization, the
CA8 $\cPqt$ tagging scale factor, and the subjet $\cPqb$ tagging scale factor
are given in Table~\ref{tab:bkg_rates},
together with the prior and posterior uncertainties.

\begin{table*}[tbp]
  \topcaption{Parameters for the background normalization of the given processes, the
  scale factor for CA8 $\cPqt$-tagged jets and the subjet $\cPqb$ tagging scale factor.
  The value of each
  parameter is obtained from a maximum likelihood fit.
  Also shown are the prior assumptions on the
  rate uncertainties and the posterior uncertainties obtained by the fit.
  In case of the subjet $\cPqb$ tagging scale factor, the best fit value and the posterior uncertainty
  are given in units of the standard $\cPqb$ tagging scale factor and uncertainty.
  \label{tab:bkg_rates}}
\centering
    \begin{scotch}{l | c c c}
     \multirow{2}{*}{Process}          & Best fit   & Prior                    & Posterior  \\
                                                    & value      & uncertainty &         uncertainty \\[0.1cm] \hline \hline \rule{0pt}{12pt}
     \ttbar                                           &  0.99 &  15\% &  2.1\% \\ [.1cm] \rule{0pt}{12pt}
     \PW+jets (light flavor)                &  0.99 &   9\% &  5.0\% \\ [.1cm] \rule{0pt}{12pt}
     \PW+jets (c flavor)                 &  1.06 &  23\% & 21\%   \\ [.1cm] \rule{0pt}{12pt}
     \PW+jets (b flavor)                 &  0.95 &  23\% & 18\%   \\ [.1cm] \rule{0pt}{12pt}
     Single top quark                                    &  0.83 &  23\% & 22\%   \\ [.1cm] \rule{0pt}{12pt}
     $\Z$+jets                                     &  1.72 &  50\% & 36\%   \\ [.1cm] \rule{0pt}{12pt}
     Diboson                                       &  1.02 &  20\% & 19\%   \\ [.1cm] \rule{0pt}{12pt}
     CA8 $\cPqt$-tagged jets scale factor      &  0.94 &  unconstrained & 3\%   \\ [.1cm] \rule{0pt}{12pt}
     CA8 subjet b tagging scale factor      &  1.3 &  unconstrained & 1.5\\
    \end{scotch}
\end{table*}

The number of expected and observed events after the
maximum likelihood estimation is shown in
Tables~\ref{tab:num_events_dilepton}--\ref{tab:num_events_htt} for all
categories in the three channels.

\begin{table}[tbp]
 \topcaption{Signal efficiency and number of
 	events in the $\Pe\Pe$, $\Pe\Pgm$, and $\Pgm\Pgm$ channels.
          The yield of each MC background is obtained from NLO+NNLL calculations,
          multiplied by a scale factor derived from the likelihood fit.
          The uncertainty given for each background process includes
          the MC statistical uncertainty added in quadrature with all systematic uncertainties.
          The resonance relative decay width $\rWzp$ is indicated by $w$.
          \label{tab:num_events_dilepton}}
\centering
  \begin{scotch}{l | r r r }
   & $\Pe\Pe$ channel & $\Pe\Pgm$ channel & $\Pgm\Pgm$ channel \\ [.1cm] \hline \hline \rule{0pt}{12pt}
   & \multicolumn{3}{c}{{Efficiency}} \\ [.1cm] \hline \rule{-2pt}{12pt}
$\PZpr\,$ {\footnotesize ($M=1$\TeV, $w=1\%$)}  & 0.22\% & 0.47\% & 0.28\% \\ [.1cm]
$\PZpr\,$ {\footnotesize ($M=2$\TeV, $w=1\%$)}  & 0.34\% & 0.84\% & 0.55\% \\ [.1cm]
$\PZpr\,$ {\footnotesize ($M=3$\TeV, $w=1\%$)}  & 0.25\% & 0.61\% & 0.54\% \\ [.1cm]
$\PZpr\,$ {\footnotesize ($M=1$\TeV, $w=10\%$)} & 0.18\% & 0.44\% & 0.28\% \\ [.1cm]
$\PZpr\,$ {\footnotesize ($M=2$\TeV, $w=10\%$)} & 0.31\% & 0.69\% & 0.49\% \\ [.1cm]
$\PZpr\,$ {\footnotesize ($M=3$\TeV, $w=10\%$)} & 0.27\% & 0.60\% & 0.37\% \\ [.1cm]
$\gKK\,$ {\footnotesize ($M=1$\TeV)}       & 0.18\% & 0.41\% & 0.21\% \\ [.1cm]
$\gKK\,$ {\footnotesize ($M=2$\TeV)}       & 0.25\% & 0.51\% & 0.35\% \\ [.1cm]
$\gKK\,$ {\footnotesize ($M=3$\TeV)}       & 0.16\% & 0.42\% & 0.28\% \\ [.3cm]
\hline\rule{0pt}{12pt}
& \multicolumn{3}{c}{{Number of events}} \\ [.1cm] \hline \rule{0pt}{12pt}
$\ttbar$          & $834 \pm 43$    & $2955 \pm 148$   & $1696 \pm 86$        \\ [.1cm]
Single top quark   & $25 \pm 7$      & $67 \pm 18$      & $40 \pm 11$          \\ [.1cm]
$\Z$+jets          & $39 \pm 23$     & $9 \pm 8$        & $158 \pm 82$         \\ [.1cm]
Diboson             & $1 \pm 1$       & $2 \pm 1$        & $3 \pm 1$            \\ [.1cm]
Total background    & $898 \pm 67$    & $3032 \pm 167$   & $1897 \pm 170$       \\ [.1cm] \hline \rule{0pt}{12pt}
Data                & 832           & 3006           & $1813$\\
 \end{scotch}
\end{table}
\begin{table*}[tbp]
 \topcaption{Signal efficiency and number of events in the  \Pe+jets and \Pgm+jets channels.
          The yield of each MC background is obtained from NLO+NNLL calculations,
          multiplied by a scale factor derived from the likelihood fit.
          The uncertainty given for each background process includes
          the MC statistical uncertainty added in quadrature with all systematic uncertainties.
          The resonance relative decay width $\rWzp$ is indicated by $w$.
          \label{tab:num_events_ljets}}
\centering
\resizebox{\linewidth}{!}{
  \begin{scotch}{l | r r r | r r r }
   & \multicolumn{3}{c|}{\Pe+jets channel} & \multicolumn{3}{c}{\Pgm+jets channel} \\ [.1cm]
   & 0-\PQt, 0-\PQb & 0-\PQt, 1-\PQb & 1-\PQt
   & 0-\PQt, 0-\PQb & 0-\PQt, 1-\PQb & 1-\PQt \\ [.1cm] \hline \hline \rule{0pt}{12pt}
   & \multicolumn{6}{c}{{Efficiency}} \\ [.1cm] \hline \rule{-2pt}{12pt}
   $\PZpr\,$ {\footnotesize ($M=1$\TeV, $w=1\%$)}  & 0.5\% & 2.3\% & 0.5\% & 0.4\% & 2.2\% & 0.4\% \\ [.1cm]
$\PZpr\,$ {\footnotesize ($M=2$\TeV, $w=1\%$)}  & 1.1\% & 2.5\% & 2.4\% & 1.1\% & 2.4\% & 2.3\% \\ [.1cm]
$\PZpr\,$ {\footnotesize ($M=3$\TeV, $w=1\%$)}  & 1.5\% & 2.3\% & 1.8\% & 1.7\% & 2.3\% & 2.0\% \\ [.1cm]
$\PZpr\,$ {\footnotesize ($M=1$\TeV, $w=10\%$)} & 0.5\% & 2.1\% & 0.4\% & 0.4\% & 2.1\% & 0.4\% \\ [.1cm]
$\PZpr\,$ {\footnotesize ($M=2$\TeV, $w=10\%$)} & 0.9\% & 2.3\% & 1.9\% & 0.9\% & 2.2\% & 1.8\% \\ [.1cm]
$\PZpr\,$ {\footnotesize ($M=3$\TeV, $w=10\%$)} & 0.9\% & 2.1\% & 1.4\% & 1.0\% & 1.8\% & 1.4\% \\ [.1cm]
$\gKK\,$ {\footnotesize ($M=1$\TeV)}       & 0.5\% & 1.8\% & 0.3\% & 0.4\% & 1.8\% & 0.3\% \\ [.1cm]
$\gKK\,$ {\footnotesize ($M=2$\TeV)}       & 0.7\% & 1.9\% & 1.3\% & 0.7\% & 1.8\% & 1.2\% \\ [.1cm]
$\gKK\,$ {\footnotesize ($M=3$\TeV)}       & 0.6\% & 1.5\% & 0.8\% & 0.6\% & 1.4\% & 0.9\% \\ [.3cm]
\hline\rule{0pt}{12pt}
& \multicolumn{6}{c}{{Number of events}} \\ [.1cm] \hline \rule{-2pt}{12pt}
$\ttbar$ & $3127 \pm 254$ & $11345 \pm 796$ & $499 \pm 47$ & $3322 \pm 255$ & $11634 \pm 749$ & $491 \pm 49$ \\ [.1cm]
$\PW$+jets (light flavor) & $4790 \pm 955$ & $222 \pm 57$ & $30 \pm 9$ & $4891 \pm 875$ & $207 \pm 50$ & $29 \pm 8$ \\ [.1cm]
$\PW$+jets (c flavor) & $1128 \pm 397$ & $303 \pm 110$ & $8 \pm 4$ & $1186 \pm 405$ & $333 \pm 117$ & $11 \pm 5$ \\ [.1cm]
$\PW$+jets (b flavor) & $111 \pm 33$ & $250 \pm 76$ & $3 \pm 1$ & $102 \pm 29$ & $243 \pm 70$ & $2 \pm 1$ \\ [.1cm]
Single top quark & $244 \pm 63$ & $667 \pm 169$ & $8 \pm 3$ & $238 \pm 61$ & $702 \pm 178$ & $8 \pm 3$ \\ [.1cm]
$\Z$+jets & $485 \pm 123$ & $90 \pm 23$ & $3 \pm 1$ & $606 \pm 153$ & $110 \pm 28$ & $4 \pm 1$ \\ [.1cm]
Diboson & $123 \pm 25$ & $29 \pm 6$ & $1 \pm 1$ & $134 \pm 27$ & $ 27 \pm 6$ & $1 \pm 1$ \\ [.1cm]
Total background & $10007 \pm 1422$ & $ 12906 \pm 1062$ & $552 \pm 53$ & $10479 \pm 1407$ & $13256 \pm 1049$ & $545 \pm 54$ \\ [.1cm] \hline \rule{-2pt}{12pt}
Data & 10204 & 12157 & 465 & 10099 & 12510 & $493$\\
  \end{scotch}
}
\end{table*}

\begin{table*}[tbp]
 \topcaption{Signal efficiency and number  of
 	  events in the high-mass all-hadronic channel.
          The yield of the \ttbar background is obtained from NLO+NNLL calculations,
          multiplied by a scale factor derived from the likelihood fit.
          The multijet background is obtained from sideband regions in data.
          The uncertainty given for each background process includes
          the statistical uncertainty added in quadrature with all systematic uncertainties.
         The resonance relative decay width $\rWzp$ is indicated by $w$.
          \label{tab:num_events_cmstt}}
\centering
  \begin{scotch}{l | r r r | r r r }
   & \multicolumn{3}{c|}{$\abs{\Delta y} < 1.0$} & \multicolumn{3}{c}{$\abs{\Delta y} > 1.0$} \\ [.1cm]
   & 0~\PQb-tag & 1~\PQb-tag & 2 \PQb-tags
   & 0~\PQb-tag & 1~\PQb-tag & 2~\PQb-tags \\ [.1cm] \hline \hline \rule{0pt}{12pt}
   & \multicolumn{6}{c}{{Efficiency}} \\ [.1cm] \hline \rule{-2pt}{12pt}
$\PZpr\,$ {\footnotesize ($M=1$\TeV, $w=1\%$)} & 0.1\% & 0.3\% & 0.4\% & 0.0\% & 0.0\% & 0.0\% \\ [.1cm]
$\PZpr\,$ {\footnotesize ($M=2$\TeV, $w=1\%$)} & 0.8\% & 1.8\% & 1.1\% & 0.6\% & 1.7\% & 1.3\% \\ [.1cm]
$\PZpr\,$ {\footnotesize ($M=3$\TeV, $w=1\%$)} & 0.6\% & 1.0\% & 0.4\% & 0.8\% & 1.6\% & 0.9\% \\ [.1cm]
$\PZpr\,$ {\footnotesize ($M=1$\TeV, $w=10\%$)} & 0.1\% & 0.3\% & 0.4\% & 0.0\% & 0.0\% & 0.0\% \\ [.1cm]
$\PZpr\,$ {\footnotesize ($M=2$\TeV, $w=10\%$)} & 0.6\% & 1.6\% & 1.0\% & 0.3\% & 1.3\% & 0.9\% \\ [.1cm]
$\PZpr\,$ {\footnotesize ($M=3$\TeV, $w=10\%$)} & 0.4\% & 0.9\% & 0.5\% & 0.4\% & 0.9\% & 0.6\% \\ [.1cm]
$\gKK\,$ {\footnotesize ($M=1$\TeV)}       & 0.1\% & 0.2\% & 0.2\% & 0.0\% & 0.0\% & 0.0\% \\ [.1cm]
$\gKK\,$ {\footnotesize ($M=2$\TeV)}       & 0.3\% & 1.0\% & 0.8\% & 0.2\% & 0.7\% & 0.6\% \\ [.1cm]
$\gKK\,$ {\footnotesize ($M=3$\TeV)}       & 0.2\% & 0.6\% & 0.4\% & 0.2\% & 0.5\% & 0.4\% \\ [.3cm]
\hline\rule{0pt}{12pt}
& \multicolumn{6}{c}{{Number of events}} \\ [.1cm] \hline \rule{-2pt}{12pt}
$\ttbar$       & $59 \pm 11$    & $243 \pm 30$   & $262 \pm 32$  & $29 \pm 5$      & $112 \pm 11$  & $109 \pm 12$ \\ [.1cm]
QCD multijet             & $1984 \pm 68$  & $678 \pm 31$   & $68  \pm 8$   & $1465 \pm 54$   & $456 \pm 24$  & $41 \pm 7$    \\ [.1cm]
Total background & $2043 \pm 70$  & $921 \pm 46$   & $330 \pm 33$  & $1493 \pm 55$   & $568 \pm 28$  & $150 \pm 14$ \\ [.1cm] \hline \rule{-2pt}{12pt}
Data             & 1956         & 933          & 305         & 1523          & 604         & 143\\
  \end{scotch}
\end{table*}
\begin{table*}[tbp]
 \topcaption{Signal efficiency and number of
 	  events in the low-mass all-hadronic channel.
          The yield of the \ttbar background is obtained from NLO+NNLL calculations,
          multiplied by a scale factor derived from the likelihood fit.
          The multijet background is obtained from sideband regions in data.
          The uncertainty given for each background process includes
          the statistical uncertainty added in quadrature with all systematic uncertainties.
         The resonance relative decay width $\rWzp$ is indicated by $w$.
 \label{tab:num_events_htt}}
\centering
\resizebox{\linewidth}{!}{
  \begin{scotch}{l | r r r | r r r }
   & \multicolumn{3}{c|}{$\HT < 800\GeV$} & \multicolumn{3}{c}{$\HT > 800\GeV$} \\ [.1cm]
   & 0~\PQb-tag & 1~\PQb-tag & 2~\PQb-tags
   & 0~\PQb-tag & 1~\PQb-tag & 2~\PQb-tags \\ [.1cm] \hline \hline \rule{0pt}{12pt}
   & \multicolumn{6}{c}{{Efficiency}} \\ [.1cm] \hline \rule{-2pt}{12pt}
$\PZpr\,$ {\footnotesize ($M=0.75$\TeV, $w=1\%$)}  & 0.17\% & 0.64\% & 0.70\% & 0.01\% & 0.04\% & 0.03\% \\ [.1cm]
$\PZpr\,$ {\footnotesize ($M=1$\TeV, $w=1\%$)}     & 0.13\% & 0.54\% & 0.56\% & 0.16\% & 0.61\% & 0.66\% \\ [.1cm]
$\PZpr\,$ {\footnotesize ($M=2$\TeV, $w=1\%$)}     & 0.04\% & 0.09\% & 0.07\% & 0.08\% & 0.26\% & 0.18\% \\ [.1cm]
$\PZpr\,$ {\footnotesize ($M=0.75$\TeV, $w=10\%$)} & 0.15\% & 0.62\% & 0.64\% & 0.01\% & 0.06\% & 0.05\% \\ [.1cm]
$\PZpr\,$ {\footnotesize ($M=1$\TeV, $w=10\%$)}    & 0.12\% & 0.54\% & 0.54\% & 0.13\% & 0.49\% & 0.50\% \\ [.1cm]
$\PZpr\,$ {\footnotesize ($M=2$\TeV, $w=10\%$)}    & 0.04\% & 0.18\% & 0.15\% & 0.07\% & 0.27\% & 0.21\% \\ [.1cm]
$\gKK\,$ {\footnotesize ($M=0.7$\TeV)}        & 0.11\% & 0.37\% & 0.42\% & 0.01\% & 0.04\% & 0.04\% \\ [.1cm]
$\gKK\,$ {\footnotesize ($M=1$\TeV)}          & 0.11\% & 0.47\% & 0.49\% & 0.08\% & 0.40\% & 0.32\% \\ [.1cm]
$\gKK\,$ {\footnotesize ($M=2$\TeV)}          & 0.06\% & 0.24\% & 0.19\% & 0.07\% & 0.23\% & 0.21\% \\ [.1cm] \hline \rule{0pt}{12pt}
& \multicolumn{6}{c}{{Number of events}} \\ [.1cm] \hline \rule{-2pt}{12pt}
$\ttbar$        & $851 \pm 216$      & $3238 \pm 716$   & $3009 \pm 644$   & $196 \pm 65$    & $698 \pm 203$   & $583 \pm 165$ \\ [.1cm]
QCD multijet              & $55932 \pm 1598$   & $18687 \pm 613$  & $1933 \pm 92 $   & $8544 \pm 331$  & $3080 \pm 168$  & $311 \pm 30$    \\ [.1cm]
Total background  & $56781 \pm 1633$   & $21926 \pm 984$  & $4942 \pm 655$   & $8740 \pm 341$  & $3778 \pm 268$  & $894 \pm 168$ \\ [.1cm] \hline \rule{-2pt}{12pt}
Data              & 57118            & 22485          & 5381           & 8920          & 3935          & 891\\
  \end{scotch}
}
\end{table*}

\section{Results}
\label{sec:results}

No significant excess of data over the expected SM background is observed.
A Bayesian statistical method~\cite{bayesbook, theta} is used to derive
95\% confidence level (CL) upper limits on the cross section times branching fraction for
$\PZpr \to \ttbar$ production.
The limits are derived employing a template based
evaluation that uses the invariant mass distribution of the reconstructed $\ttbar$ pair.
A likelihood fit is used to compare the signal and SM background expectations.
To build the likelihood function, a Poisson probability is calculated in
each bin of the mass distribution for each category in each channel.
The parameters representing
the Poisson mean of the signal strength and the background processes
are determined in the fit.
Pseudoexperiments are performed to extract expected limits under a background-only
hypothesis. The systematic uncertainties discussed in Section~\ref{sec:systematics}
are taken into account through nuisance parameters.
These are randomly varied within their ranges of validity
using log normal distributions as the probability density function.
Correlations between the systematic uncertainties across all channels are taken into account.
The statistical uncertainties of simulated samples are treated as an additional Poisson
nuisance parameter in each bin of the mass distribution.

The median of the distribution of the upper limits at 95\% CL in the pseudoexperiments
and the central 68\% (95\%) interval define the expected upper limit and
$\pm 1\sigma$ ($\pm 2\sigma$) bands, respectively.
Upper limits for three benchmark signal hypotheses are calculated:
a topcolor \PZpr boson~\cite{theory_jainharris} with relative widths $\rWzp = 1.0\%$ or
$\rWzp = 10\%$, and a Randall--Sundrum KK gluon with coupling as described in
Ref.~\cite{theory_agashe}. Resonance masses between 0.75 and 3\TeV are considered 
using simulated samples with mass values given in Section~\ref{sec:datasets}.
Above mass values of 3\TeV, narrow-width signals would have cross sections
below 1\fb, placing them beyond the reach of Run 1 at the LHC, and signals with relative
widths larger than 10\% would show no resonance structure at the collision energy of 8\TeV.
All limits are given at 95\% CL.

\begin{figure}[b]
\centering
\includegraphics[width=0.48\textwidth]{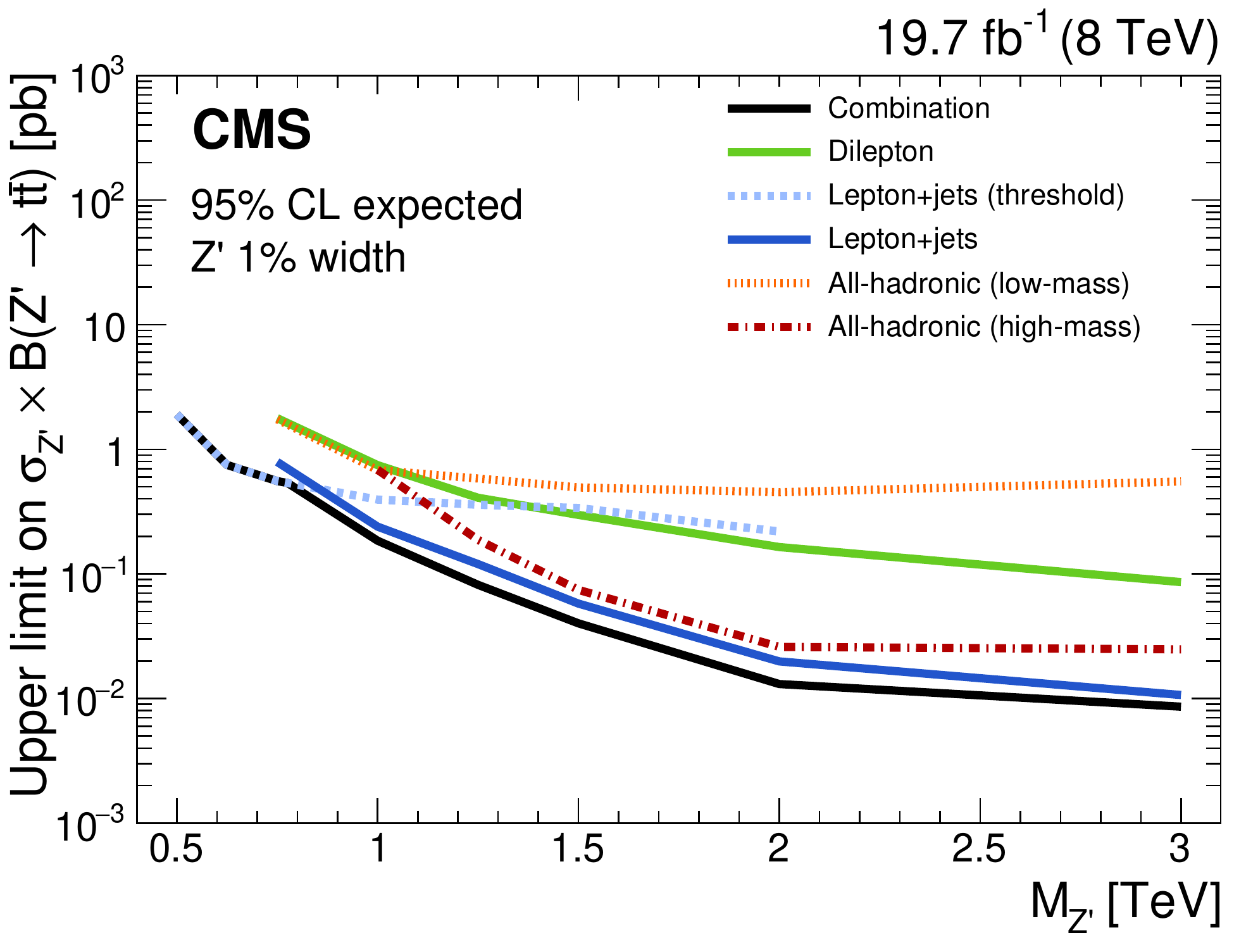}
\includegraphics[width=0.48\textwidth]{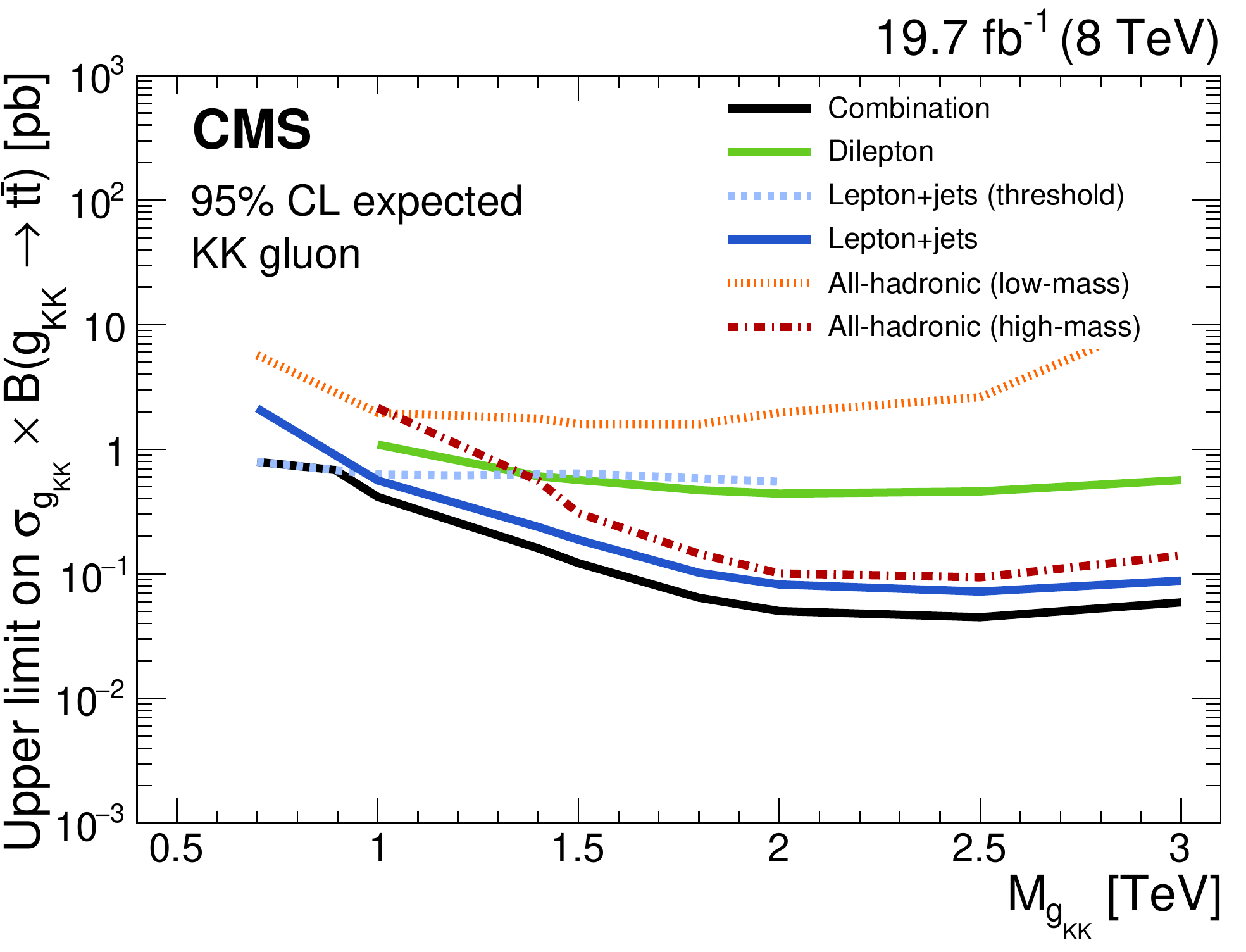}
\caption{Expected 95\% CL upper limits on the production cross section times branching fraction
         for a \PZpr boson decaying to $\ttbar$ with 1\% width (\cmsLeft) and
         a KK gluon in the RS model (\cmsRight).
         The limits obtained from the individual channels are shown separately,
         together with the result from the combination.
         Also shown are results from a threshold analysis in the lepton+jets channel~\cite{cms_ttbar_resonance4},
         optimized for low mass values.
         \label{fig:comp_expected}}
\end{figure}
A comparison of the expected limits obtained from the individual channels is shown in
Fig.~\ref{fig:comp_expected}.
Also shown are the results from a search optimized for threshold production of the \ttbar pair
in the lepton+jets channel~\cite{cms_ttbar_resonance4}.
This channel has the best sensitivity for resonance masses below 0.75\TeV.
Above this value, the combination of the boosted analyses described in this paper places
better limits on the production cross section times branching fraction.
The best overall sensitivity is obtained in the lepton+jets channel.
The high-mass selection of the all-hadronic channel has comparable sensitivity in the mass region above 2\TeV.

\begin{figure*}[tbp]
\centering
\includegraphics[width=0.48\textwidth]{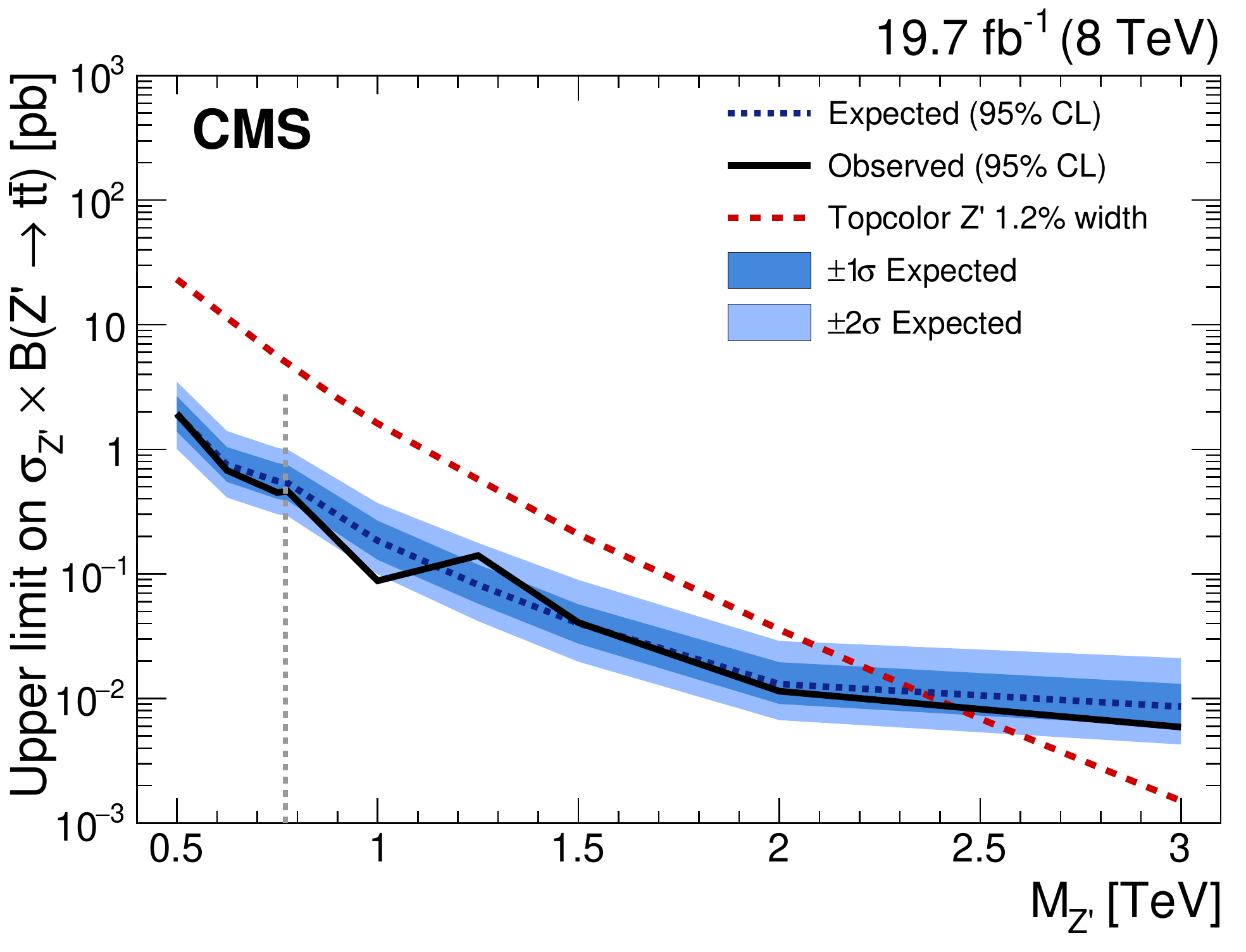}
\includegraphics[width=0.48\textwidth]{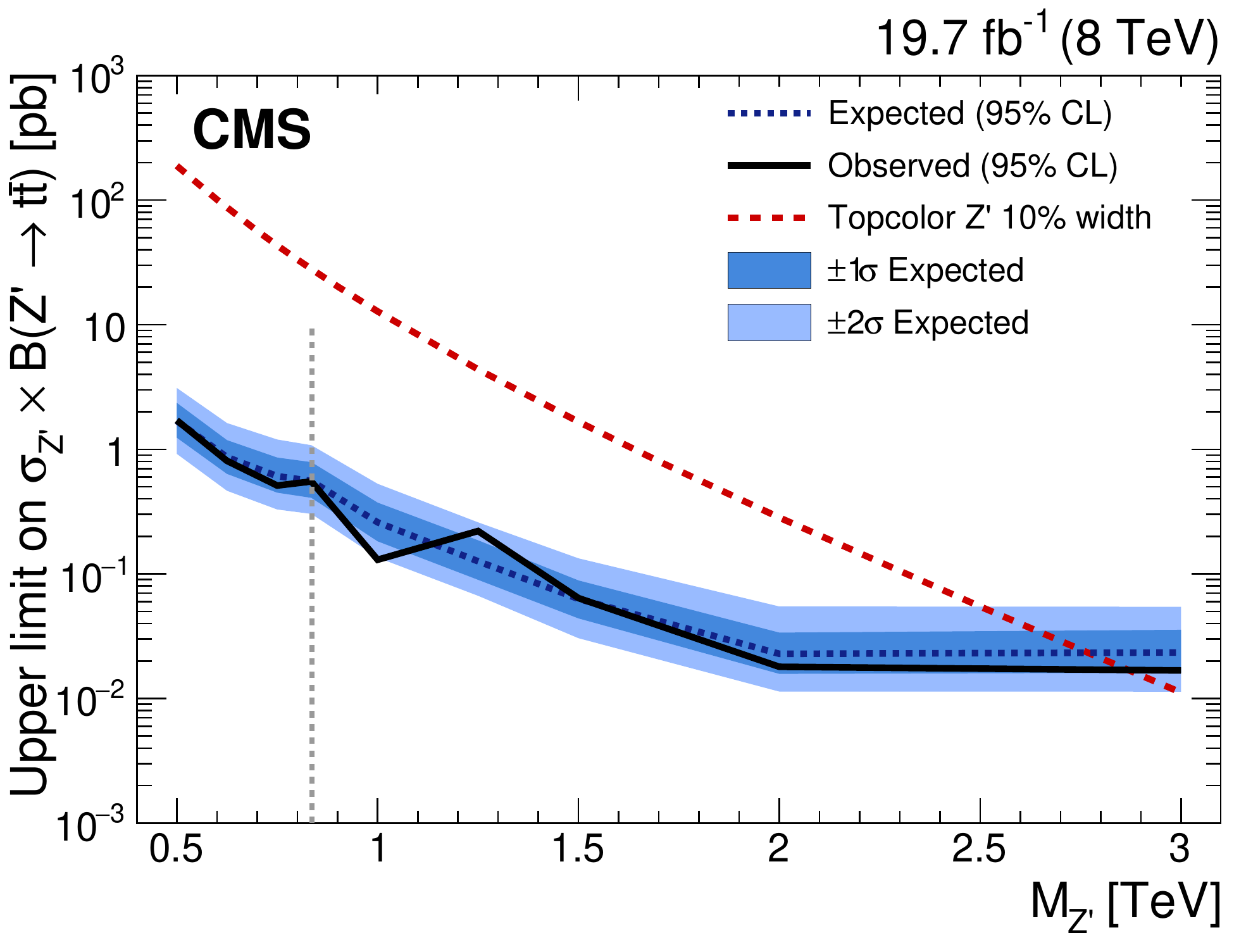}
\includegraphics[width=0.48\textwidth]{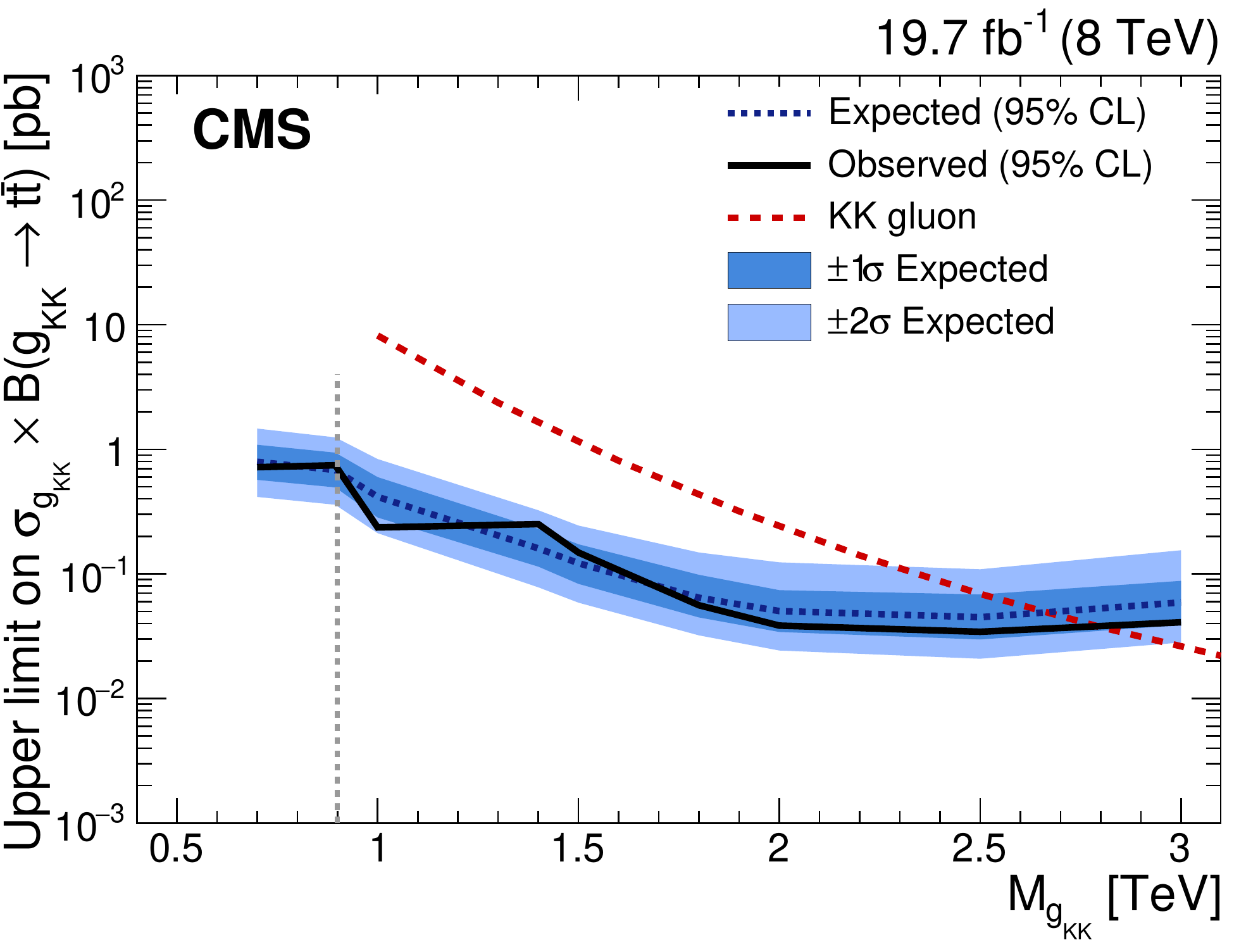}
\caption{Upper limits at 95\% CL on the production cross section times branching fraction
         for a \PZpr boson decaying to $\ttbar$ with narrow width (upper left), with 10\% width (upper right) and
         a KK gluon in the Randall--Sundrum model decaying to $\ttbar$ (bottom).
         The vertical dashed line indicates the transition from a threshold analysis~\cite{cms_ttbar_resonance4}
         to the combination,
         in providing the best expected limit. Below this dashed line, only the results of the low-mass analysis with resolved jets are quoted;
         above this line, the results from the combination of the boosted channels are given.
         The limits are shown as a function of the resonance mass and are compared to
         predictions for the cross section of a \PZpr boson with relative width of 1.2\% and
         10\%~\cite{theory_jainharris} and the prediction for the production of a KK gluon~\cite{theory_agashe}.
         The predictions are multiplied by a factor of 1.3 
         to account for higher-order corrections~\cite{ZprimeKfactor}.
         \label{fig:limits_combo}}
\end{figure*}
Figure \ref{fig:limits_combo} shows the results for each of the three signal hypotheses.
The cross section limits for the narrow signal hypothesis are compared
to the cross section for the production of a \PZpr boson with 1.2\% width.
This width is chosen for comparison with theoretical results and previous measurements.
Resonances with masses up to 2.4\TeV (2.4\TeV expected) for the narrow \PZpr hypothesis are excluded.
These cross section limits are model independent, meaning that they are valid for any resonance decaying
to \ttbar, with a width well below the experimental resolution of about 10\%.
Wide resonances with 10\% width are excluded up to 2.9\TeV (2.8\TeV expected). The better limits
with respect to narrow resonances are due to the higher production cross section of the wider \PZpr resonance.
Randall--Sundrum KK gluons decaying to \ttbar are excluded with masses below 2.8\TeV
(2.7\TeV expected).
This model exhibits the weakest upper limits on the production cross section,
because of the long tails towards low resonance masses present in the
predicted $\Mttbar$ distribution. These tails are introduced by the
interplay between the large natural width of the KK gluons and the
parton luminosity, causing masses that are far below the resonance mass to
have a larger probability than events near the resonance itself.
The expected and observed exclusion limits for different resonance masses are given in
Table \ref{tab:masslimits}.
\begin{table*}[tbp]
\topcaption{Expected and observed lower mass limits for the three benchmark models.
  Mass limits are given for the dilepton analysis, the lepton+jets analysis, the combination of the two all-hadronic analyses and
  the full combination of all four analyses. All limits are given at 95\% CL.
  \label{tab:masslimits}}
  \centering
\resizebox{\linewidth}{!}{
  \begin{scotch}{l | c c | c c | c c | c c}
                  \multicolumn{9}{c}{{\hspace{3cm}Mass limit [\TeVns{}]}} \\[0.2cm]
                  \cline{2-9}
                  \multicolumn{1}{c}{\rule{-2pt}{12pt}}& \multicolumn{2}{c |}{Dilepton channel} & \multicolumn{2}{c |}{Lepton+jets channel} & \multicolumn{2}{c |}{All-hadronic channels} & \multicolumn{2}{c}{Combined} \\
                    \multicolumn{1}{c}{}&  \quad Expected\  &  Observed\  & \quad Expected\  & Observed\  & \quad Expected\  & Observed\  & Expected\     & Observed\  \\[0.2cm] \hline \hline \rule{-2pt}{20pt}
 \PZpr, $\rWzp = 1.2\%$       &  \quad 1.4    &  1.5    & \quad 2.2    &  2.3   & \quad 2.1    &  2.1   &  2.4      & 2.4    \\[0.2cm] \rule{-2pt}{12pt}
 \PZpr, $\rWzp = 10\%$        &  \quad 2.1    &  2.2    & \quad 2.7    &  2.8   & \quad 2.5    &  2.5   &  2.8      & 2.9    \\[0.2cm] \rule{-2pt}{12pt}
 RS KK gluon                  &  \quad 1.8    &  2.0    & \quad 2.5    &  2.5   & \quad 2.4    &  2.3   &  2.7      & 2.8 \\
 \end{scotch}
}
\end{table*}

\begin{table*}[tbp]
 \topcaption{Expected and observed limits
          for the production cross section times branching fraction of a
          \PZpr boson decaying to $\ttbar$ with a width of 1\%, 10\%  and a KK gluon in the RS model.
          All limits are given at 95\% CL.
          \label{tab:limits_combo}}
 \centering
\resizebox{\linewidth}{!}{
  \begin{scotch}{c | c  rcl  rcl  c}
   \multicolumn{1}{c}{}&\multicolumn{8}{c}{$\PZpr, \rWzp = 1\%$} \\ \cline{2-9}
 \multicolumn{1}{c}{$\Mzp$ ($\TeVns{}$) }&  \rule{0pt}{12pt} Expected (pb) & \multicolumn{3}{c}{Expected range $({\pm}1\sigma)$ (pb)} & \multicolumn{3}{c}{Expected range $({\pm}2\sigma)$ (pb)} & Observed (pb) \\ [.1cm] \hline \hline \rule{0pt}{12pt}
0.75 & 0.61 & \quad\qquad 0.89 & \NA & 0.43 & \quad\qquad 1.3 & \NA & 0.32 & 0.86 \\ [.1cm]
1.0 & 0.18 & 0.27 & \NA & 0.13 & 0.37 & \NA & 0.099 & 0.088 \\ [.1cm]
1.25 & 0.082 & 0.12 & \NA & 0.058 & 0.18 & \NA & 0.042 & 0.14 \\ [.1cm]
1.5 & 0.04 & 0.057 & \NA & 0.028 & 0.089 & \NA & 0.02 & 0.041 \\ [.1cm]
2.0 & 0.013 & 0.02 & \NA & 0.009 & 0.029 & \NA & 0.0067 & 0.011 \\ [.1cm]
3.0 & 0.0086 & 0.013 & \NA & 0.0059 & 0.021 & \NA & 0.0043 & 0.0059\\
\hline
   \multicolumn{1}{c}{}&\multicolumn{8}{c}{\rule{0pt}{9pt}$\PZpr, \rWzp = 10\%$} \\ \cline{2-9}
   \multicolumn{1}{c}{$\Mzp$ ($\TeVns{}$)} &  \rule{0pt}{12pt} Expected (pb) & \multicolumn{3}{c}{Expected range $({\pm}1\sigma)$ (pb)} & \multicolumn{3}{c}{Expected range $({\pm}2\sigma)$ (pb)} & Observed (pb) \\ [.1cm] \hline \hline \rule{0pt}{12pt}
0.75 & 0.83 & \quad\qquad 1.2 & \NA & 0.57 & \quad\qquad 1.8 & \NA & 0.42 & 0.89 \\ [.1cm]
1.0 & 0.26 & 0.37 & \NA & 0.18 & 0.53 & \NA & 0.14 & 0.13 \\ [.1cm]
1.25 & 0.13 & 0.19 & \NA & 0.09 & 0.26 & \NA & 0.067 & 0.22 \\ [.1cm]
1.5 & 0.063 & 0.089 & \NA & 0.044 & 0.13 & \NA & 0.03 & 0.064 \\ [.1cm]
2.0 & 0.023 & 0.034 & \NA & 0.016 & 0.055 & \NA & 0.011 & 0.018 \\ [.1cm]
3.0 & 0.023 & 0.036 & \NA & 0.016 & 0.055 & \NA & 0.011 & 0.017 \\
\hline
   \multicolumn{1}{c}{}&\multicolumn{8}{c}{{RS KK gluon}} \\ \cline{2-9}
   \multicolumn{1}{c}{\MgKK ($\TeVns{}$)} & \rule{0pt}{12pt}  Expected (pb) & \multicolumn{3}{c}{Expected range $({\pm}1\sigma)$ (pb)} & \multicolumn{3}{c}{Expected range $({\pm}2\sigma)$ (pb)} & Observed (pb) \\ [.1cm] \hline \hline \rule{0pt}{12pt}
0.7 & 1.7 & \quad\qquad 2.5 & \NA & 1.2 & \quad\qquad 3.8 & \NA & 0.84 & 3.5 \\ [.1cm]
1.0 & 0.42 & 0.6 & \NA & 0.28 & 0.84 & \NA & 0.21 & 0.24 \\ [.1cm]
1.4 & 0.16 & 0.23 & \NA & 0.11 & 0.32 & \NA & 0.078 & 0.25 \\ [.1cm]
1.5 & 0.12 & 0.17 & \NA & 0.083 & 0.24 & \NA & 0.059 & 0.15 \\ [.1cm]
1.8 & 0.064 & 0.098 & \NA & 0.045 & 0.15 & \NA & 0.032 & 0.056 \\ [.1cm]
2.0 & 0.05 & 0.074 & \NA & 0.034 & 0.12 & \NA & 0.024 & 0.038 \\ [.1cm]
2.5 & 0.045 & 0.068 & \NA & 0.03 & 0.11 & \NA & 0.021 & 0.034 \\ [.1cm]
3.0 & 0.059 & 0.088 & \NA & 0.039 & 0.15 & \NA & 0.028 & 0.041 \\
  \end{scotch}
}
\end{table*}
The upper limits on the production cross section times branching fraction into \ttbar are given
in Table \ref{tab:limits_combo}, for different resonance masses.
The upper limits on the cross sections show improvements of about 50\% with respect to a previous
combination of results from a search in the lepton+jets and all-hadronic channels~\cite{cms_ttbar_resonance4}.
These improvements are mostly due to the use of $\cPqt$~tagging in the lepton+jets channel, and
the application of $\cPqb$ tagging on subjets in the all-hadronic channel.
The limits for $\Mzp<1\TeV$ are improved with the addition of the dilepton channel and the
low-mass selection in the all-hadronic channel.

\section{Summary}
\label{sec:conclusions}
A search has been performed for the production of heavy \ttbar
resonances in final states including two, one, or
no leptons.
The analysis is based on a data sample corresponding to an integrated
luminosity of 19.7\fbinv recorded in 2012 with the CMS detector in proton-proton
collisions at $\sqrt{s}=8\TeV$ at the LHC. No evidence is found for
a resonant \ttbar component beyond the standard model \ttbar continuum production.
Model-independent cross section limits are set on the production of such resonances
that have widths well below the experimental resolution of about 10\%.

Cross sections times branching fractions above 11\fb are excluded at 95\% confidence level (CL)
for the process $\Pp\Pp \to \PZpr \to \ttbar$
with a \PZpr resonance~\cite{theory_jainharris} with mass of 2\TeV and
width $\rWzp = 1\%$.
The corresponding 95\% CL expected cross section limit is $13\fb$.
The 95\% CL observed lower mass limit
for a topcolor narrow \PZpr resonance with $\rWzp = 1.2\%$ corresponds to 2.4\TeV,
which agrees with the expected limit.

Observed and expected 95\% CL upper limits of
18\fb (23\fb) are set for a \PZpr boson~\cite{theory_jainharris} with a mass of 2\TeV and
width $\rWzp = 10\%$. The respective
95\% CL observed and expected lower mass limits are
2.9 and 2.8\TeV for a wide topcolor \PZpr resonance.

For the production of Kaluza--Klein gluon excitations
$\Pp\Pp \to \gKK \to \ttbar$ predicted
in Randall--Sundrum models~\cite{theory_agashe},
an upper limit on the cross section of 38\fb is observed (50\fb expected)
at 95\% CL for a mass of $2\TeV$.
The observed and expected lower mass limits are 2.8 and 2.7\TeV.

These mass limits represent significant improvements over previous ones set at
$\sqrt{s} = 7\TeV$~\cite{cms_ttbar_resonance1, cms_ttbar_resonance2, cms_ttbar_resonance3}.
An improvement by about 50\% on the 95\% CL upper limits with respect to
an earlier search optimized for high masses at $\sqrt{s} = 8\TeV$~\cite{cms_ttbar_resonance4}
is achieved by the application of additional jet substructure information
and the addition of the dilepton channel.
The results presented provide the most stringent constraints on resonant
\ttbar production to date.

\begin{acknowledgments}
\label{sec:acknowledgements}
\hyphenation{Bundes-ministerium Forschungs-gemeinschaft Forschungs-zentren} We congratulate our colleagues in the CERN accelerator departments for the excellent performance of the LHC and thank the technical and administrative staffs at CERN and at other CMS institutes for their contributions to the success of the CMS effort. In addition, we gratefully acknowledge the computing centers and personnel of the Worldwide LHC Computing Grid for delivering so effectively the computing infrastructure essential to our analyses. Finally, we acknowledge the enduring support for the construction and operation of the LHC and the CMS detector provided by the following funding agencies: the Austrian Federal Ministry of Science, Research and Economy and the Austrian Science Fund; the Belgian Fonds de la Recherche Scientifique, and Fonds voor Wetenschappelijk Onderzoek; the Brazilian Funding Agencies (CNPq, CAPES, FAPERJ, and FAPESP); the Bulgarian Ministry of Education and Science; CERN; the Chinese Academy of Sciences, Ministry of Science and Technology, and National Natural Science Foundation of China; the Colombian Funding Agency (COLCIENCIAS); the Croatian Ministry of Science, Education and Sport, and the Croatian Science Foundation; the Research Promotion Foundation, Cyprus; the Ministry of Education and Research, Estonian Research Council via IUT23-4 and IUT23-6 and European Regional Development Fund, Estonia; the Academy of Finland, Finnish Ministry of Education and Culture, and Helsinki Institute of Physics; the Institut National de Physique Nucl\'eaire et de Physique des Particules~/~CNRS, and Commissariat \`a l'\'Energie Atomique et aux \'Energies Alternatives~/~CEA, France; the Bundesministerium f\"ur Bildung und Forschung, Deutsche Forschungsgemeinschaft, and Helmholtz-Gemeinschaft Deutscher Forschungszentren, Germany; the General Secretariat for Research and Technology, Greece; the National Scientific Research Foundation, and National Innovation Office, Hungary; the Department of Atomic Energy and the Department of Science and Technology, India; the Institute for Studies in Theoretical Physics and Mathematics, Iran; the Science Foundation, Ireland; the Istituto Nazionale di Fisica Nucleare, Italy; the Korean Ministry of Education, Science and Technology and the World Class University program of NRF, Republic of Korea; the Lithuanian Academy of Sciences; the Ministry of Education, and University of Malaya (Malaysia); the Mexican Funding Agencies (CINVESTAV, CONACYT, SEP, and UASLP-FAI); the Ministry of Business, Innovation and Employment, New Zealand; the Pakistan Atomic Energy Commission; the Ministry of Science and Higher Education and the National Science Centre, Poland; the Funda\c{c}\~ao para a Ci\^encia e a Tecnologia, Portugal; JINR, Dubna; the Ministry of Education and Science of the Russian Federation, the Federal Agency of Atomic Energy of the Russian Federation, Russian Academy of Sciences, and the Russian Foundation for Basic Research; the Ministry of Education, Science and Technological Development of Serbia; the Secretar\'{\i}a de Estado de Investigaci\'on, Desarrollo e Innovaci\'on and Programa Consolider-Ingenio 2010, Spain; the Swiss Funding Agencies (ETH Board, ETH Zurich, PSI, SNF, UniZH, Canton Zurich, and SER); the Ministry of Science and Technology, Taipei; the Thailand Center of Excellence in Physics, the Institute for the Promotion of Teaching Science and Technology of Thailand, Special Task Force for Activating Research and the National Science and Technology Development Agency of Thailand; the Scientific and Technical Research Council of Turkey, and Turkish Atomic Energy Authority; the National Academy of Sciences of Ukraine, and State Fund for Fundamental Researches, Ukraine; the Science and Technology Facilities Council, UK; the US Department of Energy, and the US National Science Foundation.

Individuals have received support from the Marie-Curie programme and the European Research Council and EPLANET (European Union); the Leventis Foundation; the A. P. Sloan Foundation; the Alexander von Humboldt Foundation; the Belgian Federal Science Policy Office; the Fonds pour la Formation \`a la Recherche dans l'Industrie et dans l'Agriculture (FRIA-Belgium); the Agentschap voor Innovatie door Wetenschap en Technologie (IWT-Belgium); the Ministry of Education, Youth and Sports (MEYS) of the Czech Republic; the Council of Science and Industrial Research, India; the HOMING PLUS programme of Foundation for Polish Science, cofinanced from European Union, Regional Development Fund; the Compagnia di San Paolo (Torino); the Thalis and Aristeia programmes cofinanced by EU-ESF and the Greek NSRF; and the National Priorities Research Program by Qatar National Research Fund.
\end{acknowledgments}
\clearpage
\bibliography{auto_generated}

\appendix
\section{Uncertainty on the background estimate in all-hadronic channel}
\label{sec:appendix_allhad_errors}
As mentioned in Section~\ref{sec:selection}, in the high-mass all-hadronic analysis,
the mistag rate $r$ is parameterized
as a function of the jet $\pt$, the $N$-subjettiness ratio $\tau_{32}$, and
the $\cPqb$-tagging discriminant $\beta$, of
the jet ``$a$''.  In the low-mass analysis, the
mistag rate is parameterized as a function of jet $\pt$ and $\beta$ only, but
the procedure is otherwise identical.

Taking the high-mass analysis as an example, the non-top-quark multijet background
arising from events with mistagged light jets is estimated by dividing the
data into bins along four dimensions: jet $\pt$, $\tau_{32}$, $\beta$, and
the variable of interest $\alpha$ (in this case ${\Mttbar}$, although the
procedure is applicable to any other variable).

The expected background yield for this 4-dimensional bin is obtained
by multiplying the events in this bin before the application of $\cPqt$ tagging,
$N(\Mttbar, \pt, \tau_{32}, \beta) = N_{\alpha, i, j, k}$,
by the mistag rate $r(\pt, \tau_{32}, \beta) = r_{i, j, k}$.  Summing
all the predictions along indices $i$, $j$, and $k$ then yields the total
prediction for the bin $\alpha$:
\[
N_\alpha = \sum_{a=1}^{N_\text{jets}} N_{\alpha,i,j,k} \, r_{i,j,k},
\]
where $N_\text{jets}$ is the number of jets in the event.
The four-dimensional parameterization properly accounts for correlated
and uncorrelated statistical uncertainties.
The uncertainty in each
bin of the predicted mistagged distribution $\sigma(m_\alpha)$ has two parts:
one arises from the misidentification probability ($\sigma(r_{i,j,k})$), and the other from the
number of jets in the ensemble $(\sqrt{N_{\alpha,i,j,k}})$ :
\[
\sigma(m_\alpha) = \sqrt{\sum_{a=1}^{N_\text{jets}} \left(
    \left(N_{\alpha,i,j,k} \, \sigma(r_{i,j,k})\right)^2 +
    \left(\sqrt{N_{\alpha,i,j,k}} \, r_{i,j,k} \right)^2  \right) },
\]
The first term accounts for all uncertainties in the $i,j,k$-th bin of the mistag probability.
The second term accounts for statistical uncertainties in the jet ensemble. Both of these
terms are individually added linearly since they are fully correlated within each bin.
The two pieces are added in quadrature since they are fully uncorrelated.

\cleardoublepage \section{The CMS Collaboration \label{app:collab}}\begin{sloppypar}\hyphenpenalty=5000\widowpenalty=500\clubpenalty=5000\textbf{Yerevan Physics Institute,  Yerevan,  Armenia}\\*[0pt]
V.~Khachatryan, A.M.~Sirunyan, A.~Tumasyan
\vskip\cmsinstskip
\textbf{Institut f\"{u}r Hochenergiephysik der OeAW,  Wien,  Austria}\\*[0pt]
W.~Adam, E.~Asilar, T.~Bergauer, J.~Brandstetter, E.~Brondolin, M.~Dragicevic, J.~Er\"{o}, M.~Flechl, M.~Friedl, R.~Fr\"{u}hwirth\cmsAuthorMark{1}, V.M.~Ghete, C.~Hartl, N.~H\"{o}rmann, J.~Hrubec, M.~Jeitler\cmsAuthorMark{1}, V.~Kn\"{u}nz, A.~K\"{o}nig, M.~Krammer\cmsAuthorMark{1}, I.~Kr\"{a}tschmer, D.~Liko, T.~Matsushita, I.~Mikulec, D.~Rabady\cmsAuthorMark{2}, B.~Rahbaran, H.~Rohringer, J.~Schieck\cmsAuthorMark{1}, R.~Sch\"{o}fbeck, J.~Strauss, W.~Treberer-Treberspurg, W.~Waltenberger, C.-E.~Wulz\cmsAuthorMark{1}
% ...
\vskip\cmsinstskip
% ...
1:~~Also at Vienna University of Technology, Vienna, Austria\\
2:~~Also at CERN, European Organization for Nuclear Research, Geneva, Switzerland\\
% ...

\end{sloppypar}
\end{document}